\documentclass[3p,11pt]{elsarticle}
\usepackage{lineno}
\modulolinenumbers[1]
\usepackage{geometry}
\usepackage{graphicx}
\usepackage[usenames,dvipsnames]{color}
\usepackage[english]{babel}
\usepackage{epstopdf}
\usepackage{latexsym}
\usepackage{array}
\usepackage{url}
\usepackage{amstext}
\usepackage{amssymb}
\usepackage{amsmath,amsthm}
\usepackage{pifont}
\usepackage{ulem}
\usepackage[pagebackref=false]{hyperref}
\hypersetup{
    colorlinks=true,
    linkcolor=blue,
    filecolor=magenta,
    urlcolor=blue,
}
\graphicspath{ {Images/} }
\usepackage{times}
\usepackage{wrapfig}
\usepackage{float}
\usepackage{tikz}
\usepackage{caption}
\usepackage{subcaption}

\DeclareMathOperator*{\argmin}{\arg\min}   
\biboptions{sort&compress}

\newcommand{\norm}[1]{\left\lVert #1 \right\rVert}
\usepackage{tabularray}
\usepackage{longtable}
\begin{document}

\begin{frontmatter}

\title{Nonlocal Models in Biology and Life Sciences:\\ Sources, Developments, and Applications}

% \author[inst1]{Swadesh Pal\corref{cor1}}
% \ead{spal@wlu.ca}

\author[inst1]{Swadesh Pal}
\ead{spal@wlu.ca}
\author[inst1,inst2]{Roderick Melnik}
\ead{rmelnik@wlu.ca}

% \cortext[cor1]{Swadesh Pal}
\address[inst1]{MS2 Discovery Interdisciplinary Research Institute, Wilfrid Laurier University, Waterloo, Canada}
\address[inst2]{BCAM - Basque Center for Applied Mathematics, E-48009, Bilbao, Spain}

\begin{abstract}
Mathematical modelling is one of the fundamental techniques for understanding biophysical mechanisms in developmental biology. It helps researchers to analyze complex physiological processes and connects like a bridge between theoretical and experimental observations. Various groups of mathematical models have been studied to analyze these processes, and the nonlocal models are one of them. Nonlocality is important in realistic mathematical models of physical and biological systems when local models fail to capture the essential dynamics and interactions that occur over a range of distances (e.g., cell-cell, cell–tissue adhesions, neural networks, the spread of diseases, intra-specific competition, nanobeams, etc.). This review illustrates different nonlocal mathematical models applied to biology and life sciences. The major focus has been given to sources, developments, and applications of such models. Among other things, a systematic discussion has been provided for the conditions of pattern formations in biological systems of population dynamics. Special attention has also been given to nonlocal interactions on networks, network coupling and integration, including brain dynamics models that provide an important tool to understand neurodegenerative diseases better. In addition, we have discussed nonlocal modelling approaches for cancer stem cells and tumor cells that are widely applied in the cell migration processes, growth, and avascular tumors in any organ. Furthermore, the discussed nonlocal continuum models can go sufficiently smaller scales, including nanotechnology, where classical local models often fail to capture the complexities of nanoscale interactions, applied to build biosensors to sense biomaterial and its concentration. Piezoelectric and other smart materials are among them, and these devices are becoming increasingly important in the digital and physical world that is intrinsically interconnected with biological systems. Additionally, we have reviewed a nonlocal theory of peridynamics, which deals with continuous and discrete media and applies to model the relationship between fracture and healing in cortical bone, tissue growth and shrinkage, and other areas increasingly important in biomedical and bioengineering applications. Finally, we provided a comprehensive summary of emerging trends and highlighted future directions in this rapidly expanding field. 
\end{abstract}


\end{frontmatter}

\section{Introduction}{\label{intro}}

Nonlocal interactions are among the fundamental aspects of many natural processes and phenomena studied in the life sciences, e.g., cell-cell communication, brain function, disease transmission, ecological population dynamics, etc. Mathematical models assist such studies with a rigorous foundation and, together with computational models and techniques, provide additional vital insights, complementing experimental investigations. In 1923, Dr. W. M. Feldman introduced the term ``biomathematics'' in the book title, a highly relevant discipline nowadays \cite{feldman1923}. He was a physician and was more interested in the numerical data of some of his patients' more common dynamics. In 1938, a theoretical physicist Dr. N. Rashevsky, published the first book on mathematical biology and mathematical biophysics entitled ``Mathematical Biophysics: Physico-Mathematical Foundations of Biology'' \cite{rashevsky1938}. Due to many technological advances since then, biomathematics has evolved from Feldman's simple ``mathematical principles for biology students'' to one of the most promising tools for biology and medicine. Mathematical models help break down their dynamic processes into individual elements to study a natural biological process. Moreover, these models can bridge the gap between proposed molecular interactions inside and between cells and their tissue-level effects. As they developed, the hybrid field of ``biostatistics'' emerged and helped to analyze different scientific issues mathematically, e.g.,  biodiversity, agriculture, medicine, etc. Alan Turing also contributed to biomathematics by discovering the condition for particular shapes of each organism, and he is known as ``the founder of contemporary Mathematical Biology''. In addition, his works included three main ingredients: modelling, differential equations, and numerical solutions \cite{amturing}.

Mathematical models are used to understand, predict, and analyze complex biological systems and phenomena by translating biological processes into mathematical terms. They deal with one or more equations (continuous- and fractional-time ordinary and partial differential equations) depending on the component(s) of the system \cite{jdmurray, edelstein2005, tomlin2007, ellis2019}. In addition, one or more parameters may be involved in the model to capture its different characteristics \cite{vandermeer2010, mittelbach2019}. Discrete-time models governed by difference equations are more appropriate for organisms with non-overlapping generations, e.g., an insect population with one generation per year \cite{azuaje2011,azizi2021}. On the other hand, ordinary and partial differential equations are used for continuous-time models and are appropriate for organisms with overlapping generations \cite{cappuccino1995}. In addition, fractional-time models are mainly used to incorporate the memory effect of the organisms in time. In this article, we cover both continuous- and fractional-time models.

The model described by an ordinary differential equation (ODE) queries the state of an entity at a given time; hence, only the time derivatives are present in such systems. For instance, the simplest SIR model in epidemiology helps to predict different things, such as how a disease spreads, the total number infected, and the duration of an epidemic \cite{ross1916,ross1917, kermack1927}. These models also deal with cell signalling pathways, population growth, enzymatic inhibitor reactions, and ecological models. In studying the temporal dynamics of a system, the major concern is shifting one equilibrium to another equilibrium or non-equilibrium (e.g., limit cycle) by a small perturbation around it \cite{jdmurray,jones2009,perko2013}. The temporal dynamics of a model are described by its reaction terms, which mainly include birth, death, and interactions between individuals. Sometimes, a small perturbation may completely change the dynamics of a system, known as butterfly effects in chaos theory \cite{lorenz1963}. Along with the ODE models, fractional-order differential equation (FODE) models also play a crucial role in biological systems. The fractional differentials are interpreted in terms of Riemann-Liouville \cite{miller1993}, Caputo \cite{miller1993}, Caputo-Fabrizio \cite{caputo2015}, and Riesz \cite{cheng2019} fractional derivatives. FODE provides a powerful tool for describing the memory of different substances and the nature of the inheritance, and memory and genetic properties are the two key factors in biological modelling, e.g., membranes of cells have fractional order electrical conductance \cite{cole1933}, cell rheological behaviour \cite{djordjevic2003}. FODE is also studied in the epidemic model of SIR type to understand the memory effect on disease propagation \cite{saeedian2017}. 

The use of spatio-temporal models has increased in the scientific literature, and it has additional benefits over temporal studies \cite{auchincloss2012,yu2018,byun2021}. It can capture localized clusters or heterogeneity of the space-time interactions between the components of a system. The word heterogeneity describes the difference or diversity in a system, which means ``lots of different things''. Such differences can occur in space and time, and in ecology, this term is mainly used to describe the land cover or habitat \cite{fryxell2014}. On the other hand, a homogeneous system is a single-phase system of continuous variables, and each of these variables is distributed uniformly throughout the domain of definition. This homogeneity or heterogeneity depends on the scale considered \cite{salevin}. For example, one heterogeneous habitat can be seen fifty meters from its top, but it can look like a homogeneous habitat from five hundred meters or above. Looking at the blood with a bare eye seems homogeneous, but it has different components like red blood cells, plasma, and platelets under the microscope. In reality, heterogeneity comes when every member has different values for the characteristics of our interests. Well-known reaction-diffusion (RD) equations can capture this type of heterogeneous phenomenon. These equations are conventionally used in chemical interactions to describe the concentrations of the chemicals at different times in a system but are later applied to different fields \cite{page2003pattern,colizza2007reaction,song2019spatial, wuyts2019tropical,kholodenko2003}. The Gierer-Meinhardt model and its relevant modifications capture different biological processes such as animal skin patterns, nerve cells and their activities, organ formation, etc. \cite{meinhardt1982,meinhardt2009}. In chemistry, the well-known CIMA reaction can generate different types of spatial patterns, explained by RD modelling \cite{castets1990,de1991,lengyel1991}. These spatio-temporal models also describe health outcomes, such as contracting a disease, in different locations and spatial points in time \cite{diggle2019}. 

Starting with Fisher's (1937) seminal work on gene spread, researchers have developed various ways of modelling spatial interactions in biology. This manuscript focuses on different nonlocal models based on individual interactions, which can help biologists capture essential features of spatial phenomena. The concept of nonlocality exists everywhere, and it is essential in realistic mathematical models of many aspects of the physical world \cite{silling2019, lohrey2020}. These models are useful for capturing spatially distributed phenomena in biological systems, such as movement, diffusion, or interaction among organisms, cells, or chemicals not confined to adjacent interactions about their positions. Several key factors are involved when analyzing nonlocal models in biology. For instance, the strength of interactions between two points is determined by how interactions occur over a distance as a function of the distance between them. Many biological systems have boundaries that affect behaviour at a distance, such as skin or organ membranes, influencing diffusion and interaction \cite{krause2021}. In epidemiology, nonlocal models are employed to simulate the transmission of diseases over extensive geographical areas, where persons from far locations may come into contact, hence spreading illnesses through travel or migration \cite{chen2021}. Neuronal nonlocal models in neuroscience represent the interactions between physically distant neurons yet functionally linked by long-range synaptic connections \cite{thompson2020,pal2022sr}. In an ecosystem, organisms compete for resources to survive, grow, and reproduce. Animals compete for food, shelter, water, and space, while plants compete for resources, including air, water, sunlight, and space. Therefore, in ecology, there is no real justification for assuming the interactions as local since (i) the population always competes for shared resources, (ii) the individuals in a population communicate either visually or through chemical means \cite{furter1989}. In addition, the species sense their surroundings for making different decisions, e.g., the predator finds the prey or the prey avoids the predator \cite{eftimie2007, levine2000, saha2024a, saha2025}. High population density at a spatial location impacts the same spatial point and neighbouring locations, so considering the impact of long-range effects is important. Many particles sense signals over extended regions, e.g., filopodia in cells detect signals at multiple cell diameters away \cite{chen2020, kornberg2014}. Along with these factors, dispersal distances may also be nonlocal, e.g., seeds can be transported significantly from the source. The spatial convolution is more realistic than the usual diffusion for many ecological populations \cite{sherratt2016}. Different methods have been used for kernel estimation, including dispersive individuals \cite{nathan2012}.

The usual RD model in interacting species can capture only the spatial interactions of neighbourhood locations, generally called short-range interactions. On the other hand, the nonlocal term incorporates the behaviour of a distribution function at several points, and a system of integro-differential equations describes it. Generally, these nonlocal models deal with long-range interactions in the interacting populations. Along with these nonlocal RD models, nonlocal advection-RD models have also gained more attention in recent years. These models can include cell-cell adhesion interactions for tissue dynamics \cite{chen2020, painter2024}. Nonlocal mathematical models are also used to investigate the progress of an epidemic, including age-structured models \cite{kendall1965,kang2021}. Furthermore, a system of integro-differential equations is used to describe the general nonlocal continuum theory, which helps to make bridge-base sensors and is applied to biological sciences \cite{eringen2003, lavrik2004, ekinci2005, lu2006}. In addition, the spatial dispersal mechanism of cells or organisms is one of the central topics in biology and ecology. Random dispersal has been used to describe the movement of organisms between adjacent spatial locations. However, for some organisms, the movements can occur between non-adjacent spatial locations, and the nonlocal dispersal captures such behaviours \cite{hutson2003}.

Multiscale modelling approaches of materials can capture one or more physical phenomena at different lengths and time scales \cite{nieminen2002,shaat2020}. Materials generally exhibit and secrete different phenomena at different lengths and time scales \cite{shaat2020}. Due to the lack of measures and length scales in classical mechanics, the material's microstructural features (e.g., micro-slipping and twinning), geometries, and topologies (e.g., band-gap) are unpredictable. Hierarchical and hybrid are the two types of multiscale modelling approaches studied in the literature. In the case of the hierarchical multiscale model, the material is broken down into multiple scales, and models at the different scales are run independently. In the end, coupling parameters are introduced to couple those scales. On the other hand, all different scales work together as a single package for hybrid multiscale models. The material consists of interconnected multiple subdomains over short and long ranges, and the nonlocal continuum mechanics model can capture their mechanics. 

Nonlocal models are capable of capturing some behaviour of biological materials such as cortical bone fracture \cite{deng2009}, the rupture in lipid membranes \cite{taylor2016}, tumor shrinkage \cite{lejeune2021}, etc. Peridynamics is one of the main theoretical and computational frameworks that unify the mechanics of discrete and continuous media \cite{lejeune2021}. Since the first papers were published, peridynamics methodology has been used to model discontinuous and long-range forces using pair-wise interactions between particles \cite{silling2000}. This methodology is nonlocal because it considers the long-range interaction between particles. As it deals with continuous and discrete media, it is a compelling model of biological tissue on the cell population scale. In addition, piezoelectric and other smart materials are the most commonly used medical materials. For instance, ultrasound scanners use a piezoelectric transducer; they use the converse piezoelectric effect to send out sound vibrations reflected by muscles, organs, etc. \cite{manjon2018}. 

The rest of the paper is organized as follows. Different types of local and nonlocal models are described in Sect \ref{SE2}. We start with local RD systems applied to ecological and chemical interactions. In particular, how the nonlocal Fisher-KPP equation comes in a single species population is discussed. Then, the nonlocal models capture different kinds of nonlocal phenomena. For instance, nonlocal interactions in space-time, multi-body systems, cell biology, multi-stability, ecology, peridynamics, etc. The applications of these nonlocal models are discussed in Sect \ref{SE3}. Different emerging fields and some possible extensions of the nonlocal models are discussed in Sect \ref{SE4}, followed by the conclusions in Sect \ref{SE5}.

\section{Life Sciences as a Rich Source of Nonlocal Phenomena and Processes: Mathematical and Computational Models}{\label{SE2}}

Mathematical and computational models play a pivotal role in unravelling the intricate and dynamic processes within the life sciences. These models provide a systematic framework to analyze, simulate, and predict biological phenomena, aiding researchers and practitioners in understanding the underlying principles governing life at various levels of complexity. Life sciences encompass various disciplines, from molecular biology and genetics to ecology and neuroscience. Understanding the behaviour and interactions of biological systems requires integrating data and theories into mathematical and computational frameworks. These models enable the abstraction of biological processes into mathematical equations, algorithms, or simulations, allowing for a quantitative and often more precise representation of biological phenomena.

\subsection{Homogeneous and heterogeneous systems}

Homogeneous and heterogeneous systems are fundamental concepts in mathematical biology that play crucial roles in understanding various biological phenomena. Homogeneous systems are the baseline for understanding the fundamental dynamics of biological systems before incorporating more complex factors. For instance, patterns are observed in nature and have visible regularities. These patterns recur in different contexts and can sometimes be modelled mathematically. According to Turing \cite{amturing}, ``when two chemical species with different diffusion rates react with each other, the spatially homogeneous state may become unstable, thereby leading to a nontrivial spatial structure". This looks counter-intuitive as it is expected to be a uniform distribution of the chemical substances through the diffusion term. Diffusion is the process by which particles spread from high-concentration regions to low-concentration regions. Under most straightforward assumptions, Fick’s law guides this physical process of random motions for molecules. However, in some applications, other physical laws that account for relaxation processes may be required (e.g., \cite{singh2024}). In general, the diffusion term in RD equations spreads the concentrations throughout the spatial domain, and the linear or nonlinear reaction terms come from the mass action laws \cite{jdmurray,volpert2009}. Without loss of generality, we can consider a general form of RD equations with $m$-substances $\mathbf{u} (\mathbf{x},t) = (u_{1}(\mathbf{x},t),\ldots,u_{m}(\mathbf{x},t))^{T}$ at time $t>0$, and at $\mathbf{x}$ in a spatial domain $\Omega \subseteq \mathbb{R}^{n}$ as \cite{jdmurray}:
\begin{equation}{\label{GRDE}}
        \frac{\partial \mathbf{u}}{\partial t} = \nabla\cdot(\mathbf{D}\nabla \mathbf{u}) + \mathbf{f}(\mathbf{u}),
\end{equation}
with non-negative initial conditions $\mathbf{u}(\mathbf{x},0) = \mathbf{u}_{0}(\mathbf{x})$ and appropriate boundary conditions, e.g., Dirichlet, no-flux, periodic, etc. Different types of boundary conditions have been used in RD equations, depending on the nature of the problems; no-flux and periodic boundary conditions are among the most familiar. Here, $\mathbf{D}$ is a diagonal matrix called the matrix of diffusivity. The term $\nabla\cdot(\mathbf{D}\nabla \mathbf{u})$ is the self-diffusion of the substances, taking care of the rate of change of the concentrations at a particular point in the spatial domain. The matrix $\mathbf{D}$ is not a diagonal for the case of cross-diffusion. The initial conditions for all the species are assumed to be non-negative since the concentration is always non-negative. The vector-valued function $\mathbf{f}(\mathbf{u})$ represents the reactions between the species in the spatial domain $\Omega$. These reactions could be the variables' source, sinks, and interactions. Therefore, modelling heterogeneous systems is essential for understanding the impact of individual variability on population dynamics, disease spread, and other biological processes. 

Depending on the parameters and interactions between organisms, the RD equation (\ref{GRDE}) exhibits different types of dynamic non-homogeneous solutions, such as travelling waves, wave-trains, spatio-temporal chaos, etc. In addition, reaction-diffusion systems can produce non-homogeneous stationary patterns such as spots, stripes, and target patterns \cite{jdmurray}. In 1937, Fisher and Kolmogorov, Petrovskii, and Piskunov introduced the scalar RD equation \cite{kolmogorov1937}:
\begin{equation}{\label{SRDE}}
        \frac{\partial u}{\partial t} = d\frac{\partial^{2} u}{\partial x^{2}} + g(u),
\end{equation}
and studied the existence of the travelling wave solutions along with the propagation of the wave speed. The equation (\ref{SRDE}) is known as the Fisher-KPP equation. Researchers have been studying and applying this Fisher-KPP model in various fields, e.g., physics, chemistry, biology, and medicine. A particular nonlinear form $g(u) = u(1-u)$ of the RD equation (\ref{SRDE}) has been taken a lot of attention in capturing different phenomena through mathematical modelling, e.g., spatial spreading of cell populations \cite{simpson2013}, wound-healing cell migration \cite{sherratt1990, maini2004}, protein misfolding \cite{fornari2019, schafer2021}, biological invasion \cite{petrovskii2005, skellam1951, levin2003}, etc.

As discussed earlier, the RD equations (\ref{GRDE}) are capable of producing different types of stationary and non-stationary non-homogeneous patterns. Several mechanisms are available for getting non-stationary patterns \cite{jdmurray}. Turing suggested a mechanism to obtain some of the non-homogeneous stationary patterns mathematically, called the Turing pattern \cite{amturing}. One of the main conditions for existing Turing patterns requires at least two interacting populations for the RD equations. Therefore, the usual Turing mechanism can not be applied to the single-species local RD model (\ref{SRDE}). However, it can be applied to a single species RD model with nonlocal interactions. We first describe a nonlocal version of the Fisher-KPP RD model for a single species population (\ref{SRDE}), then move to the multi-species populations and other nonlocal interactions.

\subsection{Nonlocal interactions in a single species population}

Without loss of our generality, we assume that $u$ is a dimensionless population density, and it is defined in the one-dimensional spatial domain $\Omega = [-L, L]\subset \mathbb{R}$, $L>0$. In the considered case $g(u) = u(1-u)$, the population has the reproduction term, which is proportional to its density and available resources $(1-u)$ \cite{volpert2009}. For the local interaction, the population's consumption rate at any given spatial point $x\in\Omega$ is proportional to its population's density at the same spatial point with the total amount of available resources unity. In real-world systems, individuals do not always interact only with their immediate neighbours; instead, they may interact over longer distances due to long-range communication or interactions (e.g., chemical signalling in biological systems \cite{vasilopoulos2012}), long-range dispersal (e.g., seeds blown by the wind or animals migrating over long distances \cite{viana2016}), long-range delay \cite{li2024a}, and nonlocal competition or cooperation between species in ecology \cite{furter1989, simoy2023, banerjee2022review}. In all these cases, the individual located at a spatial point $x$ interacts with others in its neighbourhood locations \cite{furter1989, banerjee2022review, billingham2020, pal2018bmb, tian2021, banerjee2017ec}. We assume that the individuals located at the spatial point $y$ access the resources of the point $x$ with the probability density function $\phi (x-y)$. In this case, the RD equation (\ref{SRDE}) modified into an integro-differential equation 
\begin{equation}{\label{LIDE}}
        \frac{\partial u}{\partial t} = d\frac{\partial^{2} u}{\partial x^{2}} + u( 1- \phi\ast u),
\end{equation}
with non-negative initial and periodic boundary conditions over the domain $[-L, L]$. Here $\phi\ast u$ is the convolution of $u$, and we use $`\ast$' as the convolution operator for the rest of this manuscript. The convolution $\phi\ast u$ in one spatial dimension is defined by 
$$(\phi\ast u)(x,t) = \int_{-\infty}^{\infty} \phi (x-y)u(y,t)dy.$$

In the RD equation (\ref{SRDE}), the diffusion term contributes to the net density change in the spatial locations and is much slower than the involvement of the consumption resources. But in the nonlocal model (\ref{LIDE}), the convolution term captures the long-range interaction in the spatial domain, one of the key factors missing in the local model. The kernel function $\phi$ satisfies the normality condition to share the same homogeneous steady-state solution corresponding to the local model, and the condition is
\begin{equation*}{\label{NCKF}}
    \int_{-\infty}^{\infty}\phi(y)dy = 1.
\end{equation*}
This kernel function represents how individuals at position $y$ influence those at position $x$. For instance, if the kernel is Gaussian, it represents the strongest interactions at close distances but tapers off smoothly. In physics, it defines systems in which individuals can influence others locally and globally through long-range interactions. In addition, the RD equation (\ref{SRDE}) fails to incorporate such long-range interactions \cite{banerjee2017ec}. The nonlocal model (\ref{LIDE}) has two homogeneous stationary solutions $u=0$ and $u=1$. In the case of the local model, the solution $u=1$ is stable under heterogeneous perturbations, but it may not be stable for the nonlocal model \cite{volpert2009}. A brief analysis of this single species nonlocal model is given in the next section. 

\subsection{Nonlocal interactions in time and space}

Researchers have studied nonlocal RD models for the triangular, parabolic kernel functions to account for different biological phenomena \cite{dornelas2021, pal2019mbe, segal2013, pal2020amm, sherratt2016, merchant2011, pal2019ijbc, bian2017, kavallaris2023}. These kernel functions are known as the truncated kernel. Such non-truncated kernel functions have been the subject of intensive studies. For instance, Britton introduced an advanced version of the single species model (\ref{LIDE}) as:
\begin{equation}{\label{BSPM}}
        \frac{\partial u}{\partial t} = d\Delta u + u( 1+\alpha u -(1+\alpha)\psi\ast u),
\end{equation}
where $\alpha u$ ($\alpha>0$) represents the local aggregation and $-(1+\alpha)\psi\ast u$ is the disadvantage in the global population as the resources may go into depletion of being high population density  \cite{britton1989}. The author studied the nonlocal model with the Laplacian kernel function \cite{sherratt2016}. Sometimes, species take time to move; in this case, the correct average population density is a spatio-temporal weighted average toward the current time and position. Taking this spatio-temporal factor into account, Britton introduced a space-time nonlocality in the single species model (\ref{BSPM}) as \cite{britton1990}:
\begin{equation}{\label{NFBNM}}
        \frac{\partial u}{\partial t} = d\Delta u + u( 1+\alpha u -(1+\alpha)\psi\ast\ast~u),
\end{equation}
where the double asterisk represents the convolution in space and time and is given by $$(\psi\ast\ast~u)(\mathbf{x},t) = \int_{-\infty}^{t}\int_{\Omega} \psi (\mathbf{x}-\mathbf{y},t-s)u(\mathbf{y},t)d\mathbf{y}ds.$$
Likewise, the spatial kernel function $\phi$ in (\ref{LIDE}), the spatio-temporal kernel function $\psi$ satisfies some conditions mentioned in \cite{britton1990,pal2020mbe}. In some cases, these types of nonlocal models with non-truncated kernels can be handled by converting them into coupled RD equations \cite{britton1989,britton1990}. However, in many other cases, this may not be possible. Indeed, nonlocal time-dependent interaction occurs frequently in physics and engineering, with nonlocal time affecting the dynamics of open quantum systems \cite{laine2012, tarasov2021}, stress depending on the current strain and its history \cite{li2006}, along with many other situations where nonlocality has to be accounted for.

\subsection{Nonlocal interactions in cell biology and genomics}

In cell biology, nonlocal interactions are interactions between cellular components that extend beyond immediate neighbours and involve long-distance signalling or communication. For instance, mechanical forces inside tissues influence cell migration and division over considerable distances from the point of force application, as tension and compression are conveyed through the cytoskeleton and extracellular matrix \cite{chanet2014}. These forces travel like waves or through elastic deformation. The transmission of forces follows principles of elasticity, where cells act like soft materials that deform under force. Additionally, tumors are abnormal tissue masses of cells with different phenotypes, such as cancer stem cells \cite{al2003, bonnet1997, dick2003}, and they are heterogeneously distributed in the body. Therefore, nonlocal interactions are crucial in cellular processes and behaviours such as cell migration, pattern formation, and collective cell responses. Motivated by the paradoxical findings of an agent-based model \cite{enderling2009}, Hillen et al. have proposed a mathematical model \cite{hillen2013}:
\begin{equation}{\label{NLCM}}
    \begin{aligned}
         \frac{\partial u (x,t)}{\partial t} &= D_{u}\Delta u +\delta\gamma\int_{\Omega}\psi (x,y,p(x,t))u(y,t)dy,\\
          \frac{\partial v(x,t)}{\partial t} &= D_{v}\Delta v + (1-\delta)\gamma \int_{\Omega}\psi (x,y,p(x,t))u(y,t)dy -\alpha v +\rho \int_{\Omega}\psi (x,y,p(x,t))v(y,t)dy,
    \end{aligned}
\end{equation}
where $u(x,t)$ and $v(x,t)$ are the densities of cancer stem cells and non-stem cancer cells at the location $x$ and time $t$. Here, $p(x,t) = u(x,t)+v(x,t)$ is the total tumour density, and $\gamma$ and $\rho$ are the number of cell cycles times per unit time of cancer stem cells and non-stem cancer cells, respectively. The parameter $\delta$ describes the fraction of cancer stem cell divisions, and $\alpha$ is the tumour cell death rate. $D_{u}$ and $D_{v}$ are the usual diffusion coefficients of cancer stem cells and non-stem cancer cells, respectively. The kernel function $\psi (x,y,p)$ describes the rate of progeny contribution to the location $x$ from a cell at location $y$, per ``cell cycle time'' and satisfies the normality condition
$$\int_{\Omega}\psi (x,y,p(x,t))dx \leq 1.$$

Gene silencing and RNA interference are exciting fields of study that have promising implications in health and industry \cite{yang2007}. For instance, a synthetic RNA molecule is designed to selectively target and mute the production of one particular gene in small interfering RNA technology. When it enters cells, it attaches itself to the target mRNA, causing it to be degraded and stopping the creation of new proteins. There are some more key technologies and models associated with gene silencing and RNA interference: short hairpin RNA \cite{lambeth2013}, micro RNA \cite{macfarlane2010}, CRISPR/Cas9 technology \cite{redman2016}, Antisense Oligonucleotides \cite{rinaldi2018}, Dicer Knockout models \cite{guo2020}, etc. It is essential to model genetic switches and networks to comprehend the complex regulatory processes within cells \cite{shomar2020}. The dynamics of genetic switches and networks have been simulated and examined using various computational and mathematical models. Some common approaches are Boolean networks, ordinary differential equations, stochastic models, agent-based models, and machine learning approaches. Furthermore, nonlocal models are important beyond DNA, RNA, and protein modelling. The use of microarray technology for placing the full human genome on chips is one such application.

A multi-scale modelling technique is required to represent a cell as a composite structure incorporating microtubules and organelles. Several models have been created to simulate these components' interactions and dynamics, such as continuum mechanics models \cite{singh2020, singh2020iw, singh2020ic}. Understanding the methods by which a cell develops, duplicates its components, and divides into two daughter cells is fundamental to studying cell cycles. Modelling cell cycles with fractional derivatives presents a mathematical technique reflecting cellular dynamics' nonlocal and memory-dependent character. In addition, rheology and hysteresis are more frequently linked with materials and physical systems, and they may also be used to describe some aspects of cellular activity, including cell mechanics and cycles \cite{melnik2009}. Understanding the interaction of grain shape, material characteristics, and nonlocal rheology is critical for forecasting and regulating the behaviour of dense granular flows \cite{fazelpour2022}.

Different types of biological rhythms occur in various physiological and behavioural processes in living organisms, and they are essential for maintaining homeostasis and optimizing physiological functions in organisms. In addition, angiogenesis is the physiological process that results in forming new blood vessels from pre-existing vessels \cite{carmeliet2005}. This complex mechanism is required for a variety of physiological activities, including embryonic development and wound healing. However, angiogenesis has been linked to several pathological disorders, including cancer and inflammatory illnesses. Furthermore, biological cells are complex structures with distinct rheological properties, and understanding their mechanical responses is critical for a variety of applications, including disease diagnostics, medication delivery, and tissue creation. Studying these rhythms provides insights into how living organisms adapt to their environments and regulate various processes over time.

\subsection{Nonlocal dispersal}

Nonlocal dispersal in biology refers to the movement or migration of individuals, cells, or organisms over extended distances, often beyond their immediate vicinity. This phenomenon plays a crucial role in shaping populations' distribution, dynamics, and persistence in various ecological and biological systems. Species expand their range of interactions through dispersal, which has many forms. Till now, we have considered the simplest type of dispersal as random diffusion, which captures the movement of organisms between adjacent spatial locations. A different type of dispersal that gives rise to nonlocality is studied in the literature, where organisms can travel some distance \cite{li2014}. In \cite{sherratt2016}, the author considered the Rosenzweig-MacArthur model with nonlocal dispersal as:
\begin{equation}{\label{DKRM}}
    \begin{aligned}
         \frac{\partial u }{\partial t} &= \mathcal{P}u + u(1-u)-\frac{\alpha uv}{1+\alpha u},\\
          \frac{\partial v}{\partial t} &= \mathcal{P}v + \frac{\alpha uv}{\beta (1+\alpha u)}-\frac{v}{\beta \gamma},
    \end{aligned}
\end{equation}
where $$\mathcal{P}u = \frac{1}{l}\int_{-\infty}^{\infty}\phi \bigg{(}\frac{x-y}{l}\bigg{)}u(y)dy - u(x)$$
denotes the nonlocal dispersal corresponding to $u$ at a spatial point $x$ with $l$ as the spatial scale representing the dispersal distance \cite{li2014}; $u(x,t)$ and $v(x,t)$ are the population densities of the prey and predator, respectively, at the spatial point $x$ and at time $t>0$. This type of nonlocal model is the derivative-free integral equation for diffusion \cite{khodabakhshi2021}. To study the oscillatory dynamics of the model, the author converted this model into normal form near the Hopf bifurcation using the complex Ginzburg-Landau equation \cite{garcia2012}. He estimated the wavelength and amplitude of the periodic travelling wave generated by the invasion of the coexisting steady-state for the parameter values close to the Hopf bifurcation threshold. Assuming specific properties for kernel function $\phi$ (non-negativity, being even, and normality), the nonlocal dispersal operator can be approximated for small $l$ as \cite{li2014}:
$$\mathcal{P}u = \frac{l^{2}}{2}\bigg{(}\int_{-\infty}^{\infty}\phi (z)z^{2}dz\bigg{)}\frac{\partial^{2}u}{\partial x^{2}}+O(l^{3}).$$
On the other hand, for $l\rightarrow \infty$ and for a periodic $u$ with periodicity $L$, the operator $\mathcal{P}$ can be expressed as \cite{li2014}:
$$\mathcal{P}u =  \frac{1}{L}\int_{0}^{L}u(y)dy - u(x).$$
In conventional RD models, particle flux is typically determined by local concentration gradients (Fick's laws). However, the nonlocal dispersal is a generalization of random walks where the particles can jump to further locations instead of stepping to nearby sites \cite{kao2010}. This type of dispersal can be facilitated by various natural phenomena, including wind and water currents. For instance, seeds may be blown across great distances by the wind or migratory birds. 

\subsection{Fractional calculus and nonlocal models}

Fractional nonlocal models have gained significant attention in biology due to their ability to capture the memory effect, anomalous transport, and long-range interactions, which are common in biological systems. Fractional calculus has been employed to study drug release from polymeric matrices used in drug delivery systems. These models account for drug diffusion's nonlocal and memory effects within the polymer, providing a more accurate representation of drug release kinetics. Different nonlocal RD models have been used in biology and other fields. Researchers have used fractional nonlocal models to understand cancer cell populations' dynamics in cancer diseases. Before going to the fractional model, we first consider a modified nonlocal RD model (\ref{LIDE}) for nonlocal consumption of resources as \cite{zhan2022, bian2017, kavallaris2023}:
\begin{equation}{\label{MLIDE}}
    \frac{\partial u}{\partial t} = d\Delta u +au^{m}(1-b\phi\ast u^{n})-cu.
\end{equation}
The parameters $m$ and $n$ are positive integers, $m=1, n=1$ and $m=2, n=1$ correspond to the asexual and sexual reproductions, respectively. This model is an advanced version of the nonlocal model kinds of (\ref{LIDE}). Here, $a$, $b$ and $c$ are the parameters, and $\phi\ast u^{n}$ is the convolution. The last term in (\ref{MLIDE}) represents the mortality with a rate constant $c$. In cancer disease, abnormal changes in cell density cause significant morbidity and mortality. Numerous mathematical models in the literature account for this disease propagation \cite{alarcon2003}. The fractional RD model is one of them, and many researchers have been focused on it. The time-fractional RD equation is used in the mathematical model to incorporate such an evolution process, and it is given by \cite{zhan2022}:
\begin{equation}{\label{TFCE}}
    D^{\alpha}u = d\Delta u +au^{2}(1-b\phi\ast u)-cu,
\end{equation}
where $D^{\alpha}u$ is the Caputo fractional derivative, and it is defined by
\begin{equation*}
    D^{\alpha}u(\mathbf{x},t) = \frac{1}{\Gamma (1-\alpha)}\int_{0}^{t}\frac{u'(\mathbf{x},s)}{(t-s)^{\alpha}}ds,~~ 0<\alpha<1.
\end{equation*}
Here, $u'(\mathbf{x},s)$ is the first-order partial derivative of $u(\mathbf{x},t)$ with respect to $t$ and evaluated at $t=s$. This type of time-fractional model incorporates memory effects where the current state of a system depends not only on its local conditions but also on its historical states \cite{pal2024a, zhou2025}. In addition, fractional nonlocal models are used to describe anomalous diffusion, where the spread of particles does not follow classical Brownian motion \cite{du2023}. The key feature of anomalous diffusion is that the mean square displacement scales as a nonlinear power law in time $\left(\left< \delta x^{2}\right> =  t^{\beta}\right)$. Superdiffusion refers to faster than linear scaling, i.e., $\beta>1$, while subdiffusion refers to slower than linear scaling, i.e., $\beta<1$ \cite{li2019}.

\subsection{Nonlocal epidemiological models and nonlocality in immunology}

Nonlocal epidemiological models have been used to study the spread of infectious diseases in biological systems, considering interactions beyond immediate neighbours and capturing the effects of long-range dispersal and spatial dependencies. In 1927, Kermack and McKendrick first proposed a deterministic SIR model to capture a transmitted viral or bacterial agent in a closed population \cite{kermack1927,ruan2007}. Let $S(t)$ be the density of a population susceptible to a disease but not yet infected at the time $t\geq 0$. Suppose $A(\theta)$ is the expected infectivity of an individual infected $\theta$ time ago. Then, the Kermack-McKendrick model follows the integral differential equation \cite{kermack1927}:
\begin{equation}{\label{KMIDE}}
    \frac{dS}{dt} = S(t)\int_{0}^{\infty} A(\theta)\frac{dS}{dt}(t-\theta)d\theta.
\end{equation}
Based on the linearization of the integral differential equation (\ref{KMIDE}), Kermack and McKendrick derived invasion criteria for which the disease invades individuals. To derive those criteria, we suppose that the whole population is susceptible at the epidemic's beginning and $S(0) = S_{0}$. Also, we assume that the linearized solution at $S_{0}$ is of the form $c_{0}\exp (rt)$, and in this case, the characteristic equation is given by
\begin{equation*}
     1 = S_{0}\int_{0}^{\infty} A(\theta)e^{r\theta}d\theta,
\end{equation*}
and the number $$R_{0} = S_{0}\int_{0}^{\infty} A(\theta)d\theta $$ describes the epidemic's growth in the initial phase of the disease propagation. The invasion criteria of the epidemic is $R_{0}$, which is greater than one. For the special choice $A(\theta)=\beta\exp (-\gamma \theta)$, where $\beta,\gamma>0$, and the quantity $$I(t) = -\frac{1}{\beta}\int_{0}^{\infty} A(\theta)\frac{dS}{dt}(t-\theta)d\theta$$
represents the number of infected individuals at the time $t$. If $R(t)$ is the number of individuals who recovered from the infection and may be infected again, then we can simplify all these equations in the following ODE system
\begin{equation}{\label{SIRM}}
    \begin{aligned}
         \frac{dS}{dt} &= -\beta S(t)I(t),\\
         \frac{dI}{dt} &= \beta S(t)I(t) -\gamma I(t), \\
         \frac{dR}{dt} &= \gamma I(t),
    \end{aligned}
\end{equation}
with non-negative initial conditions. This dynamical system provides changes in density among susceptible, infected and recovered compartments over time. Here, the individuals can be treated as the particles in a system where their interactions lead to collective behaviour, e.g., the spread of disease. In 1965, Kendall extended this SIR model (\ref{SIRM}) into a nonlocal epidemic model \cite{kendall1965}. Let $S(x,t)$, $I(x,t)$, and $R(x,t)$ be the densities of the susceptible, infected and removed individuals at the spatial location $x\in \mathbb{R}$ and time $t\geq 0$ such that their sum $S+I+R$ remains time-independent. We assume that the infected individuals are infectious and that the rate of infection is $$\beta\int_{-\infty}^{\infty}I(y,t)\Psi (x-y)dy,$$ where $\beta$ is a positive constant and $\Psi$ is the normalized kernel function in $\mathbb{R}$. This mechanism considers that the infection rate at a location is influenced by contributions from distant regions, and the kernel defines the interaction radius or probability of long-range infection spread. This function takes various forms, such as exponential decay, where the influence of an infected individual decreases with distance. Taking these into the SIR model, the modified model becomes
\begin{equation}{\label{SIRM1}}
    \begin{aligned}
         \frac{\partial S}{\partial t} &= -\beta S(x,t)\int_{-\infty}^{\infty}I(y,t)\Psi (x-y)dy,\\
         \frac{\partial I}{\partial t} &= \beta S(x,t)\int_{-\infty}^{\infty}I(y,t)\Psi (x-y)dy -\gamma I(x,t), \\
         \frac{\partial R}{\partial t} &= \gamma I(x,t).
    \end{aligned}
\end{equation}
In classical RD models, the infections spread locally and gradually to adjacent regions. However, the nonlocal interaction leads to wave-like infection propagation due to the fact that distant regions can become infected simultaneously or even ahead of nearby regions, depending on the strength and range of nonlocal interactions (e.g., travel networks). 

Nonlocal epidemiological models have been applied in biology to study the spread of infectious diseases in spatially extended and heterogeneous environments, such as urban areas with varying population densities or regions with different levels of healthcare access \cite{chen2021, chang2023, wang2024, li2024b}. These models incorporate long-range interactions and spatial dependencies, allowing for a more realistic representation of disease transmission dynamics. The COVID-19 pandemic was a global health crisis caused by the SARS-CoV-2 virus in a heterogeneous environment \cite{sohrabi2020}. This disease spreads rapidly through respiratory droplets when an infected person talks, coughs, or sneezes, and new variants of the virus continuously emerge, which causes concern among health experts worldwide. Different types of measurements have been implemented to control the transmission of the virus, including targeted lockdowns, ramped-up testing, and widespread vaccination efforts. In terms of mathematical modelling, numerous mathematical frameworks have been proposed to investigate the transmission dynamics of the COVID-19 pandemic \cite{biswas2020,shi2023,xu2023}, and we highlight one of them where the spatial domain is considered. The novel coronavirus infects contacts during cold-chain transmission, so studying the characteristics of COVID-19 dynamics can be more accurate in considering long-range population diffusion. Suppose $S(x,t)$, $V(x,t)$, $E(x,t)$, $I_{a}(x,t)$, and $I_{s}(x,t)$ are respectively for the populations of susceptible, fully vaccinated, latent infection, asymptomatic infected, and symptomatic infected at a point $x\in\mathbb{R}$ and time $t>0$, with $B(x,t)$ as the SARS-CoV-2 virus in the environment. Following \cite{shi2023}, we consider the spatio-temporal nonlocal mathematical model for COVID-19 as:
\begin{equation}{\label{NMC19}}
    \begin{aligned}
         \frac{\partial S(x,t)}{\partial t} &= d_{s} \mathcal{D}[S(x,t)] - (\mu (x) +\eta (x))S(x,t) +\Lambda(x)-S(x,t)f_{s}(x,t),\\
         \frac{\partial V(x,t)}{\partial t} &= d_{v} \mathcal{D}[V(x,t)] + \eta(x)S(x,t) -(\mu(x)+p(x))V(x,t) -V(x,t)f_{v}(x,t),\\
         \frac{\partial E(x,t)}{\partial t} &= d_{e} \mathcal{D}[E(x,t)] + V(x,t)f_{v}(x,t) + S(x,t)f_{s}(x,t) -(\mu(x)+\gamma(x))E(x,t),\\
         \frac{\partial I_{a}(x,t)}{\partial t} &= d_{a} \mathcal{D}[I_{a}(x,t)] + \epsilon(x)\gamma(x)E(x,t) - (\mu(x)+r_{a}(x)+\zeta_{a}(x))I_{a}(x,t),\\
         \frac{\partial I_{s}(x,t)}{\partial t} &= d_{s}\mathcal{D}[I_{s}(x,t)] + (1-\epsilon(x))\gamma(x)E(x,t) - (\mu(x)+r_{s}(x)+\zeta_{s}(x))I_{s}(x,t),\\
         \frac{\partial B(x,t)}{\partial t} &= d_{b} \mathcal{D}[B(x,t)] + \delta_{e}(x)E(x,t)+\delta_{a}(x)I_{a}(x,t)+\delta_{s}(x)I_{s}(x,t) -c_{s}(x)B(x,t),
    \end{aligned}
\end{equation}
with non-negative initial conditions and Neumann boundary conditions. Here, $\mathcal{D}$ is the dispersal convolution operator defined by $$\mathcal{D}[u(x,t)] = \int_{\Omega}\psi(x-y)[u(y)-u(x)]dy,$$
where $\psi(x-y)$ is the probability that an individual at position $y$ moves to position $x$ in the space $\Omega$. This type of nonlocal diffusion accounts for jump-like interactions, where infection can jump across large distances (such as through transportation networks or migration). The functions $f_{s}(x,t)$ and $f_{v}(x,t)$ are defined as 
$$f_{s}(x,t) = \int_{\Omega}[\beta_{e}(x,y)E(x,t)+\beta_{a}(x,y)I_{a}(x,t)+\beta_{s}(x,y)I_{s}(x,t)+\beta_{b}(x,y)B(x,t)]dy,$$
$$f_{v}(x,t) = \int_{\Omega}[\alpha_{e}(x,y)E(x,t)+\alpha_{a}(x,y)I_{a}(x,t)+\alpha_{s}(x,y)I_{s}(x,t)+\alpha_{b}(x,y)B(x,t)]dy,$$
where $\beta_{k}(k = e, a, s)$ are the rate of infection of infected people to susceptible people and $\alpha_{k}(k = e, a, s)$ are the rate of infection of infected people to those vaccinated, and $\beta_{b}$ and $\alpha_{b}$ are the rate of infection of the virus in the population. The other parameters used in the model (\ref{NMC19}) are given in \cite{shi2023}. The authors have shown that the effective vaccination rate is one of the best measures to prevent and control the transmission of COVID-19 under current medical conditions.

Nonlocality in immunology refers to the phenomena in which immune system interactions and signalling processes extend beyond the immediate area of the immune cells involved. T-cell receptor (TCR) activation, for example, refers to the assumption that T-cell responses are not limited to the local contact between the TCR and its antigen. T-cell activation, on the other hand, is a networked and dynamic process that includes interactions at remote places \cite{smith2009}. It can lead to systemic effects, influencing immune responses and inflammation throughout the body, which has implications for autoimmune diseases and chronic inflammatory conditions \cite{bachmann2010}. Furthermore, nonlocal TCR activation is important in cancer immunotherapy because activated T cells may target cancer cells in distant metastatic locations. \cite{rosenberg2014}. In T memory cell homeostasis, for example, the bone marrow serves as a niche and interacts with stromal cells, and cytokines contribute to their survival and nonlocal homeostasis \cite{tokoyoda2009}. These T cells grow and contract in distinct stages, including interactions and signals that reach beyond the T cells' local surroundings. During T cell activation and proliferation, cytokines such as interleukins are generated not just locally at the site of activation but also systemically. These cytokines assist in the nonlocal regulation of T-cell proliferation \cite{boyman2012}.

\subsection{Relaxation-time models in life sciences}

Relaxation-time models are pretty interesting, especially in the context of life sciences. These models help us understand the dynamics of biological systems and provide insights into the time-dependent behaviour of various processes. For instance, the nonlocal model in space and time can be considered the relaxation-time nonlocal model (\ref{NFBNM}) in space and time. A similar type of extension can be applied to the model (\ref{SIRM1}). We first include the spatial variable into the model (\ref{SIRM1}) to account for the random movement of individuals as \cite{de1979}:
\begin{equation*}{\label{SSIRM1}}
    \begin{aligned}
         \frac{\partial S}{\partial t} &= d_{1}\Delta S -\beta S(x,t)\int_{-\infty}^{\infty}I(y,t)\Psi (x-y)dy,\\
         \frac{\partial I}{\partial t} &= d_{2}\Delta I + \beta S(x,t)\int_{-\infty}^{\infty}I(y,t)\Psi (x-y)dy -\gamma I(x,t), \\
         \frac{\partial R}{\partial t} &= d_{3}\Delta R + \gamma I(x,t),
    \end{aligned}
\end{equation*}
subject to the Neumann boundary condition. Here $d_1$, $d_2$, and $d_3$ are the diffusion rates for the susceptible, infective and removed individuals. Studying a time delay to account for the latent period on the spread of the disease, we have to consider the following model:
\begin{equation}{\label{SDSIRM1}}
    \begin{aligned}
         \frac{\partial S}{\partial t} &= d_{1}\Delta S -\beta S(x,t)\int_{-\infty}^{t}\int_{-\infty}^{\infty}I(y,s)\Psi (x-y,t-s)dyds,\\
         \frac{\partial I}{\partial t} &= d_{2}\Delta I + \beta S(x,t)\int_{-\infty}^{t}\int_{-\infty}^{\infty}I(y,s)\Psi (x-y, t-s)dyds -\gamma I(x,t), \\
         \frac{\partial R}{\partial t} &= d_{3}\Delta R + \gamma I(x,t),
    \end{aligned}
\end{equation}
where the kernel function describes the interaction between the infective and the susceptible individuals at location $x$ and the present $t$, which occurred at location $y$ and at earlier instance $s$. This kernel function satisfies some conditions mentioned in \cite{ruan2007}. In this context, the term ``relaxation'' refers to the process by which the population returns to equilibrium after a disruption, such as an epidemic outbreak. In a nonlocal SIR model, the relaxation process includes both local recovery rates and nonlocal temporal impacts. For example, after a local epidemic has subsided, a population may continue to be infected because adjacent regions continue to have high infection rates, delaying the return to equilibrium. In contrast, a local model would predict a faster return to equilibrium once the local infection subsides.

The nonlocal effects and time delay may come in the model when the mature death and diffusion rates are age-independent. To derive such a delay model, we follow the work \cite{liang2003} where the author considered a single-species population model with two age classes and a fixed maturation period living in a spatial transport field. Suppose $u(x,a,t)$ is the population density at a spatial point $x\in\Omega\subset \mathbb{R}$ and time $t\geq 0$ with age $a\geq 0$. Following \cite{liang2003}, the single-species model given by
\begin{equation}{\label{SSDM}}
    \frac{\partial u}{\partial t} + \frac{\partial u}{\partial a} = \frac{\partial}{\partial x}\bigg{(} D(a)\frac{\partial u}{\partial x}+Bu \bigg{)} + d(a)u,
\end{equation}
with the natural boundary condition. Here, $D(a)$ and $d(a)$ are the diffusion and death rates at the age $a$, respectively, and $B$ is the transport velocity of the field. For simplicity, we can assume that the population has two age stages: immature and mature. We also assume that $\tau \geq 0$ is the maturation time and $r$ is the life limit of an individual species. Then, the total matured population density is given by
$\int_{\tau}^{r}u(x,a,t)da,~~x\in\Omega,~~t\geq 0,$
and the equation (\ref{SSDM}) can be transformed into \cite{liang2003}:
\begin{equation}{\label{TSSDM}}
    \frac{\partial v^{s}}{\partial t}  = \frac{\partial}{\partial x}\bigg{(} D(t-s)\frac{\partial v^{s}}{\partial x}+Bv^{s} \bigg{)} + d(t-s)v^{s},~~s\geq 0,~~s\leq t\leq s+\tau,
\end{equation}
where $v^{s}(x,t) = u(x,t-s,t)$. The dynamics of a population can be heavily influenced by age-specific behaviours \cite{noftle2010, hoy2020, ripoll2023}. For instance, younger individuals may disperse more than older ones, older individuals often have higher mortality rates, etc.

\subsection{Nonlocal problems and conservation laws}

Nonlocal problems and conservation laws are particularly relevant in modelling biological transport and reaction-diffusion processes. They describe the movement of substances or entities within biological systems and the interactions between different components. Traditional conservation laws (e.g., conservation of mass, momentum, energy) express that certain quantities remain constant within a closed system. They are typically local, depending only on the properties and states of nearby points. Based on the mass conservation law, many mathematical models have been studied in cellular populations \cite{chauvet1993, buttenschon2018}. For example, if $u(\mathbf{x},t)$ and $J(\mathbf{x},t)$ are the density and flux of a population, then the conservation equation is given by 
\begin{equation}{\label{CONE}}
    u_{t}(\mathbf{x},t) = -\nabla\cdot J(\mathbf{x},t).
\end{equation}
An appropriate choice of flux helps capture different biological phenomena. As an example, the flux $J$ can be divided into the random motion ($J_{d}$) and the adhesion ($J_{a}$), i.e.,
$$J(\mathbf{x},t) = J_{d}(\mathbf{x},t)+J_{a}(\mathbf{x},t).$$
We assume Fick's law for the diffusive flux, i.e., $J_{d} = -d\nabla u$. On the other hand, following Armstrong et al. \cite{armstrong2006}, the adhesion flux is proportional to the population density and adhesion force. It is inversely proportional to the cell size, so
$$J_{a} = \frac{c}{L}u(\mathbf{x},t)F(\mathbf{x},t),$$
where $c$ is the proportionality constant, $F$ is the adhesion force, and $L$ is the cell size. Using Stoke's law, the total adhesion force is given by
$$F(\mathbf{x},t) = \int_{\mathbb{B}^{n}(R)}g(u(\mathbf{x}+r,t))P(r)dr,$$
where $g(\cdot)$ and $P(\cdot)$ represent the nature of the force and its direction, respectively. Taking all these factors in the conservation equation (\ref{CONE}), we obtain
\begin{equation}{\label{CONEA}}
    u_{t}(\mathbf{x},t) = d\Delta u(\mathbf{x},t)-\nabla\cdot (u(\mathbf{x},t)(\Phi\ast u)(\mathbf{x},t)),
\end{equation}
where $$(\Phi\ast u)(\mathbf{x},t) = \frac{c}{L} \int_{\mathbb{B}^{n}(R)}g(u(\mathbf{x}+r,t))P(r)dr.$$
Analyzing the stability and well-posedness of solutions to nonlocal conservation laws is more complex than local laws, often requiring specialized techniques to ensure unique and stable solutions \cite{sherratt2009, colombo2018, vo2024}. The authors in \cite{sherratt2009} considered the nonlocal model (\ref{CONEA}) with an additional term (logistic growth) in the reaction kinetics and derived sufficient conditions for the boundedness of the solutions. However, these conditions do not satisfy some parameter values numerically, and hence, there are still some mathematical challenges.

\subsection{Nonlocal variational problems}

In traditional variational problems, the goal is to find an extremum of a function that depends on local derivatives of a function. Nonlocal variational problems are mathematical optimization problems that involve nonlocal operators in their formulations. These formulations have been used to derive and study nonlocal partial differential equations (PDEs) in diverse physical and biological systems. The general form of a nonlocal variational problem can be expressed as follows:
\begin{equation}{\label{NVP}}
    \mbox{Min or Max}~J(u) = \int_{\Omega} L(\mathbf{x},\mathbf{y},u(\mathbf{x}),u(\mathbf{y})) d\mathbf{x}d\mathbf{y},
\end{equation}
subject to boundary conditions and constraints on the function $u(\mathbf{x})$ over the domain $\Omega$. Here, the function $L(\mathbf{x},\mathbf{y}, u(\mathbf{x}),u(\mathbf{y}))$ is the nonlocal kernel that characterizes the interaction between the points $\mathbf{x}$ and $\mathbf{y}$ in the domain $\Omega$, and $u(\mathbf{x})$ represents the unknown function being optimized.

Nonlocal mathematical model also arises from crop-raiding of large-bodied mammals living in the biodiversity-rich tropics \cite{hsu2014}. Let $\Omega$ be a spatial domain with $\overline{\Omega}_{0}\subset \Omega$ and $u(\mathbf{x},t)$ be the population density of the mammal at position $\mathbf{x}$ and time $t$. Here, we assume that the forest is safe for the mammal species but is of poor resources, while the farm has rich resources but is dangerous. Furthermore, mammals cannot survive if they only stay in the forest and do not attempt to go out to search for food; they cannot produce offspring on the farm because it is not a safe place for reproduction. In this case, the model takes the form \cite{hsu2014}:
\begin{equation}{\label{LVP}}
    u_{t} = d \Delta u + \gamma v(\mathbf{x})u\int_{\Omega \smallsetminus \Omega_{0}}u(\mathbf{y},t)d\mathbf{y} - u^{p},~~\mbox{in}~\Omega\times (0,T), 
\end{equation}
with Dirichlet boundary condition on $\partial\Omega$ and positive initial condition. Here, $\gamma>0$ and $v(\mathbf{x})$ is the characteristic function of $\Omega_{0}$, which is defined by
\begin{equation*}
   v(\mathbf{x}) = \begin{cases}
    1~~\mbox{if}~\mathbf{x}\in \Omega_{0},\\
    0~~\mbox{if}~\mathbf{x}\in \Omega\smallsetminus\Omega_{0}.
\end{cases}
\end{equation*}
The parameter $d$ is the diffusivity and $p\geq 2$ considers the population's overcrowding effects in $\Omega$. The non-negative steady-states for the model (\ref{LVP}) are the nonnegative solutions of the semilinear nonlocal elliptic problem 
\begin{equation*}
    -d \Delta u = \gamma v(\mathbf{x})u\int_{\Omega \smallsetminus \Omega_{0}}u(\mathbf{y},t)d\mathbf{y} - u^{p},~~\mbox{in}~\Omega\times (0,T).
\end{equation*}
A similar type of Kirchhoff-Carrier type equation boundary value problems studied in the literature \cite{xu2022}:
\begin{equation*}
    -(a + b\alpha (||u||,|u|_{\gamma}) )\Delta u = \lambda \bigg{(} a_{1}+b_{1} \bigg{(}\int_{\Omega_{0}}c(x)|u|^{q_1} \bigg{)}^{\gamma_{1}} \bigg{)}u^{q_2}+f(u),
\end{equation*}
where $|u|_{\gamma} = (\int_{\Omega}|u|^{\gamma} )^{1/\gamma}$, $||u|| = (\int_{\Omega}|\nabla u|^{2})^{1/2}$, $\alpha$ is a non-negative function, and $c$ is a non-negative continuous function in $\Omega$. The conditions on $f$ and the other parameters are given in \cite{xu2022}. A weak solution to this problem is a function $u$ in a Hilbert space $H_{0}^{1}(\Omega)$ such that
\begin{equation*}
    (a + b\alpha (||u||,|u|_{\gamma}))\int_{\Omega}\nabla u \cdot \nabla h = \lambda \bigg{(} a_{1}+b_{1} \bigg{(}\int_{\Omega_{0}}c(x)|u|^{q_1} \bigg{)}^{\gamma_{1}} \bigg{)}\int_{\Omega}u^{q_2}h + \int_{\Omega}f(u)h,~~\forall h\in H_{0}^{1}(\Omega).
\end{equation*}
Furthermore, the variational structure $J_{\lambda}(u)$ has derivative operator given by 
\begin{equation*}
   J_{\lambda}'(u)h =  \bigg{(}\int_{\Omega_{0}}c(x)|u|^{q_1} \bigg{)}^{\gamma_{1}}\int_{\Omega}u^{q_2}h,~~\forall u, h\in H_{0}^{1}(\Omega).
\end{equation*}

Structure-preserving characteristics and variational integrators are fundamental notions in numerical analysis, particularly in the simulation of dynamical systems \cite{leimkuhler2004}. These approaches are intended to retain fundamental aspects of the underlying physical system, such as energy conservation, ensuring that the numerical solution closely resembles the genuine system's behaviour over long periods. These methodologies are used in various scientific areas, ranging from celestial mechanics to molecular dynamics simulations. 

Sometimes, the complete information of a system's states may be inaccessible, but only partial information about the states is observable due to limitations in experimental techniques. In this case, observability is a critical aspect of partially observed dynamics, highlighting which parts of a system are accessible to measurement \cite{luenberger1971}. The reduction techniques focus on accounting for hidden variables and incorporating partial observations into the system's modelling. Stochastic models and Bayesian inference are employed to handle uncertainties associated with partially observed dynamics, and these frameworks provide probabilistic representations of hidden states and update beliefs based on observed data.

\subsection{Nonlocal models in pattern formations}

Pattern analysis was essential to human survival and advancement in the past. To make sense of the world, ancient civilizations studied patterns in astronomy, agriculture, social behaviour, and nature. Seasonal trends and natural cycles were intimately related to agricultural operations. Farmers rotated the crops to maintain soil fertility and minimize depletion, and they grew in a given field. Numerous ancient societies, like the Greeks, Egyptians, and Mesopotamians, depended on the yearly inundation of rivers like the Nile and Tigris-Euphrates to replenish the soil with nutrients and used astronomical patterns as a guide. The early agricultural communities may not have had the advanced maps that we have today, but they still effectively planned and organized their agricultural activities using basic spatial techniques. For example, the ancient city of Mohenjo-Daro dates back to around 2600 BCE and was part of the historic Indus Valley civilization. Cuneiform tablets uncovered in ancient Mesopotamia, one of the first centres of civilization, date to approximately 2300 BCE and show primitive maps of land property.

History demonstrates that the time of agricultural activities was greatly influenced by the locations of the sun, moon, and stars. A Greek mathematician and astronomer, Hipparchus (190–120 BCE), is frequently acknowledged for having made some of the first attempts at star mapping. By mapping the stars using a coordinate system, he popularized the ideas of celestial longitude and latitude. Hipparchus's suggestion of a method to measure the apparent brightness of stars is one of his most important contributions. Babylonian mathematicians also investigated natural numerical patterns. Pythagorean triples are listed on the Plimpton 322 clay tablet from ancient Babylonia, which dates to approximately 1800 BCE and illustrates mathematical patterns in ancient times. In addition, traditional medical professionals and healers examined trends in disease and symptomology. They helped to shape the early forms of herbal medicine and other medical practices by observing the impact of specific plants or ceremonies on health. Ancient civilizations did not have the advanced technology that we use today, but they developed their methodologies to understand the world. Their models may not have been explicitly nonlocal in the way we think about it today, but they certainly reflected an understanding of interconnectedness and patterns in the world.    

The physics of ecological pattern formation explores the complex dynamics between consumer-resource interactions, which often leads to various spatial and temporal structures \cite{rietkerk2008}. Nonlocal interactions in ecological systems can significantly impact such spatio-temporal pattern formations, e.g., the development of spatial patterns in plants. Nonlocal interactions between plants, such as resource competition or facilitation, can result in the creation of regular patterns such as stripes, spots, or waves. These patterns frequently emerge as a result of self-organization in which interactions between individuals on a broader scale give rise to organized patterns on a landscape level. Nonlocal interactions in predator-prey relationships can result in the development of predator fronts, in which predators pursue prey concentrations over longer distances, altering the spatial dynamics of both populations. Nonlocal predator-prey interactions are generally two types: intra- and inter-specific. First, we consider a general nonlocal predator-prey model with nonlocal interaction in the intra-specific prey competition in a two-species population as
\begin{equation}{\label{TSINLM}}
    \begin{aligned}
        \frac{\partial u}{\partial t} &= d_{1}\frac{\partial^{2} u}{\partial x^{2}} + \alpha u\bigg{(}1-\frac{\phi\ast u}{\kappa}\bigg{)} - f(u)v,\\
        \frac{\partial v}{\partial t} &= d_{2}\frac{\partial^{2} v}{\partial x^{2}} + (ef(u)-g(v))v,
    \end{aligned}
\end{equation}
subjected to non-negative initial conditions and appropriate boundary conditions of prey ($u$) and predator ($v$) populations. The parameters $\alpha$ and $\kappa$ are intrinsic growth rate and environmental carrying capacity, respectively, and $e$ $(0<e\leq 1)$ is the conversion efficiency from prey biomass into predator biomass. If $e=0$, then there is no interaction between the species. For simplicity, we have considered a one-dimensional spatial domain $\Omega\subset \mathbb{R}$, and $d_{1}$ and $d_{2}$ are the self-diffusion coefficients. As defined earlier, $\phi\ast u$ is the convolution term. The kernel function $\phi$ satisfies all the assumptions mentioned earlier, i.e., normalized, and has compact support in $\mathbb{R}$. For an unbiased movement of the species in both spatial directions, we assume $\phi$ as an even function. Still, sometimes, species do not have the same area for nonlocal consumption of resources in both spatial directions. This generally happens for a bounded spatial domain, e.g., studying a nonlocal interaction for the populations living in a river \cite{sherratt2016,sun2019,lutscher2005}. The function $f(u)$ is called a functional response, and it satisfies three conditions \cite{myerscough1996,pal2021cnsns}: (i) $f(0) = 0$, (ii) $f$ is an increasing function and (iii) there exists a finite $M > 0$ such that $\lim_{u\rightarrow \infty} f(u) = M$. Many biologically relevant functional responses are available in the literature, e.g., Holling type II and III, Ivlev function, etc. \cite{myerscough1996,pal2021cnsns,real1977,pal2024,yadav2023, skalski2001, saha2025, pal2025}. The function $g(v)$ represents the per capita death rate of the predator population \cite{berryman1992}. In \cite{banerjee2017ec}, authors have studied a particular form of the nonlocal model (\ref{TSINLM}). Nevertheless, the local model corresponding to that nonlocal model does not produce any nonhomogeneous stationary pattern, but the nonlocal model does.

Researchers have also studied the pattern formation for nonlocal models with more than two species populations. Generally, the three components of the local or nonlocal model in ecology add another feature to the two-component model \cite{pal2024b}. It allows us to study the dynamics of the predator-prey model in the presence of an additional species. These three components of the model may be (a) food chain; (b) two predators–one prey; (c) one predator–two prey; (d) food chain with omnivory; and (e) food chain with cycle \cite{hsu2015}. Autry et al. \cite{autry2018} have considered a three-species nonlocal RD model in a food chain system with ratio-dependent functional responses as: 
\begin{equation}{\label{SPNLM}}
    \begin{aligned}
        \frac{\partial u}{\partial t} &= d_{1}\frac{\partial^{2} u}{\partial x^{2}} +  u\bigg{(}1- \phi\ast u-\frac{a_{1}v}{u+v}\bigg{)},\\
        \frac{\partial v}{\partial t} &= d_{2}\frac{\partial^{2} v}{\partial x^{2}} + v\bigg{(} -\mu_{1}+\frac{m_{1}u}{u+v}-\frac{a_{2}w}{v+\psi\ast w} \bigg{)},\\
        \frac{\partial w}{\partial t} &= d_{3}\frac{\partial^{2} w}{\partial x^{2}} + w\bigg{(} -\mu_{2}+\frac{m_{2}v}{v+\psi\ast w} \bigg{)},
    \end{aligned}
\end{equation}
where $u$, $v$, and $w$ represent the concentration of prey, predator, and super-predator, respectively. All the convolution terms are defined as before, and $\phi$ and $\psi$ are the kernel functions. Here, the prey and super-predator interact directly with the intermediate predator and only interact indirectly with the intermediate predator. Authors in \cite{autry2018} have studied two aspects of the nonlocality: C-type and P-type. The C-type nonlocality occurs when the prey species competes with itself. This type of nonlocal interaction usually occurs due to the overlapping ranges of the species, e.g., extensive root systems or a shared water table in plant species. On the other hand, P-type stands for pest type, which arises when one species nonlocally interacts with the other species, and this nonlocal interaction only affects the encounter rate, which appears in the functional response term. In addition, one can consider the same type of nonlocal term in the functional response for the first and second equations to study the nonlocal effect of high-mobile intermediate predator species $v$ on the crops or prey $u$.

In population biology, age structure models and pattern formation provide insights into the dynamics of populations and their structural changes over time. Age structure refers to the distribution of individuals in a population across different age groups, and understanding its patterns is crucial for demographic analysis and prediction. McKendrick and Von Foerster developed this concept \cite{m1925, von1959}. Suppose $n(a,t)$ is the population at age $a$ and time $t$. Following McKendrick-Von Foerster, the population $n(a,t)$ satisfies the following equation \cite{trucco1965}:
\begin{equation*}
    \frac{\partial n}{\partial t} + \frac{\partial n}{\partial a} + \mu(a)n(a,t) = 0,
\end{equation*}
with the initial condition $n(a,0) = f(a)$ and the birth boundary condition:
$$n(0,t) = \int_{0}^{\infty}b(a)n(a,t)da,$$ where $b(a)$ is the birth rate of the individuals which depends on the age of the adult population. Here, $\mu(a)$ is the death rate of individuals. These types of age structure models often use population pyramids to represent the distribution of age groups in a population visually.

In reaction-diffusion systems such as the Gierer-Meinhardt system, a limiting situation known as the ``Shadow System" occurs when one component (usually the inhibitor) diffuses significantly faster than the other \cite{kavallaris2021, marciniak2017, marciniak2013}. In this case, the fast-diffusing component instantly enters a quasi-steady state, resulting in a reduced system closely matching the original system's behaviour. Researchers have shown that the Gierer–Meinhardt system can be controlled by only the dynamics of the activator itself given by the nonlocal problem \cite{kavallaris2017, duong2021}:
\begin{equation}{\label{SRDS}}
    \frac{\partial u}{\partial t} = \Delta u - u + u^{p}\bigg{/}\bigg{(}\int_{\Omega}u^{r}dx \bigg{)}^{\gamma}
\end{equation}
with positive initial and no-flux boundary conditions. Furthermore, the authors obtained global-in-time existence and blow-up solutions for the nonlocal problem in both finite and infinite time. The nonlocal model predicts the regulated expansion of the activator, which is ensured by global-in-time existence results; yet, the activator's unlimited growth is relevant to finite- and infinite-time blow-up results. 

In pattern formation, morphogenesis is a process by which spatial patterns and structures arise during the development of organisms \cite{gierer1972}. The creation of the various and complex forms found in organisms depends on this complex process. It involves a cascade of molecular and cellular processes that coordinate their orientation, polarity, and assembly of cells to generate distinct morphologies and configurations. Examples include (i) the development of an embryo into a fully formed organism, (ii) the differentiation process by which cells in developing organisms take on certain destinies and functions, and (iii) precise control of gene expression. The multidisciplinary study of morphogenesis aims to understand the fundamental processes that give rise to the astounding diversity of biological forms. It draws on the disciplines of genetics, developmental biology, and biophysics, among others. 

Spatio-temporal patterns in epidemiology reveal the dynamic interplay of place, time, and disease distribution within populations. Environmental, social, and biological factors all have an impact on these patterns \cite{tadic2023,mitrovic2023}. Epidemics can follow seasonal trends, exhibit cyclic behaviour, or demonstrate irregular patterns driven by various factors, such as host susceptibility and environmental conditions, referred to as temporal patterns. However, spatio-temporal models deal with many spatial heterogeneity factors, such as higher or lower incidence rates in certain areas \cite{elliot2000}. 

Nonlocal models are essential for describing the development patterns in various biological systems, especially when it comes to cell communication, cell adhesion, and bacterial aggregation. Cell adhesion models take into account distant adhesive forces, enabling cells to recognize and respond to the spatial distribution of adhesion molecules \cite{armstrong2006, zhigun2024}. It can capture the collective behaviour of bacteria by addressing long-range interactions, such as quorum detection, where bacterial cells communicate and coordinate their activities over longer distances \cite{miller2001}.

Nonlocal models derived from quantum confinement are intriguing in characterizing pattern generation at the nanoscale and have several applications in materials research, quantum mechanics, and device manufacturing. Quantum confinement is the phenomenon in which the electronic and optical characteristics of materials are significantly affected by their size, particularly when limited to dimensions equivalent to the characteristic length scale of electrons \cite{otten2003}. The nonlocal models derived from quantum mechanics are frequently used to characterize the electrical and optical behaviour of quantum dots, nanowires, and other nanostructures \cite{michalek2015, taghipour2017}. The discrete energy levels and wave functions of the confined electrons produce distinct patterns in their characteristics. Quantum dot solar cells, light-emitting diodes, and quantum dot transistors are some of the uses \cite{schornbaum2015, kahmann2020}. Nonlocal models can assist in explaining the development of energy bands and electronic patterns in quantum dot arrays, which influence their optical and electrical characteristics \cite{freeney2022}. Furthermore, quantum confinement can cause nonlocal electron interactions and transport behaviours, affecting the operation of nanoelectronic devices such as sensors, catalysis, and quantum computing.

An interesting perspective on the emergence of complex patterns in many physical and biological systems may be gained from nonlocal models that describe pattern development resulting from symmetry-breaking processes \cite{ishihara2018, denk2020}. Symmetry-breaking is a basic idea in which complex patterns are formed when a system changes from a symmetrical to an asymmetrical state. These occurrences, prompted by disturbances or outside factors, start to establish patterns. Fluid dynamics, chemical reactions, and biological processes are examples of systems where this may happen \cite{crawford1991, islas2005, li2010, tanzy2013, mcglinchey2022}.

\subsection{Complex nonlocal systems, multi-stability, and control}

Nonlocal interactions enhance the species' coexistence in the Lotka-Volterra competition model, and it supports how two competing species can coexist in the same resource \cite{hutchinson1961}. Different mechanisms have been proposed to explain such coexistence in ecological communities with the competitional environment \cite{levins1971,chesson2000,amarasekare2003,kartzinel2015,maciel2018}. Authors in \cite{maciel2021} proposed a nonlocal competition model as
\begin{equation}{\label{NLLVCM}}
    \begin{aligned}
        \frac{\partial u_{1}}{\partial t} &= d_{1}\frac{\partial^{2} u_{1}}{\partial x^{2}} + b_{1}u_{1} \bigg{(} 1- \frac{a_{11}u_{1}+a_{12}u_{2}}{K_{1}} - h_{11}\overline{u}_{11} -h_{12}\overline{u}_{12} \bigg{)},\\
        \frac{\partial u_{2}}{\partial t} &= d_{2}\frac{\partial^{2} u_{2}}{\partial x^{2}} + b_{2}u_{2} \bigg{(} 1- \frac{a_{21}u_{1}+a_{22}u_{2}}{K_{2}} - h_{21}\overline{u}_{21} -h_{22}\overline{u}_{22} \bigg{)},
    \end{aligned}
\end{equation}
subjected to non-negative initial conditions and appropriate boundary conditions. The first term on the right-hand side of each equation represents the dispersal of the species at rate $d_{i}$ ($i=1,2$). For the species $i$ ($i=1,2$), $b_{i}$ is the intrinsic growth rate, $a_{ij}$ ($j=1,2$) is the coefficient corresponding to the local competition, and $K_{i}$ is the environmental carrying capacity. The parameters $h_{ij}$ represent the intensity of nonlocal competition with the same species and its competitor. $\overline{u}_{ij}$ is the average density of all $j$-individuals interact with the individual $i$ in a neighbourhood centred at $x$. For spatial isotropy, $\overline{u}_{ij}$ is:
\begin{equation*}
    \overline{u}_{ij}(x,t) = \int \Phi_{ij}(|x-y|)u_{j}(y,t)dy,
\end{equation*}
where $\Phi_{ij}$ is a kernel function and represents the influence of $j$-individuals on $i$-individuals when they are at a distance $|x-y|$ \cite{jdmurray}. In this case, a normalized top-hat kernel function can be considered, and the authors in \cite{maciel2021} presented differently to incorporate spatial scales. For two species, it is given by:
\begin{equation*}
    \Phi_{ij}(r) = \Pi_{2R_{ij}} = \begin{cases}
        \frac{1}{2R_{ij}} &r\leq R_{ij}\\
        ~~0&\mbox{otherwise},
    \end{cases}
\end{equation*}
where $R_{ij}$ is the interaction range and is a function of the species-specific ranges of competition, $R_{i}$ and $R_{j}$. Depending on these spatial scales $R_{i}$ and $R_{j}$, two possibilities can occur: additive and non-additive influence ranges. For the additive influence ranges, both the ranges of the competition overlap with each other \cite{tarnita2017}. In this case, the range of interspecific competition is $R_{i}+R_{j}$, and the ranges of intra-specific competitions are $2R_{i}$ and $2R_{j}$. This leads to a change in the kernel function ($\Phi_{ij}(r) = \Phi_{ji}(r) = \Pi_{2(R_{i}+R_{j})}(r)$, $\Phi_{ii}(r)$ = $\Pi_{2R_{i}}(r)$, and $\Phi_{jj}(r)$ = $\Pi_{2R_{j}}(r)$). On the other hand, for non-overlapping ranges of competitions \cite{bais2003, granato2019}, the kernel function remains the same for inter- and intra-specific competitions ($\Phi_{ij}(r) = \Pi_{2R_{i}}(r)$, $\Phi_{ji}(r) = \Pi_{2R_{j}}(r)$, $\Phi_{ii}(r)$ = $\Pi_{2R_{i}}(r)$, and $\Phi_{jj}(r)$ = $\Pi_{2R_{j}}(r)$). The range of the kernel function is crucial to the nonlocal competition. A narrow range shows that competition is mostly local, while a wide range suggests that individuals compete over a broader area. These types of nonlocal models produce stable spatial structures similar to physical systems where repulsive forces lead to the separation of particles. In addition, they are used to study phenomena such as species segregation, biodiversity, and the evolution of social behaviours.  

A mixed hyperbolic-parabolic problem is studied in the predator-prey model in accounting for the flow of predators' direction \cite{colombo2014}. We consider the mathematical problem as
\begin{equation}{\label{MHPE}}
    \begin{aligned}
        &\frac{\partial u}{\partial t} -d\Delta u = (\gamma-\delta v)u,\\
        &\frac{\partial v}{\partial t} + \nabla \cdot (v\xi (u)) = (\alpha u-\beta)v,
    \end{aligned}
\end{equation}
where $u(\mathbf{x},t)$ and $v(\mathbf{x},t)$ are the prey and predator densities, respectively, at the position $\mathbf{x}\in \Omega \subset \mathbb{R}^{n}$ and time $t\in \mathbb{R}^{+}$. Here, the term $v\xi (u)$ represents the flow preferred predators' direction, and the velocity function $\xi (u)$ is generally nonlocal and nonlinear. A typical choice is \cite{colombo2014}:
\begin{equation*}
    \xi (u) = \kappa \frac{\nabla (\phi\ast u)}{\sqrt{1+||\nabla (\phi\ast u)||^{2}}}.
\end{equation*}
Here, the denominator $\sqrt{1+||\nabla (\phi\ast u)||^{2}}$ is the normalization factor. Prey species spread in all directions in this model, but predators have a direct movement that migrates towards higher prey density zones. Furthermore, the radius of support of $\phi$ specifies how far predators can detect the presence of prey and, as a result, the direction in which they move.  

Another type of parabolic nonlocal model where the nonlocal term appears as the master equation's mean field limit when reaction radius is used, which is given by \cite{ichikawa2012, kavallaris2013, kavallaris2014, kavallaris2021}:
\begin{equation}{\label{NHE}}
    \begin{aligned}
        &\frac{\partial q_{A}}{\partial t} = D_{A}\Delta q_{A} - \frac{k_{B}}{|B(\cdot, R)|}\int_{B(\cdot, R)\cap\Omega}q_{B}dy\cdot q_{A},\\
        &\frac{\partial q_{B}}{\partial t} = D_{B}\Delta q_{B} - \frac{k_{A}}{|B(\cdot, R)|}\int_{B(\cdot, R)\cap\Omega}q_{A}dy\cdot q_{B},
    \end{aligned}
\end{equation}
with non-negative initial and no-flux boundary conditions in a bounded domain $\Omega\subset \mathbb{R}^{n}$. Here, $q_{A}$ and $q_{B}$ are the probabilities of the molecules $A$ and $B$ located at position $\mathbf{x}$ and at time $t$, respectively, and $B(\mathbf{x}, R) = \{ \mathbf{y}\in \mathbb{R}^{n}: |\mathbf{y}-\mathbf{x}|< R \}$. In a chemical reaction, this $R$ represents the reaction radius of the fundamental reaction process. In this case, the assumption is that the chemical reaction occurs if and only if a pair of two different chemical particles is in the distance $R$. The mean-field limit approach is commonly used in statistical mechanics and kinetic theory, where it simplifies complex many-body systems by averaging interactions over a large number of particles \cite{golse2003}.

In evolutionary game dynamics, a key component is to describe how a population of players adjusts their strategies throughout the game in response to the strategies' level of success. Nonlocal models have gained prominence in studying such dynamics, notably replicator dynamics. These models account for interactions that are not only local neighbours but can also involve distant people or entire communities, which is critical for accurately portraying the complexity of real-world settings. For a finite-dimensional strategy space $S = \{1,2,\ldots,m\}$, the replicator dynamics equation can be written as \cite{kravvaritis2010, kavallaris2017a}:
\begin{equation}{\label{NLGT1}}
    \frac{dp_{i}(t)}{dt} = \bigg{(}\sum_{j=1}^{m}a_{ij}p_{j}(t) + \sum_{k=1}^{m}\sum_{j=1}^{m}a_{kj}p_{j}(t)p_{k}(t)\bigg{)}p_{i}(t),~~ i=1,2,\ldots,m,
\end{equation}
where $a_{ij}$ is the payoff for players 1 and 2 with strategies $i$ and $j$. Here, the set $p = (p_{1},\ldots, p_{m})$ satisfies the conditions $p_{i}\geq 0$, $i=1,\dots,m$, and $\sum_{i=1}^{m}p_{i} = 1$. If the strategy space consists of an infinite number of strategies, then the replicator dynamics take the form:
\begin{equation}{\label{NLGT2}}
    \frac{dp_{i}(t)}{dt} = \bigg{(}\sum_{j\in\mathbb{Z}}a_{ij}p_{j}(t) + \sum_{k\in\mathbb{Z}}\sum_{j\in\mathbb{Z}}a_{kj}p_{j}(t)p_{k}(t)\bigg{)}p_{i}(t),~~ i=1,2,\ldots,
\end{equation}
with $p_{i}(t)\geq 0$ and $\norm{p}_{l^{1}} = 1$. Furthermore, we can rewrite the replicator dynamics equation in terms of absolutely continuous Lebesgue measure with density $u(t,x)$ as:
\begin{equation*}
    \frac{\partial u}{\partial t} = \bigg{(} \int_{S}f(x,y)u(t,y)dy - \int_{S}\int_{S}f(z,y)u(t,y)u(t,z)dydz \bigg{)}u(t,x),
\end{equation*}
where $f(x,y)$ is the payoff for players 1 and 2 play $x$ and $y$, respectively. Additional details can be found in  \cite{kravvaritis2010}. 

Bio-cells participate in complex, dynamic interactions encompassing various molecular, cellular, and tissue activities. These interactions are essential for living things to operate normally and have significance in several physiological and pathological situations. Cells, for example, communicate using complex signalling networks. Hormones, growth factors, and neurotransmitters connect to cell surface receptors, triggering a chain reaction of events within the cell \cite{alberts2017}. Several approaches are used to simulate and model these cellular processes, including metabolism. In flux balance analysis, we operate the cellular processes under steady-state conditions and use constraints based on mass balances, thermodynamics, and other factors to predict metabolic flux distributions \cite{orth2010}. The kinetic model involves individual reactions with kinetic reactions represented by ordinary or partial differential equations and studying how the concentrations change over time. In the cellular automata model, each cell follows some rules governing its interactions, and these rules determine its state, reflecting biological behaviours. In addition, agent-based models simulate individual agents (cells or molecules) and their interactions and help in studying emergent properties and spatial considerations. 

Components of complex systems interact in sophisticated ways, and their behaviours tend to be nonlocal. Nonlocal interaction models can assist in describing how a change in one section of the system affects the entire network. Tsallis entropy, in contrast to the standard Shannon entropy, is a non-extensive entropy that considers long-range correlations in a system. It has been used to account for nonlocal interactions and power-law distributions in many domains, including statistical mechanics and information theory. Moreover, nonlocal models bridge the gap between multiple scales, like zooming in from the cosmic scale to the quantum scale, seamlessly connecting the vastness of the universe to the tiniest building blocks of matter. 

Nonlocal PDEs with nonlinear diffusion terms are effective mathematical tools for understanding nonlocal interactions and complex diffusion processes. These equations extend typical diffusion models by accounting for impacts from distant locations, and they generate patterns in biological systems, such as population dispersion or the generation of spatial patterns in ecosystems. Complexity and pattern production in biophysical systems is a diverse field that spans several scales, from molecular interactions to tissue and organism-level structures \cite{wolpert2015}. Furthermore, microbial systems are exceptionally diversified and complex, ranging from intricate interactions within microbial communities to involvement in environmental processes and human health \cite{madsen2015}. 

Modelling many-particle systems is also complex and has applications in physics, chemistry, biology, and other sciences. Newton's equations of motion are frequently used in classical physics to describe the motion of particles in a many-body system \cite{goldstein2002}. In addition, statistical mechanics gives a statistical description of a vast number of particles \cite{feynman2018}. In kinetic theory, mathematical models involve the construction of equations that explain particle distribution and behaviour. Furthermore, network and graph-based models give a solid foundation for understanding the structure, dynamics, and behaviour of complex interactions in various systems \cite{melnik2012, antoniouk2013, melnik2015}. 

Self-organized criticality (SOC) has been studied in various disciplines, including evolutionary biology and dynamic models with complex relationships \cite{jensen1998, tadic2017, tadic2021a, tadic2023}. SOC refers to the tendency of complex systems to grow towards a critical state in which minor events can lead to large-scale cascades and emergent behaviour, e.g., in the forest-fire model \cite{grassberger1993}, in the brain \cite{plenz2021}, etc. Complex systems naturally evolve to a critical state, and SOC provides a valuable framework for understanding the dynamics of evolutionary and complex interactive processes.

Complex systems comprise many linked components with nonlinear interactions, frequently resulting in emergent phenomena that cannot be easily derived from individual components. When just a subset of the system's variables or states are available for observation, it introduces uncertainty in precisely understanding and forecasting the system's behaviour \cite{rupe2022}. The integration of these concepts is critical for acquiring insights into real-life occurrences that are characterized by both partial observations and complex interactions. Modelling such systems is an interdisciplinary process that requires understanding mathematics, statistics, and the specific application domain \cite{melnik1997, melnik1998a}.

Control problems involving human variables are common in many fields, including automation, transportation, and healthcare, where humans interact with complex systems. These systems incorporate human factors that examine cognitive, perceptual, and physical elements of human interaction with technology \cite{russell2019}. The objective is to create systems that account for human skills, limits, and behaviours, maximizing performance and reducing errors. These models are also used in neurorehabilitation and robotics to combine muscle mechanics and control, which logically leads to planning and behaviour \cite{burdet2013}. 

Nonlocal high-order connectivity and interactions are the studies of long-distance effects and correlations between distant locations in a system, emphasising higher-order links beyond close neighbours. This notion is especially important in a wide range of scientific areas, from physics and materials science to neurology and complex systems analysis \cite{battiston2020, bick2023}. Exploration of nonlocal high-order connectivity in the context of neurodegenerative diseases has become increasingly important in understanding the complex dynamics of brain function, cognition, information processing, and the progression of conditions such as Alzheimer's, Parkinson's, and others \cite{pal2022sr, herzog2022, uzel2022}.

\subsection{Nonlocal models in mechanobiology and biomechanics}

Nonlocal models are essential in mechanobiology and biomechanics for capturing mechanical interactions of the spatial and temporal aspects in biological tissues. These models provide a more comprehensive and realistic representation of mechanical phenomena, offering valuable insights for research and medical applications. For instance, (i) how mechanical signals are transmitted within cells and tissues, (ii) the influence of mechanical forces on cells and extracellular matrix components, considering the nonlocal effects of stress and strain distribution, (iii) blood vessel mechanics and the effects of blood flow on vessel walls, etc. 

Nonlocal beam models have been applied in various engineering and mechanical systems to study the behaviour of slender structures and address challenges that classical local beam theories cannot adequately capture \cite{reddy2007, park2005, liu2006}. They have found applications in various engineering and physics fields, where long-range interactions and size effects influence the behaviour of beams. For instance, bridge-based sensors and actuators have many applications in biological sciences \cite{lavrik2004,ekinci2005,lu2006}. The outcome of these devices depends on their dynamic properties of beamlike elements, which can be improved by studying their micro- or nano-electromechanical system (MEMS or NEMS) structures \cite{lu2006}. In addition, comprehensive investigations have been conducted for parabolic and hyperbolic models in deterministic and stochastic setups, taking into account a variety of nonlocal models that arise in the MEMS industry \cite{guo2012, kavallaris2011, kavallaris2016, drosinou2021, kavallaris2023a}. A set of integro-differential equations can describe the general nonlocal continuum theory, and different kernel functions have been used in the model to incorporate different problems \cite{eringen2003, ei2021}. For example, the equation of motion in nonlocal elasticity theory is given by
\begin{equation}{\label{NLEE}}
t_{kl,k}+f_{l} = \rho \ddot u_{l},
\end{equation}
where $\rho$ and $f_{l}$ are the mass density and the applied body forces, respectively. For a linear homogeneous elastic solid, the constitutive equations in nonlocal elasticity theory can be written as an integral over the body as \cite{lu2006, eringen2003, li2024, eringen1983}:
\begin{equation}{\label{TKL}}
t_{kl}(\mathbf{x}) = \int_{V}\alpha (|\mathbf{x}-\mathbf{y}|)\sigma_{kl}(\mathbf{y})dv(\mathbf{y}), 
\end{equation}
where $t_{kl}$ and $\sigma_{kl}$ are the stress tensors of nonlocal and local elasticities at the spatial points $\mathbf{x}$ and $\mathbf{y}$, respectively, and
$$\sigma_{kl} = \lambda \epsilon_{rr}\delta_{kl}+\mu \epsilon_{kl},~~\bigg{(}\epsilon_{kl} = \frac{u_{k,l}+u_{l,k}}{2}\bigg{)}$$
represents the conventional constitutive relations for an isotropic elastic material with $\epsilon_{kl}$ as the strain tensor and $u_{l}$ as the displacement vector. The parameters $\lambda$ and $\mu$ are material constants. The function $\alpha$ is called the kernel function, which characterizes the amount of stress at a material point influenced by the strains of all other points in the body. 

The classical continuum models cannot go sufficiently on smaller scales, which are the main part of nanotechnology. Atomic and molecular models can solve these smaller-scale problems but are computationally intensive. However, the nonlocal continuum models can take care of smaller length scales. For instance, in equation (\ref{TKL}), if $\alpha$ is the Green's function of a linear differential operator $L$: $L\alpha (|\mathbf{x}' - \mathbf{x}|,\tau) = \delta (|\mathbf{x}' - \mathbf{x}|)$, and $L$ is a differential operator with constant coefficients, one can obtain $(Lt_{kl})_{,k} = Lt_{kl,k}$. In this case, the integro-differential equation (\ref{NLEE}) reduces to the partial differential equation
\begin{equation*}
\sigma_{kl,l}+L(f_{l}-\rho\ddot u_{l}) = 0.
\end{equation*}
The operator $L = 1-(e_{0}a)^{2}\nabla^{2}$ is a differential operator for a two-dimensional kernel function $\alpha$ satisfying the Green's function \cite{eringen1983,lu2007}, and in this case, the constitutive relations simplified to 
\begin{equation}
[1-(e_{0}a)^{2}\nabla^{2}]t_{kl} = \lambda \epsilon_{rr}\delta_{kl}+\mu \epsilon_{kl}.
\end{equation}
Researchers have been studying different nonlocal beam models to incorporate different issues, e.g., the nonlocal Timoshenko and Euler beam models \cite{lu2006,weaver1991,zienkiewicz2005, wang2005,peddieson2003,zhang2005}. Flexoelectricity is one of the issues where an electromechanical coupling happens between the polarization and the strain gradient, and these flexoelectric materials play a vital role in making nanoscale sensors, actuators and energy harvesters \cite{navickas2006, su2020}. Moreover, the nonlocal elasticity theory helps capture the long-range interactions and strain gradient stress of these flexoelectric nanosensors. Taking the nonlocal elasticity theory, the stress and electric displacement tensors as \cite{su2020,lim2015,li2016}:
\begin{equation}
\begin{aligned}
     \relax[1-(e_{0}a)^{2}\nabla^{2}] \sigma_{ij} & = c_{ijkl} \epsilon_{kl} - e_{kij}E_{k},\\
     \relax[1-(e_{0}a)^{2}\nabla^{2}]\sigma_{ijk} & = -\mu_{lijk} E_{l} + g_{ijklmn}\epsilon_{lm,n},\\
    D_{i} = \kappa_{ij}E_{j} & + e_{ijk}\epsilon_{jk} + \mu_{ijkl}\epsilon_{jk,l}, 
\end{aligned}
\end{equation}
where $\sigma_{ij}$ and $\sigma_{ijk}$ are the normal and higher-order stress tensors, respectively; $D_{i}$ is the electric displacement vector; $E_{k}$ is the electric field. All the tensors' expressions are mentioned in \cite{su2020} and the references therein. Using these equations, authors have derived a cantilever beam model for the flexoelectric sensor with external loads. Motezaker and his coauthors constructed a sandwich-type nanostructure from a nanoplate covered by two smart layers under the externally applied voltage \cite{motezaker2020}. They have considered surface higher-order nonlocal piezoelasticity theory as:
\begin{equation}
\begin{aligned}
     \relax[1-\mu_{1}^{2}\nabla^{2}][1-\mu_{0}^{2}\nabla^{2}]\sigma_{ij} &= c_{ijkl} \epsilon_{kl} - e_{ijk}E_{k},\\
     \relax[1-\mu_{1}^{2}\nabla^{2}][1-\mu_{0}^{2}\nabla^{2}]D_{ij} &= e_{kli}\epsilon_{kl} + \varepsilon_{ik} E_{k},\\
    D_{i}^{S} = D_{i}^{0} +e_{kli}^{S}\epsilon_{kl} & +\varepsilon_{ik}^{S}E_{k},\\ 
    \sigma_{ij}^{S} = \tau_{ij}^{S}+ c_{ijkl}^{S}\epsilon_{kl} &-e_{ijk}^{S}E_{k},\\
    \sigma_{rz}^{S} = \tau_{iz}^{S}w_{,r},~~\sigma_{\theta z}^{S} &= \tau_{iz}^{S}w_{,\theta},
\end{aligned}
\end{equation}
where $\mu_{0}$ and $\mu_{1}$ are the nonlocal parameters; $\sigma_{ij}^{S}$ and $D_{i}^{S}$ surface stress tensor and electric displacement vector, respectively; the remaining terms are mentioned in \cite{motezaker2020} and the references therein. Using this higher-order nonlocal model, researchers have shown a decrease in the frequency and bucking load and an increase in the deflection by enhancing the nonlocal parameter \cite{motezaker2020}. The opposite happens by applying the negative external voltage.

Nonlocal plate theory is used to evaluate the axisymmetric bending of micro/nanoscale circular plates, accounting for the impact of small-scale influences \cite{duan2007}. It is especially pertinent at the micro/nanoscale, where traditional plate theories would not be able to adequately predict the behaviour, as it takes into account the nonlocal effects resulting from the tiny size of the structure. Furthermore, pyroelectricity's impact on thermally produced vibrations in piezothermoelastic plates is a complex, multidisciplinary issue. To understand the behaviour of such linked systems, researchers frequently employ experimental validation together with numerical models  \cite{krommer2004}. Variational principles provide a strong and systematic technique for developing the governing equations of motion for piezoelectric, thermopiezoelectric, and hygrothermopiezoelectric materials \cite{altay2007}. These concepts are founded on minimizing or extremizing a certain functional, typically referred to as the total potential energy or the total Lagrangian, concerning the relevant field variables. 

Sometimes, the distances between adjacent metallic structures in a metasurface might get extremely small (near the nanoscale). In these situations, the classical local response might not be sufficient when taking the spatially independent material characteristics into account. This is due to the quantum nature of the electron gas in metals at such small scales. Such circumstances can be handled via nonlocal effects by adjustments to the resonance frequencies, damping rates, and dispersion relations of the plasmonic modes. This behaviour differs from traditional predictions, especially as the gap size reduces due to the increased prominence of nonlocal influences. Nonlocal effects in plasmonic metasurfaces with nearly contacting surfaces include shifts in resonance frequencies, changes in field enhancement, and changes in near-field distribution \cite{yang2020}. These effects are critical for applications such as sensing, imaging, and improved light-matter interactions. 

Nonlinear viscoelastic solids refer to materials that exhibit time-dependent and strain-dependent behaviour under the influence of applied stress or deformation \cite{lockett1972}. In this case, the constitutive equations governing the behaviour of nonlinear viscoelastic solids often involve time-dependent and strain-dependent terms \cite{csengul2021}. A common representation is through a stress-strain relationship that includes both elastic and viscoelastic components and where the material properties are functions of time and strain. This helps us predict the performance and durability of materials in real-world applications. In addition, studying the thermoelastic behaviour along with viscoelastic solids gives us the experience of stress under the influence of both thermal and mechanical loads \cite{loret1990, rivera1998}.

\subsection{Peridynamics, fracture and damage problems, multiscale and multiphysics analysis}

Peridynamics is a nonlocal theory that has gained popularity in recent years due to its applicability in multiscale and multiphysics research. It is especially effective for modelling materials and structures having discontinuities, such as cracks and fractures, and it provides a framework for considering long-range interactions \cite{lejeune2021}. The peridynamic theory of solid mechanics substitutes conventional local stress-strain equations with integral formulations in which the force applied to a location is determined by the deformation of the surrounding area. Arguably, the first paper on peridynamics was published in 2000, and the author discussed the models' discontinuities and long-range forces using a constitutive relation based on pairwise interactions between particles \cite{silling2000}. It is being used to capture biological material in fracture, e.g., bone fracture \cite{deng2009,ghajari2014}. Here, we describe the basic formulation of peridynamics at a spatial point $\mathbf{x}$ and time $t$. Similar to the neighbourhood of a spatial point for nonlocal models, we consider a horizon $\mathcal{H}_{\mathbf{x}}$ of size $\delta_{\mathbf{x}}$ as
\begin{equation*}
    \mathcal{H}_{\mathbf{x}} = \{\mathbf{y}:||\mathbf{y}-\mathbf{x}||<\delta_{\mathbf{x}}\}.
\end{equation*}
Depending on the dimension of the spatial domain, the shape of the horizon $\mathcal{H}_{\mathbf{x}}$ is different, e.g., in one-dimension $\mathcal{H}_{\mathbf{x}}$ is an open interval $(\mathbf{x}-\delta_{\mathbf{x}},\mathbf{x}+\delta_{\mathbf{x}})$; in two dimensions, it is a circular region of radius $\delta_{\mathbf{x}}$ centred at $\mathbf{x}$, etc. In addition, the dual horizon of $\mathbf{x}$ is defined as the union of points whose horizons include $\mathbf{x}$, i.e., $\mathcal{H}_{\mathbf{x}}' = \{\mathbf{y}: \mathbf{x}\in \mathcal{H}_{\mathbf{y}}\}$. The equation of motion is formulated as an integral of interaction forces between points on the spatial domain (body) $\Omega$. Following the balance of linear momentum, we obtain \cite{lejeune2021}:
\begin{equation}{\label{PDYE}}
    \rho \ddot{\mathbf{u}}(\mathbf{x},t) = \int_{\mathbf{y}\in \mathcal{H}_{\mathbf{x}}'}\mathbf{f}_{\mathbf{x}\mathbf{y}}(\mathbf{\eta}, \mathbf{\xi})\mathbf{d}V_{\mathbf{y}} - \int_{\mathbf{y}\in \mathcal{H}_{\mathbf{x}}}\mathbf{f}_{\mathbf{y}\mathbf{x}}(-\mathbf{\eta}, -\mathbf{\xi})\mathbf{d}V_{\mathbf{y}} + \mathbf{b}(\mathbf{x},t),
\end{equation}
where $\rho$ is the density, $\ddot{\mathbf{u}}$ is the acceleration of the displacement vector $\mathbf{u}$, $\mathbf{\eta}$ is the relative displacement vector, $\mathbf{\xi} = \mathbf{y}-\mathbf{x}$ is the bond vector between $\mathbf{x}$ and $\mathbf{y}$, $\mathbf{f}_{\mathbf{x}\mathbf{y}}$ is the force density vector per volume acting on particle $\mathbf{x}$ due to particle $\mathbf{y}$, and $\mathbf{b}$ is the body force. Here, the first integration contributes the direct force term acting on the point $ \mathbf{x}$, and the second integration contributes the reaction force term. The discrete form of the balance of linear momentum can be obtained by replacing the integrals with summations \cite{lejeune2021}. The equation for force density $\mathbf{f}_{\mathbf{x}\mathbf{y}}$ is based on ordinary state-based peridynamics and is a function of collective deformation \cite{silling2007}. In computing the force density vectors for different problems, we refer the reader to the peridynamics literature \cite{silling2019,madenci2014,oterkus2015} and the references therein.

Investigating damage problems, particularly in the context of materials with flexoelectricity, adds an additional layer of complexity to the analysis. Flexoelectricity refers to the coupling between strain gradients and electric polarization, and its consideration is crucial for understanding the behaviour of certain materials. The peridynamics formulation for flexoelectricity gives insight into how this nonlocal theory could be developed to incorporate the effects of flexoelectric coupling in materials \cite{roy2019}. Many materials have structures on several length scales, and the structural hierarchy can play a significant role in defining bulk material characteristics \cite{lakes1993, kamath2006}. Furthermore, different stages and challenges have to be faced in understanding and utilizing the flexoelectricity in materials \cite{wang2019}. In addition, nonlocal continuum mechanics models are critical for capturing the impacts of long-range interactions and resolving concerns such as material failure, damage, and scale effects. Eringen introduced nonlocal elasticity and gave answers to particular situations such as screw dislocations and surface waves \cite{eringen1983}. Nonlocal models account for material inhomogeneities and the effect of gradients in the localization of deformation and fracture \cite{singh2020, aifantis1992, krishnaswamy2020c}.

\subsection{Nonlocal boundary and initial conditions and their approximations}

We have discussed so far the problems where the state of a system is specified at the boundaries of the domain or initial conditions for time-dependent problems given as values at a single point in time. An extension of the conventional local boundary and initial conditions used in solving differential equations is the nonlocal boundary and initial conditions \cite{bouziani1997, merad2015, sytnyk2021}. This is especially true for systems in which the boundary conditions are constraints based on the value of the function at the boundary and how it behaves over a larger region, or the initial conditions are based on the system's past results. These problems play a crucial role in mathematical biology \cite{ripoll2023}. In addition, they can capture the influence of interactions extended beyond the immediate neighbours of many biological processes. For instance, it could represent (i) the migration of species into or out of the study area in ecological models, (ii) the impact of neighbouring cells or tissues at the boundaries of a simulated domain, (iii) the impact of external factors on the initial distribution of a disease when the introduction of the disease happens from outside the modelled region, etc. Here, we first consider a hyperbolic integro-differential equation that models the quasistatic flexure of a thermoelastic rod as \cite{merad2015}:
\begin{equation}{\label{NLBVP}}
    \frac{\partial^{2}u}{\partial t^{2}} - \frac{\partial^{2}u}{\partial x^{2}} = \int_{0}^{t}a(t-s)u(x,s)ds,~~0<x<1,~~0<t\leq T,
\end{equation}
with the initial conditions $u(x,0) = f(x)$, $u_{t}(x,0) = g(x)$ for $0<x<1$ and the nonlocal conditions $$\int_{0}^{1}u(x,t)dx = p(t)~\mbox{and}~\int_{0}^{1}xu(x,t)dx = q(t),~~0<t\leq T,$$
where $f,g,p$, and $q$ are the given functions, $T$ is a positive constant, and $a$ is a suitable defined function that satisfies some conditions \cite{merad2015}. This is an example of a nonlocal boundary conditions problem, as the boundary conditions involve all the points in the spatial domain $[0,1]$. Next, we move to the notion of nonlocal initial value problems. We consider a general form of a mathematical model in a bounded domain $\Omega\subset\mathbb{R}^{n}$ as \cite{ntouyas2006}:
\begin{equation}{\label{NLIVP}}
    \begin{aligned}
    &Lu + c(x,t)u = F(x,t),~~x\in\Omega, 0<t<T,\\
    &u(x,t) = G(x,t),~~x\in\partial\Omega, 0<t<T, \\
    &u(x,0) + \sum_{k=1}^{N}\beta_{k}(x)u(x,t_{k}) = H(x),~~x\in\Omega,
\end{aligned}
\end{equation}
where $t_{k}\in(0,T], k=1,\ldots, N,$ and $L$ is a parabolic operator with continuous and bounded coefficients. Due to the complexity of analysis and computation, these models require sophisticated mathematical and numerical approaches.

\subsection{Physics-informed neural networks}

Physics-informed neural networks offer a reliable technique which combines physics concepts with the flexibility and learning capabilities of neural networks. These models are useful for tackling complex physical problems when obtaining analytical solutions may be difficult or impossible. This technique can solve physics-based PDEs with the help of neural network architecture. It consists of two main components: a neural network and a loss function. The neural network parameterizes the solution to the problem, while the loss function ensures that the solution is close to the governing physics equations. Here, we describe a general methodology to deal with such problems. Let us consider a nonlinear partial differential equation in $\Omega\subset\mathbb{R}^{n}$ \cite{raissi2019}:
\begin{equation}{\label{PINN}}
    u_{t}+\mathcal{N}[u] = 0,~~x\in\Omega, t\in [0,T],
\end{equation}
where $u(x,t)$ is the hidden solution and $\mathcal{N}[\cdot]$ is a nonlinear differential operator. Now we define $f(x,t):= u_t+\mathcal{N}[u]$ and proceed by using a deep neural network to approximate $u(x,t)$. The common parameters of the neural networks $u(x,t)$ and $f(x,t)$ can be trained by reducing the mean squared error loss: $$MSEL = \frac{1}{N_{u}}\sum_{j=1}^{N_{u}}|u(x_{u}^{j},t_{u}^{j})-u^{j}|^{2}+\frac{1}{N_{f}}\sum_{j=1}^{N_{f}}|f(x_{f}^{j},t_{f}^{j})|^{2},$$ 
where $\{ x_{u}^{j},t_{u}^{j},{u}^{j}\}_{j=1}^{N_{u}}$ are the initial and boundary training data on $u(x,t)$ and $\{x_{f}^{j},t_{f}^{j}\}_{j=1}^{N_{f}}$ ate the collocations points for $f(x,t)$. If we apply this methodology to Navier-Stokes equations in two dimensions, we obtain the mean squared error (MSE) loss function as \cite{raissi2019}: 
$$MSE:=\frac{1}{N}\sum_{j=1}^{N}\bigg{(} |u(x^{j},y^{j},t^{j})-u^{j}|^{2}+ |v(x^{j},y^{j},t^{j})-v^{j}|^{2} \bigg{)}+\frac{1}{N}\sum_{j=1}^{N}\bigg{(} |f(x^{j},y^{j},t^{j})|^{2}+|g(x^{j},y^{j},t^{j})|^{2} \bigg{)},$$
where $f:=u_{t}+\lambda_{1}(uu_{x}+vu_{y})+p_{x}-\lambda_{2}(u_{xx}+u_{yy})$ and $g:=v_{t}+\lambda_{1}(uv_{x}+vv_{y})+p_{y}-\lambda_{2}(v_{xx}+v_{yy})$ with $p(x,y,t)$ as the pressure and $\lambda_{1}$ and $\lambda_{2}$ are the unknown parameters. A fundamental advantage of utilizing neural networks in PINNs is their capacity to execute automated differentiation, which allows the network to effectively compute derivatives of the anticipated function (in terms of time and space). In addition, the key concept behind PINN techniques is to encode the underlying physical rule (i.e., the PDE) as prior information in the neural network. However, the PINN technique fails to produce realistic approximations of the solution when the PDE contains non-convex flow functions, such as shocks and mixed waves \cite{fuks2020}. Nevertheless, training PINNs requires careful initialization, especially when working with stiff equations or highly nonlinear systems.

\subsection{Modelling nonequilibrium phenomena and processes with nonlocal models}

Nonequilibrium phenomena occur when a system is not in equilibrium, i.e., it has net flows of energy, matter, or information \cite{de2013}. These systems often exhibit complex, time-dependent behaviour, whereas the macroscopic properties of equilibrium systems are stable and time-invariant. In addition, the entropy is constantly produced for nonequilibrium processes; however, it is maximal and constant for the equilibrium case. These nonequilibrium systems have some key characteristics, such as self-organization behaviours \cite{nicolis1977} (e.g., pattern formation in chemical reactions and ecological systems), irreversibility (e.g., processes such as diffusion, heat conduction, and chemical reactions get the system closer to equilibrium over time but never entirely achieve it), etc. 

Using nonlocal models to investigate nonequilibrium events adds an exciting dimension to understanding dynamic processes caused by long-distance interactions. Traditional equilibrium models assume that systems are in a stable state, but many real-world situations involve constant change and flux. In the context of nonequilibrium phenomena, nonlocal models can be applied to various fields, such as physics, fluid dynamics, and materials science. Nonlocal models can describe the behaviour of particles and waves in systems that deviate from equilibrium. For instance, they can be used to study the transport of heat or charge in materials with non-uniform structures. Here, we describe a nonlocal model of an electron-hole semiconductor plasma, which accounts for the nonequilibrium phenomena. An electron system's electro-hydrodynamic model is commonly expressed as \cite{blotekjaer1970,melnik2000}: 
\begin{equation}{\label{EHM}}
    \frac{\partial \mathbf{z}}{\partial t} = \zeta + \bigg{(}\frac{\partial \mathbf{z}}{\partial t}\bigg{)}_{\mbox{col}},
\end{equation}
where $\mathbf{z} = (n,\mathbf{v},W)^{T}$, $\zeta = (\mathcal{F}_{1},\mathbf{\mathcal{F}}_{2},\mathcal{F}_{3})^{T}$, $\mathcal{F}_{1} = -\nabla\cdot(n\mathbf{v})$, $\mathbf{\mathcal{F}}_{2} = -\mathbf{v}\cdot\nabla\mathbf{v} - q\mathbf{E}_{\mbox{eff}}/m_{n}-\nabla(nT_{n})/(m_{n}n)$, and $\mathcal{F}_{3} = -\nabla\cdot(\mathbf{v}W)-qn\mathbf{v}\cdot\mathbf{E}_{\mbox{eff}}-\nabla\cdot(\mathbf{v}nT_{n})-\nabla\cdot \mathbf{q}$. Here, $n$ is the electron concentration, $v$ is their average velocity, $W$ is the energy density, $\mathbf{E}_{\mbox{eff}}$ is the effective electric strain, $m_{n}$ is the effective electron mass, $T_{n}$ is the electron temperature given in energetic units, and $\mathbf{q} (\sim -k\nabla T_{n})$ is the heat flow. The simplified model of this can be investigated mathematically by developing drift-diffusion models; however, they are incompatible with technological improvements. In \cite{melnik2000}, the authors reduced the computational cost with the help of normalization and described several non-stationary physical phenomena in semiconductor devices, including carrier heating and velocity overshoot. This reduced model also accounts for the nonequilibrium and nonlocal character of electron-hole semiconductor plasma. 

\subsection{Nonlocal many-particle systems with AI and machine learning}

Nonlocal interactions in multi-body systems extend beyond the immediate neighbours of individual bodies because multiple species evolve in an ecosystem and interact with themselves and others. For instance, species sense their surroundings to make decisions, e.g., the predator finds the prey, the prey avoids the predator, and the aggregation in swarms, called nonlocal sensing \cite{eftimie2007, levine2000, mogilner1999}. This nonlocal sensing happens on different scales, affecting populations' spatial distributions leading to species aggregation, segregation, or complex patterns \cite{bastille2018, benhamou2014, martinez2020, potts2014, lee2001, jewell2023, painter2024}. Researchers have studied these kinds of biological problems mathematically by aggregation-diffusion equation \cite{giunta2021,carrillo2019,topaz2006, painter2024}. For a single species model, it leads to the following form:
\begin{equation}{\label{SSADE}}
    \frac{\partial u}{\partial t} = \Delta u^{m}-\nabla \cdot [u\nabla (\phi\ast u)],
\end{equation}
where $m$ is a positive integer. This model can generate a non-uniform stationary pattern consisting of single or multiple patches with various shapes and sizes \cite{james2015,craig2016}. A generalization of the aggregation-diffusion equation (\ref{SSADE}) for multi-species populations is \cite{giunta2021}:
\begin{equation}{\label{GADE}}
    \frac{\partial u_{j}}{\partial t} = D_{j}\Delta u_{j}^{m} - \nabla\cdot \bigg{[} u_{j}\nabla \sum_{k=1}^{N}h_{jk}\phi\ast u_{j} \bigg{]},~~ j=1,\ldots, N,
\end{equation}
where $u_{j}(x,t)$ are the densities of $N (\geq 1)$ populations at the spatial location $x$ and at time $t$. Here, $D_{j} (>0)$ is the diffusion coefficient of the population $j$ and $h_{jk}$ are constants that denote the attractive or repulsive tendencies (depending on its sign) of the population $j$ to the population $k$. Here, births and deaths have been neglected in the model because many animals make their spatial arrangement on shorter time scales. Authors in \cite{potts2019} studied the model (\ref{GADE}) for $m=1$ with the top-hat kernel function $\phi$ and incorporated the detection behaviour of the populations distributed over habitat in animal ecosystems. They have explained the detection in various forms, e.g., direct observations of the individuals at a distance, indirect communication between individuals via marking the environment, and the memory of past interactions with other populations \cite{giunta2021, potts2019, watson1973}. 

Nonlocal models for crowd dynamics come from studying many-body issues, gaining influence from the Boltzmann model and growing within a hierarchy of mathematical models \cite{thieu2023}. The Boltzmann model is a core model for understanding many-body systems and offers a platform for extending notions to crowd dynamics. In addition, for the many-body problem, each individual's behaviour impacts and is impacted by the activities of others \cite{contreras2025}. Traditional models frequently focus on local interactions, but nonlocal models incorporate effects throughout the entire population. Due to the complexities of real-world events, validating and calibrating nonlocal models for crowd dynamics is not easy. 

Data-driven nonlocal models in biology are an exciting intersection of computational methods and biological research \cite{srivastava2022}. These models leverage data to understand and simulate complex biological systems, considering interactions that occur over long distances. They can be applied to study phenomena like the spread of diseases, signalling pathways in cells, or ecological interactions in a community. By incorporating data into these models, researchers can gain a more comprehensive understanding of the underlying processes and make predictions based on real-world observations. 

Data-driven nonlocal models are effective tools for solving various inverse issues in a range of scientific fields \cite{arridge2019}. When compared to conventional approaches, they enable more robust and accurate solutions to inverse issues by utilizing the knowledge found in nonlocal relationships within the data. These models effectively reconstruct high-quality images from noisy or partial data in computer vision and medical imaging \cite{li2023}. They improve reconstruction accuracy by accounting for nonlocal similarities between image patches. In addition, the nonlocal technique in signal processing makes it possible to recover signals from undersampled or distorted measurements more effectively. The goal of various ongoing research projects is to create effective algorithms to deal with these problems. Among them is the Bayesian approach, which offers a probabilistic framework for quantifying uncertainty that combines known information with empirical data in an easy-to-use manner \cite{xie2014}. Instead of treating model parameters as fixed values, it treats them as probability distributions. The posterior distribution is then obtained by applying Bayes' theorem to update the prior distribution of beliefs and combine it with the likelihood of observed data. Because algorithms take uncertainty into account, Bayesian models are naturally resilient to small amounts of data. This is especially useful in situations when acquiring huge datasets is difficult.

AI has revolutionized all fields, including biology \cite{bellomo2024}, and helps analyze and interpret the vast amount of biological data used to develop nonlocal models. This includes tasks such as learning, reasoning, problem-solving, perception, and language understanding. On the other hand, the generative models learn the underlying patterns and structures of the data and can then create new samples that share similar characteristics with the help of machine learning or deep learning algorithms \cite{boussange2023, anstine2023, farajpour2024}. In addition, these models provide an emerging collection of tools with enhanced characteristics for recreating segregated and whole-brain dynamics \cite{ramezanian2022}. Furthermore, generative models in AI and learning processes are based on non-equilibrium dynamics \cite{carbone2023}. 

Several types of machine learning algorithms are available in the literature, which can be generally classified into three categories: supervised learning, unsupervised learning, and reinforcement learning \cite{srivastava2022}. In supervised learning, the algorithm is trained using a labelled dataset to identify the mapping function from inputs to outputs, which allows the algorithm to generate predictions or classifications on previously unknown data. In the case of unsupervised learning, the algorithm searches for patterns, correlations, or structures in the data. Agent learning algorithms are used in reinforcement learning to make decisions by interacting with the environment. The agent receives feedback in the form of benefits based on the behaviours it takes. Here, we describe a data-driven technique using an optimization-based methodology to learn nonlocal constitutive laws for stress wave propagation models. For this, we consider a high-fidelity model in a domain $\Omega\subset\mathbb{R}^{d}$ \cite{you2021}:
\begin{equation*}
    \mathcal{L}_{HF}[u] = f(x),
\end{equation*}
subject to the boundary conditions $\mathcal{B}u = g(x), x\in\partial\Omega$. The operator $\mathcal{L}_{HF}$ is a high-fidelity operator, e.g., differential or integral operator, and $f$ is the forcing term. Suppose the solution to this high-fidelity problem can be approximated by the solution to the nonlocal model:
\begin{equation}{\label{NLHF}}
    \mathcal{L}_{K}[u] = f(x),~x\in\Omega_{I},
\end{equation}
with the boundary condition $\mathcal{B}_{I}u = g(x), x\in\partial\Omega$ where $\Omega_{I}$ is an appropriate nonlocal interaction domain and $\mathcal{B}_{I}$ is the corresponding nonlocal interaction operator specifying a volume constraint \cite{you2021}. Here, $\mathcal{L}_{K}$ is a nonlocal operator defined by $\mathcal{L}_{K}[u](x) = \int_{\overline{\Omega}}K(|x-y|)(u(y)-u(x))dy$. To learn the kernel $K$, we assume that $N$ pairs of values $\mathcal{D} = \{ (u_{j},f_{j})\}_{j=1}^{N}$ are given for the forcing terms $f_{j}$ and the high-fidelity solutions $u_{j}$. In this setup, the goal is to optimize the following problem
\begin{equation*}
    K^{*} = \argmin_{K}\frac{1}{N}\sum_{j=1}^{N}||\mathcal{L}_{K}[u_{j}]-f_{j}||_{\mathcal{X}},
\end{equation*}
where $||\cdot||_{\mathcal{X}}$ an appropriate norm over $\Omega$. Authors in \cite{you2021} described a detailed algorithm to optimize this problem. In addition, this type of methodology can also be applied to hyperbolic problems.

\subsection{Multiscale and multifidelity models}

In computational science and engineering, various multilevel methods have been applied to mathematical models, such as multigrid for solving systems of equations, multilevel Monte Carlo and multilevel stochastic collocation for estimating mean solutions of partial differential equations with stochastic parameters, and so on. The common approach is using surrogate models, simplified models trained on a combination of high and low-fidelity data. These surrogate models can then be used to quickly approximate solutions to the nonlocal models. In these techniques, a parameter is involved and controls the trade-off between error and computational costs. Here, we first describe the Monte Carlo method for a general problem to determine approximate statistical information of the high-fidelity model with low-fidelity surrogate models \cite{peherstorfer2016}. 

Let $f:\mathcal{D}\rightarrow \mathcal{Y}$ be a function for the source of information, where $\mathcal{D}\subset\mathbb{R}^{d}$ for some $d\in\mathbb{N}$ and $\mathcal{Y}\subset\mathbb{R}$. Here, we consider the high-fidelity model $f^{(1)}$ as the ``truth'' model and the low-fidelity models $f^{(2)},\ldots,f^{(k)}$ as surrogate models. For $j=1,\ldots,k$, we assume the costs of evaluating each of the models $f^{(j)}$ are $w_{j}\geq 0$. For the Monte Carlo method, the authors in \cite{peherstorfer2016} draw independent and identically distributed (i.i.d) realizations $\mathbf{z}_{1},\ldots,\mathbf{z}_{m}\in \mathcal{D}$ of the random variable $Z:\Omega\rightarrow\mathcal{D}$, where $\Omega$ is a sample space, and estimated $\mathbb{E}[f^{(j)}(Z)]$ by 
$$\overline{y}_{m}^{(j)} = \frac{1}{m}\sum_{k=1}^{m}f^{(j)}(\mathbf{z}_{k}),~~j=1,\ldots,k.$$ 
Now, for $j=1,\ldots,k$, this Monte Carlo estimator is an unbiased estimator of $\mathbb{E}[f^{(j)}(Z)]$, and if the variance of $f^{(j)}(Z)$ is finite, then the MSE of the estimator $\overline{y}_{m}^{(j)}$ with respect to $\mathbb{E}[f^{(j)}(Z)]$ is $$e(\overline{y}_{m}^{(j)}) = \mathbb{E}\bigg{[} \bigg{(} \mathbb{E}[f^{(j)}(Z)]-\overline{y}_{m}^{(j)} \bigg{)}^{2}\bigg{]} = \frac{\mbox{Var}[f^{(j)}(Z)]}{m},$$
and the cost of computing the Monte Carlo estimator is $c(\overline{y}_{m}^{(j)}) = w_{j}m.$ In this case, our goal is to estimate the expectation $s = \mathbb{E}[f^{(1)}(Z)]$ with realizations of the random variable $Z$ as inputs. 

The authors in \cite{peherstorfer2016} showed the MSE for the multifidelity Monte Carlo (MFMC) method is lower than the Monte Carlo estimator with the same computational cost. The authors considered $k$ number of i.i.d realizations $\mathbf{z}_{m_{1}},\ldots,\mathbf{z}_{m_{k}}\in \mathcal{D}$ for the $k$ models $f^{(1)},\ldots,f^{(k)}$, where $m_{j}\in\mathbb{N}$ and they satisfy $0<m_{1}\leq \ldots\leq m_{k}$. In this case, the MFMC estimates $$\widehat{s} = \overline{y}_{m_{1}}^{(1)}+\sum_{j=1}^{k}\alpha_{j}(\overline{y}_{m_{j}}^{(j)}-\overline{y}_{m_{j-1}}^{(j)}),$$
where $\alpha_{2},\ldots,\alpha_{k}\in\mathbb{R}$ are the 
weights for the differences of the Monte Carlo estimators $\overline{y}_{m_{j}}^{(j)}$ and $\overline{y}_{m_{j-1}}^{(j)}$, $j=2,\ldots,k$. This MFMC method can be applied to get a lower cost without compromising accuracy for the steady-state nonlocal diffusion model \cite{khodabakhshi2021}:
\begin{equation}{\label{MSMFM}}
    \begin{aligned}
        -\int_{x-\delta}^{x+\delta}(u_{\delta}(y)-u_{\delta}(x))\Gamma (x,y)dy &= b(x),~~x\in\Omega,\\
        u_{\delta}(x) &= g(x),~~x\in\Omega_{I},
    \end{aligned}
\end{equation}
where $\Omega = (0,L)\subset\mathbb{R}$, $\Omega_{I} = \{ y\in\mathbb{R}\smallsetminus\Omega:|x-y|\leq\delta~\mbox{for some}~x\in\Omega\} = [-\delta,0]\cup [L,L+\delta]$ for a length scale $\delta>0$ and $b(x)$, $g(x)$, and $\Gamma (x,y)$ are defined on $\Omega$, $\Omega_{I}$, and $(\Omega \cup\Omega_{I})\times(\Omega \cup\Omega_{I})$, respectively. 

Data resolution is crucial for data-driven models as it affects their accuracy, interpretability, and application. Low-resolution data is often more aggregated or sparse and may not capture slight changes in the underlying phenomenon. In this situation, models trained on low-resolution data may oversimplify complicated connections, resulting in restricted prediction ability and missing critical details. High-resolution data, on the other hand, offers extensive information, catching nuanced patterns and variations within the dataset, which can produce more accurate forecasts and identify subtle linkages. They may, however, be more computationally intensive and prone to overfitting if not appropriately managed. Despite limitations on low-resolution data, it can still be useful, especially where high-resolution data is impracticable or prohibitively expensive to gather. It is frequently utilized in large-scale environmental monitoring, satellite image processing, and healthcare applications.

\subsection{Nonlocal interactions on networks, network coupling and integration}

So far, we have discussed the models using PDEs or other continuous mathematical frameworks that model the spatial distribution of variables, such as concentrations of substances, populations, or other physical quantities. Here, we discuss the network models where systems are composed of discrete entities (nodes) and the relationships (edges) between them. Networks do not require spatial continuity and can model interactions between distant entities. Nonlocal interactions on networks in biology refer to interactions between nodes in a network (such as genes, proteins, or cells) that extend beyond their direct neighbours or immediate connections. These interactions are essential for understanding the processes and characteristics of complex biological systems in a network. This nonlocality is essential in systems where connections do not follow spatially continuous patterns, such as social networks, signalling networks, ecological food webs, and neuronal networks. While space-continuous systems and network models appear fundamentally different, they can provide complementary perspectives on complex systems. Space-continuous models are adopted in capturing continuous spatial dynamics, while network models are well-suited for nonlocal, discrete interactions.

Neurons in the brain communicate and interact through synapses, which can involve long-range connections \cite{uzel2022, zitnik2024}. Nonlocal models have been employed to describe neurons' collective activity and synchronization in neural networks. In addition, Alzheimer's disease (AD) is also growing fast in the current days, and researchers have been trying to understand the mechanisms behind its propagation. Many factors are involved in AD progression, but two protein families, amyloid-beta and tau proteins, are known to be the main contributors \cite{matthaus2009, bressloff2014}. In \cite{thompson2020}, authors have considered the heterodimer model to incorporate the interaction between these two proteins, each consisting of healthy and toxic densities. Also, they derived the network model corresponding to the PDE model to apply it to AD progression in the brain connectome. Authors in \cite{pal2022sr} modified the heterodimer model by introducing nonlocal interactions into the model. They studied the following nonlocal network model
\begin{equation}{\label{HCMN}}
\begin{aligned}
\frac{du_{j}}{d t} & = -\sum_{k=1}^{N}L_{jk}u_{k}+ u_{j}(a_{0} -a_{1}u_{j}) - \frac{a_{2}u_{j}}{1+c_{u}u_{j}}\Phi_{j}\ast\widetilde{u}_{j}, \\
\frac{d\widetilde{u}_{j}}{d t} & =  -\sum_{k=1}^{N}L_{jk}\widetilde{u}_{k} -\widetilde{a}_{1}\widetilde{u}_{j} + a_{2}\widetilde{u}_{j}\Phi_{j}\ast \bigg{(} \frac{u_{j}}{1+c_{u}u_{j}}\bigg{)},\\
\frac{dv_{j}}{dt} & = -\sum_{k=1}^{N}L_{jk}v_{k} + v_{j}(b_{0} -b_{1}v_{j}) -b_{3}\widetilde{u}_{j}v_{j}\widetilde{v}_{j}- \frac{b_{2}v_{j}}{1+c_{v}v_{j}}\Phi_{j}\ast\widetilde{v}_{j},\\
\frac{d\widetilde{v}_{j}}{dt} & = -\sum_{k=1}^{N}L_{jk}\widetilde{v}_{k}  -\widetilde{b}_{1}\widetilde{v}_{j} +b_{3}\widetilde{u}_{j}v_{j}\widetilde{v}_{j} + b_{2}\widetilde{v}_{j}\Phi_{j}\ast \bigg{(} \frac{v_{j}}{1+c_{v}v_{j}}\bigg{)},
\end{aligned}
\end{equation}
where $u_{j}$ and $\widetilde{u}_{j}$ are the densities of healthy and toxic amyloid beta, and $v_{j}$ and $\widetilde{v}_{j}$ are the densities of healthy and toxic tau proteins at the node $j$ in a given graph $G$. Here, $N$ is the total number of nodes, and $L_{ij}$ are the entries of the Laplacian matrix for the given network \cite{thompson2020, pal2022sr}. Here, we define the convolution term for a given graph $G$ \cite{pal2022sr}. For a fixed node $j$ in the network, let $S_{j,1}$ be the set containing all directly connected nodes to the node $j$. We then find the nodes in the graph that are connected to the nodes in the set $S_{j,1}$ except for the node $j$, and we refer to this set as $S_{j,2}$. This process continues until we find the complete set of nodes as $S_{j,1}, \ldots, S_{j,m_{j}}$, and $S_{j}$ is the union of this complete set. Following the approach outlined in \cite{pal2022sr}, we find the weight vector $V_{j}$ of $N$ elements, and the convolution term for the node $j$ becomes:
$$\Phi_{j}\ast\widetilde{u_{j}} = \sum_{n=1}^{N}V_{j}(n)\widetilde{u}_{n}.$$
We can similarly compute the other convolution terms used in the model (\ref{HCMN}).

There are nonlocal mathematical models on neural networks, especially convolutional neural networks. The traditional neural network blocks feature representations in a local sense; however, long-range dependencies have significant practical learning problems. Here, we describe a nonlocal network in the context of traditional image classification tasks and compare different neural networks. Suppose $X = [X_{1},\ldots,X_{M}]$ is an input sample with $X_{j} (j=1,\dots,M)$ as the feature at position $j$. A nonlocal output signal at position $j$ can be defined as \cite{tao2018}:
\begin{equation*}
    Z_{j} = X_{j}+\frac{W_{Z}}{\mathcal{C}_{j}(X)}\sum_{k}\omega (X_{j},X_{k})g(X_{k}),
\end{equation*}
where $W_{Z}$ is the weight matrix, $C_{j}(X)$ is the normalization factor, $\omega$ is a pairwise function which computes a scalar between the positions $j$ and $k$, and $g$ is a given function, e.g., $g(X_{k}) = W_{g}X_{k}$. Incorporating these nonlocal blocks into a residual network, the network model can be written as:
\begin{equation}{\label{NNNM}}
    Z^{n+1} := Z^{n} + \mathcal{F}(Z^{n};W^{n}),~~n=0,1,\ldots,N,
\end{equation}
where $W^{n}$ is a parameter set with $N$ number of network blocks with the initial input sample $Z^{0} = X$. 

Genetic regulatory networks (GRNs) are important in coordinating gene expression and orchestrating complex developmental processes in cells. These networks involve transcription factors binding to particular DNA sequences, cis-regulatory elements, and other substances, resulting in a dynamic and linked web of regulatory interactions. Understanding the abnormal alterations in GRNs gives insight into several disease pathways, such as cancer and developmental disorders \cite{nalls2021}. Stochastic transcriptional regulation adds a layer of complexity to GRNs, altering cellular heterogeneity, phenotypic variability, and the general adaptability of biological systems \cite{melnik2009a,thieu2021,kim2023}. Furthermore, pattern memory in GRNs is important in cellular decision-making processes, allowing cells to respond to the same stimuli differently based on their experience.

The spatio-temporal dynamics of biomolecular networks and biochemical systems are related to the complex dynamic patterns of molecular interactions that occur both in space (inside cellular compartments) and over time \cite{kim2018}. The spatial organization impacts many biomolecular interactions in various cellular compartments, such as the nucleus, cytoplasm, and organelles. Furthermore, spatio-temporal dynamics in metabolic networks are critical for coordinating energy generation, nutrient use, and maintaining cellular homeostasis \cite{peukert2014}. Nonetheless, spatial gradients of signalling molecules contribute to generating concentration gradients, impacting cell fate and pattern formation throughout development. The dynamics of transcriptional regulation involve both local and nonlocal components, reflecting the complex interplay of molecular processes within a cell and interactions that extend beyond immediate neighbours. Local dynamics often relate to nearby activities, such as interactions between genes and transcription factors. In contrast, nonlocal dynamics involve events that affect distant genomic areas or operate across distant locations and temporal dimensions. 

Nonlocal models are effective for understanding complex dynamics that involve long-range interactions. However, their complexity and computational requirements can sometimes restrict their practical application. It is important to balance accuracy and feasibility when using these models. Here, we provide an overview of the mathematical models discussed so far by highlighting their strengths and limitations:
\begin{center}
\begin{longtable}{ | m{5.0em} | m{6.4cm}| m{6.6cm} | } 
\hline
Models & Advantages & Limitations \\ 
\hline
RD model (\ref{GRDE}) & It has a wide range of applications, such as in chemical reactions, biological pattern formation, and even ecological systems & The assumption of continuity and homogeneity might not be valid in all mediums such as cellular or molecular discreteness \\ 
\hline
Fisher-KPP model (\ref{LIDE}) & This model describes the system where interactions are not limited to immediate neighbours but extend over a broader range & The range and the mathematical form of the nonlocal interactions can be difficult to measure from empirical observations\\ 
\hline
Aggregation model (\ref{BSPM}) & The aggregation and depletion of populations can be captured by these models  & It may not be appropriate for systems where local interactions dominate\\ 
\hline
Space-time model (\ref{NFBNM}) & It describes distant interactions and the delayed responses of populations that shift in space and time & High-dimensional integrals increase the computational costs and longer simulation times\\ 
\hline
Agent-based model (\ref{NLCM}) & It gives insights into complex biological systems through the simulation of individual agents & Analyzing the nonlocal model analytically is challenging\\ 
\hline
Dispersal model (\ref{DKRM}) & It can accurately represent situations where entities are spread out over long distances, such as the spread of diseases & It is specific to the particular system and can not be generalized to other environments\\ 
\hline
Fractional model (\ref{TFCE}) & It adds memory effects, allowing these models to explain systems where the current state depends on the overall history of the system & Establishing the suitable fractional order and other parameters in these models can be difficult; especially for the noise or sparse experimental data\\ 
\hline
Epidemio-logical model (\ref{SIRM1}) & These models accurately represent disease distribution across regions and countries, which is critical for understanding global pandemics and epidemics & Extensive data on long-distance interactions, travel patterns, and network structure are required to parameterize nonlocal models effectively \\
\hline
COVID-19 model (\ref{NMC19}) & It can capture the spatial distribution of infections while the model (\ref{SIRM1}) does not & Obtaining high-quality data on population distribution and mobility patterns \\ 
\hline
Time-relaxation model (\ref{SDSIRM1}) & It represents the nonlocal dependency of the latent period in disease spread to consider memory effects; as a result, the system's past states influence its current state & It has computational and implementation challenges, especially for large complex systems, and one can avoid them if memory effects are minimal \\ 
\hline
Age-dependent model (\ref{TSSDM}) & These models can take care of age-dependent death and diffusion rates & The accuracy of the model is heavily dependent on the quality of the data used and inaccurate data can lead to unreliable model predictions \\ 
\hline
Cell population model (\ref{CONEA}) & These models can incorporate the effect of adhesion between cells on their movement & Deriving conditions for boundedness is substantially harder in higher dimensions than in one dimension \\ 
\hline
Resource-dependent model (\ref{LVP}) & These models represent crop raiding effects of large-bodied mammals living in the biodiversity-rich tropics & These approaches can not be applied if the mammals have sufficient resources everywhere in the spatial domain\\ 
\hline
Intra-specific competition model (\ref{TSINLM}) & It offers the intra-specific competition of the prey individuals when they compete for resources over larger distances & The complexity increases considerably to account for interactions over a broader range of nonlocal interactions, which raises computational demands \\ 
\hline
Intra-specific-pest competition model (\ref{SPNLM}) & These models account for the nonlocal intra-specific competition among prey individuals and pest-type nonlocal interaction with other species & The complexity increases considerably to account for interactions over a broader range of nonlocal interactions, which raises computational demands \\ 
\hline
Shadow model (\ref{SRDS}) & This process reduces the complexity of the original reaction-diffusion model by eliminating the fast-diffusing component and makes the system easier to analyze mathematically & This reduction may miss transient behaviours, such as wave propagation or oscillations, that could be important in certain biological contexts \\ 
\hline
Competition model (\ref{NLLVCM}) & It can incorporate the nonlocal intra- and inter-specific competitions of the species  & Nonlocal interaction terms introduce additional parameters related to the range and strength of the nonlocal effects, which can be difficult to estimate from empirical data \\ 
\hline
Hyperbolic-parabolic model (\ref{MHPE}) & This model accounts for the predators' flow towards the prey population  & The choice of the velocity function is crucial and depends on the nature of the predator-prey interactions \\ 
\hline
Mean field approach (\ref{NHE}) & This model incorporates the phenomena that the reaction between two chemical particles depends on the reaction radius & The effective reaction radius can change with varying temperature, pressure, or medium \\ 
\hline
Models in game theory (\ref{NLGT1}) and (\ref{NLGT2}) & These models can capture the long-range spatial heterogeneity of the strategy space  & Large strategy space increases the complexity and computational cost of the model \\ 
\hline
Models in elasticity (\ref{NLEE}) &  This theory is highly effective in modelling materials at small scales (e.g., nanomaterials) where long-range atomic interactions significantly affect mechanical behaviour & Finding appropriate kernels related to strains and stress is not straightforward and often requires empirical data \\ 
\hline
Model in materials and structures (\ref{PDYE}) & It can naturally handle discontinuities for modelling materials and structures (e.g., cracks and fractures) without requiring special techniques such as introducing singularities & It is challenging when microcrack growth isn't determined by local deformation or stress \\ 
\hline
Initial and boundary value problems (\ref{NLBVP}) - (\ref{NLIVP}) & These models can incorporate the measurements at more spatial and time points  & The mathematical and computational parts of these problems are challenging, including the weights related to long-range interactions \\ 
\hline
Physics-informed neural network (\ref{PINN}) & It can integrate the physical laws into the training process and estimate the parameter values involved in the laws & This approach struggles with problems that have discontinuities, sharp gradients, or complex geometries \\ 
\hline
Non-equilibrium model (\ref{EHM}) & This model can be simplified and mathematically tractable and can reduce the computational cost  & Only low carrier densities and small electric fields allow for the rigorous justification of these drift-diffusion models \\ 
\hline
Multispecies aggregation model (\ref{GADE}) & These models incorporate collective behaviours, including swarming, flocking, and pattern creation, which are important in understanding how animals or agents act in groups & Reaction terms need to include for capturing quantities change over time due to internal dynamics, such as population growth, competition, predation, etc. \\ 
\hline
High-fidelity model (\ref{NLHF}) & Optimisation-based approaches enable the model to learn constitutive laws from data by reducing the differences between predicted and observed patterns  & These methods can suffer from convergence issues, particularly in high-dimensional or highly nonlinear problems \\ 
\hline
Multiscale and multifidelity model (\ref{MSMFM}) &  It reduces the overall workload without significantly sacrificing accuracy by using low-cost surrogates for most computations and high-fidelity models when necessary & Choosing the appropriate surrogate model is not straightforward, which can impact substantially model accuracy, performance, and computational efficiency \\
\hline
Nonlocal network model (\ref{HCMN}) & It offers a more comprehensive understanding of the structural connectivity and functions of the brain & Integrating data from different sources (e.g., MRI, fMRI, PET scans) into nonlocal network models can be difficult\\ 
\hline
Model in neural network (\ref{NNNM}) & This nonlocal model can quantify the characteristics of the damping effect of nonlocal blocks in a network and is robust to the number of nonlocal blocks & Due to the learn long-range features in the network; it needs extra work to study the initialization strategy or to fine-tune the parameters \\
\hline
\end{longtable}
\end{center}

\section{Applications of Nonlocal Models in Life and Health Sciences}\label{SE3}

Nonlocal models find diverse applications in life and health sciences, offering a valuable framework for capturing spatial and temporal interactions that extend beyond immediate neighbours. These key applications include (i) the intraspecific competition between individuals for accessing food resources, (ii) the spatial spread of infectious diseases, (iii) tumor growth, considering the influence of nonlocal interactions on the proliferation and migration of cancer cells, (iv) spread of neural activity and the interactions between distant brain regions, (v) predicting tissue repair dynamics and optimizing regenerative medicine approaches, (vi) health informatics by predicting disease outcomes, analyzing medical images, and identifying patterns in health data, etc.

\subsection{Ecology and epidemiology}

The classical Fisher-KPP model and its extensions have been used in many applications, including cell biology, high-energy physics, statistical physics, chemistry, ecology, and epidemiology \cite{el2019}. Along with the Fisher-KPP model, different types of nonlocal models have been discussed in this article, and we have summarised their applications in this subsection. 

In ecological systems, individuals within a population often interact nonlocally due to resource competition, predator-prey interactions, or spatial dispersal. Nonlocal models have been applied to study population dynamics, species coexistence, and spatial patterns in ecological communities. The classical Fisher-KPP equation and its extensions describe the space-time evolution of population densities \cite{skellam1951,levin2003,shigesada1995,steele1998}. The nonlocal Fisher-KPP equation (\ref{LIDE}) is called the ``competitive Lotka-Volterra model'' \cite{meszena2005,pigolotti2007, perthame2008}. It is a simple nonlocal model with many applications, e.g., nonlocal intra-specific competition in the predator-prey model \cite{pal2018bmb, banerjee2017ec, pal2019mbe, pal2019ijbc, djilali2020, pal2020mbe, bian2017, kavallaris2023, zhou2025}, global consumption of resources \cite{calsina1994}, fear effects of prey species \cite{saha2023}, cooperative behaviour \cite{saha2024}. As mentioned earlier, the homogeneous solution $u=1$ can be unstable due to nonlocal interaction. For this, we linearize the nonlocal model (\ref{LIDE}) around the homogeneous solution $u=1$, and then applying the Fourier transform on both sides, we obtain the eigenvalue equation
\begin{equation}{\label{SEVE}}
    \lambda (k) = -dk^{2}-\widehat{\phi}(k),
\end{equation}
where $k$ is the wave number and $\widehat{\phi}(k)$ is the Fourier transform of the kernel function $\phi$ \cite{volpert2009, dornelas2021}. If we choose the kernel function as the Dirac $\delta$-function, then the nonlocal model (\ref{LIDE}) reduces to the local model (\ref{SRDE}), and the homogeneous solution $u=1$ becomes stable as $\lambda <0$ for all positive $k$ and $d$. The eigenvalue $\lambda$ may not maintain the negative sign for the other choices of the kernel functions. For example, suppose $\phi$ is a top-hat kernel function \cite{pal2018bmb, genieys2006}, then the eigenvalue equation (\ref{SEVE}) becomes
\begin{equation*}{\label{EVTKF}}
    \lambda (k) = -dk^{2}-\frac{\sin(k\delta)}{k\delta},
\end{equation*}
and it is not always negative for all $k>0$; rather $\lambda$ can make positive for some $k>0$ by choosing some specific value of $\delta$ and $d$ \cite{genieys2006}. For such choices of $\delta$ and $d$, the homogeneous solution $u=1$ becomes unstable, and the nonlocal model produces a non-homogeneous stationary solution. This type of linear stability analysis around the non-trivial homogeneous solution(s) predicts the existence of heterogeneous solution(s) for the nonlocal model, and this method is known as Turing instability analysis \cite{amturing}. This technique can be applied to the other nonlocal models described in this review article if the homogeneous solution satisfies the Turing instability conditions \cite{jdmurray}. 

The considered nonlocal model with top-hat kernel function has important characteristics. For a fixed value of $\delta$, we may find the critical values $k_{c}$ and $d_{c}$ by which $\lambda = 0$ holds for unique $k=k_{c}$ and $d=d_{c}$, and $\lambda >0$ holds for a range of values of $k$ with $d<d_{c}$. However, these critical values do not exist for small values of $\delta$ as $\lambda$ negative for $\delta$ approaches $0$. Hence, $\delta$ (range of nonlocal interaction) requires a minimum value to unstable the homogeneous solution $u=1$ \cite{pal2018bmb, pal2019mbe}. In addition, two length scales can be identified in the presence of the top-hat kernel function: (i) the diffusion parameter $d$ as activation and (ii) the range of nonlocal interaction $\delta$ as inhibition. We can observe a Turing instability for $d \ll \delta$ (short-range activation and long-range inhibition), and it produces a Turing pattern \cite{amturing, meinhardt1982, fuentes2004}. On the other hand, for $\delta \ll d$ (long-range activation and short-range inhibition), travelling wavefront propagation can be observed. In particular, for $\delta \rightarrow 0$, the nonlocal model reduces to the local classical Fisher-KPP equation and exhibits a travelling wave solution \cite{jdmurray,berestycki2007}. The same type of behaviours can be observed for other truncated kernel functions (triangular, parabolic, etc.), but the minimum range of the nonlocal interaction varies for different functions \cite{pal2019mbe}. 

In plant communities, the local competition term could represent the competition for light, assuming the canopy size is negligible compared to the range of the root system \cite{maciel2021}. If the opposite holds, nonlocal competition comes into the picture \cite{granato2019, schenk2006}. In this case, individuals compete to access the resources, which can redistribute itself \cite{furter1989, gourley2000}. In ecology, nonlocal models can capture many factors; e.g., Cantrell and Cosner \cite{cantrell1996} constructed a model for ladybirds feeding on aphids. They focused on the nonlocal RD model for aphid density in a particular patch that involves the whole population.

Another type of solution called `travelling wave solution' is studied for a partial differential equation that propagates with a constant speed while maintaining its shape in space. These solutions represent wave-like patterns that propagate through space while maintaining their shape and speed. Travelling wave solutions are common in biology, and numerous models in population dynamics produce biological waves. In the context of population dynamics, the travelling wave manifests itself as a wave of change in population density through a habitat, e.g., a plague that travels through a continent. This travelling wave solution also signifies the invasion of the species over the habitat. The travelling wave solutions have been studied for the nonlocal models \cite{billingham2003, yang2024}, e.g., let us consider an extension of the nonlocal FKPP model (\ref{LIDE}) as:
\begin{equation}{\label{SPNM}}
        \frac{\partial u}{\partial t} = d\frac{\partial^{2} u}{\partial x^{2}} + u\bigg{\{} 1+\alpha u- \beta u^{2} -(1+\alpha-\beta) \int_{-\infty}^{\infty} \mu \psi (\mu (x-y))u(y,t)dy \bigg{\}},
\end{equation}
$x\in \mathbb{R}$ with ``thin-tailed'' kernel functions which are even, satisfies the normalized condition and $\psi(z)\rightarrow 0$ as $z\rightarrow \pm\infty$. All the parameters involved in the equation (\ref{SPNM}) are positive and satisfy $0<\beta<1+\alpha$. 

The initial condition plays a crucial role in finding the travelling wave solutions. A small population is introduced in a localized space, which is considered the initial condition for finding the solutions to the travelling wave. In addition, the boundary conditions for this nonlocal model are considered as $u_{x}(0,t) = 0$ and $u\rightarrow 0$ for $x\rightarrow \infty$ \cite{billingham2003}. For the ``thin-tailed'' kernel function $\psi(z) = \exp{(-|z|)}/2$, the nonlocal model (\ref{SPNM}) can be converted to the coupled equations \cite{billingham2003}:
\begin{equation}{\label{CSPNM}}
    \begin{aligned}
         \frac{\partial u}{\partial t} &= d\frac{\partial^{2} u}{\partial x^{2}} + u[ 1+\alpha u- \beta u^{2} -(1+\alpha-\beta) w ],\\
          0 &= \frac{\partial^{2} w}{\partial x^{2}} + \mu^{2} (u-w),
    \end{aligned}
\end{equation}
with localized initial conditions for the species $u$ and boundary conditions $u_{x}(0,t) = 0$, $u\rightarrow 0$, $w_{x}(0,t) = 0$, and $w\rightarrow 0$ for $x\rightarrow \infty$ \cite{billingham2003}. Sometimes, this conversion technique can be applied to the nonlocal model with space-time nonlocality \cite{pal2019mbe}, e.g., the single species model considered in (\ref{NFBNM}). Along with travelling wave solutions, periodic travelling waves (or wavetrain) are also observed in local and nonlocal models, which are the periodic solutions of travelling waves. These travelling wave solutions also signify the migration of two or more sub-populations. 

The single species nonlocal model (\ref{BSPM}) in population dynamics was introduced by Britton \cite{britton1989, britton1990}, and it is an advanced version of the local FKPP model as it can be obtained by considering $\phi$ as the delta function. The nonlocal model (\ref{BSPM}) includes local aggregation that takes care of animal grouping for protective measures against predation, which could arise for many reasons, such as grassland herds, schools of fish or flocks of birds. Sometimes, species take time to move from one position to another. In this case, the nonlocal model (\ref{NFBNM}) containing a spatio-temporal average weighted toward the current time and position better fit these types of scenarios \cite{britton1990}. This nonlocal model accounts for the influence of past conditions on present ecological processes.

Nonlocal interactions are important in multi-body systems because each body's behaviour is influenced by its nearby neighbours and bodies at a distance. Such interactions are common in a wide range of physical and biological systems, including granular materials, colloidal suspensions, and biomolecular systems. In the case of population biology, species sense their surroundings to make different decisions, called nonlocal sensing, and these happen on different scales \cite{eftimie2007, levine2000, mogilner1999}. The nonlocal sensing effect in population spatial distributions is modelled by the equations (\ref{SSADE}) and (\ref{GADE}) and leads to species aggregation, segregation, or complex patterns \cite{bastille2018, benhamou2014, martinez2020, potts2014}. In addition, the nonlocal multi-species advection-diffusion model (\ref{GADE}) can capture a wide variety of patterns \cite{giunta2021, potts2019}. These types of models have been applied to predator-prey interactions \cite{fagioli2019}, cell-sorting \cite{painter2015}, and animal territoriality \cite{potts2016}. Authors in \cite{giunta2021} showed that the inclusion of self-attractive terms causes oscillatory solutions for the two-species model, and a qualitative effect on patterns occurs due to a change in the width of the spatial average. 

Nonlocal dispersal is critical in understanding population dynamics, species coexistence, and spatial patterns in ecological communities. Individuals or seeds moving over long distances can influence the colonization of new habitats, gene flow, and species interactions. This type of dispersal in biology is observed in various systems, including plants dispersing seeds, animals migrating across vast distances, and microorganisms spreading in heterogeneous environments. As researchers continue to investigate the consequences of nonlocal dispersal, it becomes apparent that it plays a pivotal role in ecological and evolutionary processes, shaping the distribution and persistence of life on Earth. 

The nonlocal dispersal used in the model (\ref{DKRM}) is important for plants and animals. It characterizes the movements of organisms in a habitat. For plants, this nonlocal dispersal happens through the seed dispersal process. In the review article, Bullock et al. fitted 11 types of probability density functions as dispersal kernel functions for 144 plant species, and all of them are good fits to the grouped data sets  \cite{bullock2017}. In addition, the dispersal-competition models have been used in the integrodifference equations, which are discrete in time and continuous in space \cite{allen1996}. For instance, the integrodifference equations of a population size $n(t,x)$ at time $t$ and position $x$ can have the form \cite{allen1996}:
\begin{equation*}
    n(t+1,x) = \int_{R}\phi (x,y)f(n(t,y))dy,
\end{equation*}
where $\phi (x,y)$ is the kernel function that satisfies the normality condition over the spatial domain $R$. In the absence of dispersal, the growth equation is of the form $n(t+1,x) = f(n(t,x))$. These models apply to the development and spread of seed-dispersed plants. Moreover, a similar model can be considered for the single-species annual plant model when seeds do not survive for more than one year \cite{allen1996}. In addition, the nonlocal dispersal is also applied to study the metapopulation dynamics. In \cite{hanski1998}, the author studied the effects of nonlocal dispersal on population persistence and the conservation of species in fragmented habitats where local populations are interconnected through dispersal, colonization, and extinction processes. It has implications for ecosystem resilience, invasive species management, and conservation biology. As research in ecology and population dynamics progresses, the study of nonlocal dispersal will continue to shed light on the intricate processes that shape biodiversity and ecological patterns. 

In \cite{hastings2005}, Hastings et al. provide insights into the spatial spread of invasions, a specific example of nonlocal dispersal in ecology. The paper discusses theoretical developments and empirical evidence related to the spatial dynamics of invasive species, highlighting the significance of nonlocal dispersal in driving the spread of invasive populations. Furthermore, in \cite{okubo2001}, the authors delve into the role of diffusion in ecological problems. It covers various aspects of dispersal and diffusion in ecology, including nonlocal dispersal and its effects on population dynamics and spatial ecology. The book provides a comprehensive and interdisciplinary perspective on diffusion in ecological systems and is an excellent resource for those interested in applying nonlocal dispersal in biology.

These applications demonstrate the versatility of fractional nonlocal models in capturing the complexities of biological systems. Fractional calculus provides a powerful framework to describe anomalous transport, long-range interactions, and memory effects, offering valuable insights into various biological processes. For instance, FODEs are used to study the memory effect on population growth or some inherent randomness at a microscopic level that regulates the macroscopic dynamics \cite{yin2018}. In ecology, FODEs have been studied in predator-prey models to account for the interactions of the population densities dependent on a certain time in the past \cite{javidi2013, rihan2015, matouk2016}. The nonlocal FODE model (\ref{TFCE}) describes the emergence and evolution of biological species in the presence of memory. Furthermore, the fractional integral representation is used in the statistical thermodynamics of confined systems to obtain generalized transport equations, distribution functions, and correlation functions that take into account the particular characteristics of confinement-induced dynamics \cite{sisman2021}.

Maciel et al. introduced nonlocal terms in the competitive terms in the Lotka-Volterra competition model (\ref{NLLVCM}). They studied the effect of the range of nonlocal interaction on each species’ influence ranges \cite{maciel2021}. In this case, we have two scenarios: additive and non-additive influence ranges. In the case of non-additive influence ranges, individuals affect competitors of the same or other species in a spatial range that depends only on the influence range of the individual itself. An application of this could be the allelopathic interactions between plants, in which the extension of the roots or branches determines the species’ influence range \cite{bais2003, bertin2003, cheng2015}. Another application could be the different types of antagonistic interactions established in microbial communities and governed by the release of chemicals into the environment \cite{granato2019, czaran2002, nadell2016}. On the other hand, for additive influence ranges, the ranges of the nonlocal interactions are a combination of the influence ranges of the species engaged in the interaction. This signifies that the plants compete for resources through their roots or home-ranging animals competing for forage over a finite region, such as central place foragers \cite{houston1985, olsson2008}. 

The mixed hyperbolic-parabolic nonlocal model (\ref{MHPE}) has two important key features \cite{colombo2014}. First, the prey moves in all directions due to the presence of Laplacian term, and predators move to regions with higher prey density. Second, the range of nonlocal interaction of nonlocal terms defines how far predators can ``feel'' the presence of prey, and they move accordingly. 

Nonlocal epidemiological models have been used to study the spread of infectious diseases in regions with diverse population densities, mobility patterns, and environmental factors. These models consider the nonlocal interactions between individuals or populations across different spatial locations, providing insights into how the disease can propagate and persist in complex environments. Furthermore, these models account for long-range dispersal, human mobility, and spatial heterogeneity, providing a more realistic representation of disease dynamics in complex biological systems.

In the classical SIR model, it is assumed that the epidemic is in a homogeneous population \cite{kermack1927}. But, generally, a disease in an epidemic spreads heterogeneously among individuals. So, a nonlocal version (\ref{SIRM1}) of the SIR model is an advancement model to investigate the propagation of the epidemic \cite{kendall1965, keyfitz1997}. Researchers have studied diffusion and nonlocal dispersal with delayed transmission in the SIR model \cite{wang2015, tian2017, liu2019, zhou2020}. In \cite{kendall1965}, Kendall proved the existence of a travelling wave solution of this nonlocal model with a minimum speed. Another type of nonlocal epidemic model has been studied in recent days \cite{kang2021, kang2020, kang2021b}, namely age-structured susceptible-infectious-susceptible (SIS). This is mainly a nonlocal spatial diffusion model, slightly different from earlier considerations. Suppose, at time $t\geq 0$ and at $x\in \Omega\subset \mathbb{R}^{N}$, $S(t,a,x)$ and $I(t,a,x)$ are the densities of susceptible and infective individuals of age $a\geq 0$. We further assume that $a^{+}$ is the maximum of the individuals and is finite. Now, for $t>0$, $a\in (0,a^{+}]$, and $x\in \Omega$, we consider an age-structured nonlocal model as \cite{kang2021b}:
\begin{equation}{\label{NLSIS}}
    \begin{aligned}
         \frac{\partial S}{\partial t} +\frac{\partial S}{\partial a} &= d(\phi\ast S - S) - \lambda (t,a,x) S-\mu (a,x)S(t,a,x)+\gamma (a,x) I(t,a,x),\\
          \frac{\partial I}{\partial t} +\frac{\partial I}{\partial a} &= d(\phi\ast I - I) + \lambda (t,a,x) S-\mu (a,x)I(t,a,x)-\gamma (a,x) I(t,a,x),
    \end{aligned}
\end{equation}
where the terms $(\phi\ast S - S)$ and $(\phi\ast I - I)$ are the nonlocal dispersal terms as defined in (\ref{DKRM}). The term $\lambda (t,a,x)$ is the force of infection of infectious individuals to susceptible individuals of age $a$, $\mu (a,x)$ is the mortality rate of the individuals, and $\gamma (a,x)$ is the recovery rate. The force of infection $\lambda (t,a,x)$ is given by
\begin{equation*}
    \lambda (t,a,x) = \int_{0}^{a^{+}} K(a,\sigma,x,y)I(t,\sigma,y)dyd\sigma,
\end{equation*}
where $K(a,\sigma,x,y)$ represents the rate of disease transmission from infective individuals of age $\sigma$ at position $y$ to susceptible individuals of age ``$a$'' at position $x$. If the dispersal of individuals follows a random walk, then the age-structured SIS model (\ref{NLSIS}) reduces to the local model
\begin{equation}{\label{LSIS}}
    \begin{aligned}
         \frac{\partial S}{\partial t} +\frac{\partial S}{\partial a} &= d\Delta S - \lambda (t,a,x) S-\mu (a,x)S(t,a,x)+\gamma (a,x) I(t,a,x),\\
          \frac{\partial I}{\partial t} +\frac{\partial I}{\partial a} &= d\Delta I + \lambda (t,a,x) S-\mu (a,x)I(t,a,x)-\gamma (a,x) I(t,a,x).
    \end{aligned}
\end{equation}
These models consider nonlocal interactions between individuals, accounting for long-range disease transmission and spatial interactions that traditional local epidemiological models do not capture. These are essential in scenarios where geographic distances influence the spread of infections. Researchers have been studying the nonlocal dispersal epidemic models with and without age structures \cite{kang2021, kang2020, kang2021b, xu2020, yang2019, yang2013}. Most of these studies have been focused on the existence of travelling wave solutions to capture the transmission of the disease. In addition, researchers have studied the memory effect on disease propagation with the help of FODE to reveal the inherent memory of the disease propagation based on the size of the initial infected population \cite{saeedian2017}. 

Different types of models can be employed to study the spread of infectious diseases, and they may include elements such as population dynamics, epidemiological parameters, and the effects of interventions like vaccination. As discussed in the previous section, the transmission of the COVID-19 virus is nonlocal because people travelling between different areas have the potential to bring the virus to new places or contribute to its spread within and between groups. One of the most important strategies for halting the COVID-19 outbreak has been vaccination \cite{shi2023}. Effective vaccinations limit the overall spread of the virus among the community by reducing the severity of the illness and promoting herd immunity. These transmissions can be simulated and predicted using mathematical models, which often incorporate differential equations. These models can account for variables, including travel patterns, vaccination coverage, recovery rates, and infection rates. Researchers use computational models to investigate various scenarios and provide guidance for public health policies. For instance, compartmental models (e.g., nonlocal SIR models discussed earlier), agent-based models, and spatial models are often used for this purpose. Policymakers can use these models to inform choices on vaccination distribution, travel bans, and other ways to stop the virus from spreading. 

\subsection{Biology, medicine and health sciences}

Nonlocal interactions in cell biology are an exciting field of study because they provide a deeper understanding of cell behaviours and tissue dynamics in a variety of healthy and pathological circumstances. It gives more insight into the cooperation and exchange of information among cells in complex biological systems as cellular imaging and computational tools advance. Nonlocal interaction refers to the long-range signalling and communication between cells that extend beyond their immediate neighbours, and it captures various biological phenomena in cell biology, such as tissue morphogenesis \cite{chen2020} and wound healing \cite{webb2022}. In addition, cancer cells interact with the extracellular matrix and surrounding tissues in tumour growth and invasion over extended distances \cite{bitsouni2018}. Therefore, these models have been employed to describe tumor growth patterns and the influence of micro-environmental factors. Furthermore, nonlocal interaction plays a significant role in coordinating the movements of cells within the migrating group, enabling efficient collective migration. 

The spatial spreading patterns of invasive cell populations have been modelled by the Fisher-KPP equation and its extensions and applied in vivo malignant spreading \cite{swanson2008, swanson2003, perez2018} and in vitro cell biology experiments \cite{vo2015, warne2019, tremel2009, sengers2007, cai2007}. The most common disease, `cancer', is a collection of diseases described by uncontrolled growth of cells and tumor development that invades the tissue of origin and distant organs \cite{hillen2013}. Researchers have suggested that intrinsic and environmental bottlenecks challenge early tumour growth and progression \cite{folkman1991, d2005,teng2008, almog2009, hanahan2011}. In the initial stage, a small subset of cancer cells called `cancer stem cells' intrinsically populates the tumour. The agent-based nonlocal model (\ref{NLCM}) describes the time evolution of cancer stem cells and tumour cells \cite{hillen2013}. This modelling approach widely applies to solid, avascular tumours in any organ. 

The nonlocal model (\ref{MLIDE}) can be applied to various aspects of cell migration and growth \cite{chen2020}. For example, this model could account for the tumour cell heterogeneity behaviour influenced by the tumour microenvironment's composition. The memory effect is one of the key factors in disease propagation, and in this case, time-fractional RD equations (e.g., the nonlocal model (\ref{TFCE})) are used in the mathematical model to incorporate such an evolution process. 

In addition, anomalous diffusion processes, characterized by fractional diffusion equations, have been observed in biological transport systems, such as moving molecules within cells or through porous media \cite{hedley2024}. These models also have been applied to study the movement of molecules, proteins, and lipids in biological membranes. The anomalous transport accounts for the nonlocal interactions and memory effects of diffusing species within the membrane, e.g., tumor growth and invasion \cite{solis2019}. These models further capture the anomalous transport of cancer cells and the interactions between tumor cells and their microenvironment, providing insights into tumor growth patterns and invasion dynamics. In addition, these models account for the nonlocal and memory effects of drug diffusion within the polymer, providing a more accurate representation of drug release kinetics \cite{yin2011}. Furthermore, the nonlocal models with fractional derivatives have been applied to describe anomalous transport of molecules in biological tissues, such as fractional Fick's law for describing fractional diffusion of solutes in biological membranes.

Fractional models are used to describe the movement of neurotransmitters and other molecules within neurons, accounting for the nonlocal interactions and subdiffusive behaviour observed in neuronal dynamics \cite{magin2010}. Furthermore, these models are used to describe the memory effects and long-range interactions in immune cell recruitment and activation during an immune response.

Biology conservation laws represent fundamental mass or energy conservation principles in biological systems. They ensure that the total amount of mass or energy within a specific domain remains constant over time, which is essential for the accurate modelling of biological processes. For instance, conservation laws in biology include mass conservation in biological transport \cite{sterner2011}, conservation of energy in metabolic processes \cite{pang2014, idumah2023}, scalar conservation law in nonlocal traffic flow model \cite{bohme2025}, and conservation of species or population densities in ecological systems \cite{santini2022}.

Nonlocal problems and conservation laws have been increasingly applied in mathematical modelling in biology. They are studied in various biological systems where the interactions between components are not limited to immediate neighbours and where the conservation of certain quantities is crucial to understanding the system's dynamics. Suppose in the absence of diffusion, a population density $u(\mathbf{x},t)$ moves with the velocity $\mathbf{p}(\mathbf{x},t)$, where $\mathbf{x},\mathbf{p} \in \mathbb{R}^{n}$ and $t\geq 0$. Here $u\geq 0$ as it represents the density. In our timescale of interest, if we neglect birth, death, immigration, and emigration of organisms, then $u$ satisfies the standard conservation equation
\begin{equation}{\label{SCE}}
    u_{t} + \nabla\cdot (\mathbf{p}u) = 0.
\end{equation}
Following the basic principles of aggregation and dispersal, we can divide the velocity $\mathbf{p}$ into two groups: individuals move towards and away from each other \cite{mogilner1999, edelstein1998, topaz2004, bertozzi2009}. Suppose $\mathbf{p}_{a}$ and $\mathbf{p}_{d}$ are the attractive and dispersive velocities of the populations at time $t\geq 0$, and hence the resultant velocity $\mathbf{p} = \mathbf{p}_{a}+\mathbf{p}_{d}$. The species' aggregation may occur due to different sensing mechanisms (e.g., smell or sight), and we can assume a spatial average of nearby populations with some weights for this sensing, i.e., nonlocal sensing. Taking all these into consideration, the attractive nonlocal velocity as \cite{topaz2006}:
$$\mathbf{p}_{a} = Vl^{n+1}\nabla (\Psi\ast u),$$
where
$$(\Psi\ast u)(\mathbf{x}) = \int_{\mathbb{R}^{n}}\Psi (\mathbf{x}-\mathbf{y})u(\mathbf{y})d\mathbf{y},$$
$V$ is the species-specific attractive speed, $l$ is the characteristic range of mechanismscal sensing, and $\alpha$ is the characteristic density. On the other hand, dispersal arises from the anti-crowding mechanism, which operates over a shorter scale \cite{mogilner1999, breder1954}, and we can consider this as spatially local. The direction of dispersal is generally opposite to the direction of population gradients, and the simplest model in describing these effects as $$\mathbf{p}_{d} = -Vrl^{2n+1}u\nabla u,$$
where $r$ is the ratio of repulsive to aggregative velocities. Taking all these factors into the equation (\ref{SCE}), we obtain
\begin{equation*}
    u_{t} + \nabla\cdot (Vl^{n+1}u\Psi\ast\nabla u -Vrl^{2n+1}u^{2}\nabla u) = 0.
\end{equation*}
Following \cite{topaz2006}, after rescaling this model, we obtain:
\begin{equation}{\label{FCSE}}
    u_{t} + \nabla\cdot (u\Psi\ast\nabla u -ru^{2}\nabla u) = 0.
\end{equation}
This dimensionless model (\ref{FCSE}) conserves the zeroth and first moments of $u$ but does not conserve higher moments \cite{topaz2006}. The zeroth moment shows that the total mass is conserved, while the first moment implies that the centre of mass is conserved. 

Now, we consider a nonlocal chemotaxis model for cell migration, which follows the mass conservation law. Suppose $p=p(t,\mathbf{x},v,\widehat{\mathbf{v}})$ is the density distribution of a cell population at a mesoscopic level at time $t>0$ and position $\mathbf{x}\in\Omega \subset \mathbb{R}^{d}$ \cite{conte2022}. Here $v$ is the statistical distribution of the speed $v\in [0,V]$ with the maximal speed $V$, and $\widehat{\mathbf{v}}\in \mathbb{S}^{d-1}$ is the polarization directions. Then, the transport equation for the cell distribution at the mesoscopic level is given by \cite{conte2022}:
\begin{equation}{\label{MMSL}}
    \frac{\partial p}{\partial t}(t,\mathbf{x},v,\widehat{\mathbf{v}}) + \mathbf{v}\cdot \nabla p(t,\mathbf{x},v,\widehat{\mathbf{v}}) = \mathcal{F}[p](t,\mathbf{x},v,\widehat{\mathbf{v}}),
\end{equation}
where $\mathcal{F}[p](t,\mathbf{x},v,\widehat{\mathbf{v}})$ is the scattering of the microscopic velocity in direction and speed. The run and tumble are the typical microscopic dynamics of a cell \cite{berg1983, block1983}. These are modelled by the velocity jump process, characterized by a turning frequency $\mu$ and transition probability $T$ \cite{conte2022, stroock1974}. Assuming the nonlocal transition probability \cite{loy2020} and its independency on the prep-orientation velocity \cite{stroock1974,othmer1988,hillen2006}, the turning operator becomes \cite{conte2022}:
\begin{equation*}
    \mathcal{F}[p](t,\mathbf{x},v,\widehat{\mathbf{v}}) = \mu (\mathbf{x})[\rho (t,\mathbf{x})T(\mathbf{x},v,\widehat{\mathbf{v}})- p(t,\mathbf{x},v,\widehat{\mathbf{v}})],
\end{equation*}
with the normality condition on $T$, i.e., $$\int_{\mathbb{S}^{d-1}}\int_{0}^{V}T(\mathbf{x},v,\widehat{\mathbf{v}})dvd\widehat{\mathbf{v}} = 1,~~\forall \mathbf{x} \in \Omega.$$
In the two-dimensional bounded domain, authors in \cite{conte2022} studied the model (\ref{MMSL}) with mass conservation. By incorporating nonlocal interactions and conservation laws, mathematical models in biology can better capture the complexities of biological systems and provide deeper insights into the behaviour of living organisms. These modelling approaches are particularly relevant in studying systems with spatial dependencies, memory effects, and long-range interactions, which are prevalent in many biological processes.

Nonlocal interactions in cell biology refer to interactions between cellular components that extend beyond immediate neighbours and may involve long-range signalling or communication. These interactions play a crucial role in cellular processes and behaviours, such as cell migration, pattern formation, and collective cell responses. Cellular adhesion is one of the fundamental forms of cell-cell interactions, and its continuum modelling remains mathematically challenging \cite{buttenschon2018, Elia2024}. The adhesion model (\ref{CONEA}) is the first continuum model, verified experimentally in Steinberg's adhesion-driven cell-sorting experiments \cite{armstrong2006}. This model has been extended and applied to many areas, e.g., cell-cell and cell-extracellular matrix (ECM) adhesion in studying the invasion of cancer cells into the ECM \cite{gerisch2008, sherratt2009, painter2010, andasari2011, chaplain2011, domschke2014}. This type of nonlocal advection-reaction-diffusion model can capture the cell-cell interactions to account for chemotactic movements due to overcrowding or the action of the stellates’ processes on hepatocytes \cite{armstrong2006, sherratt2009, mogilner1995, green2010}. The coexistence of chemotaxis occurs in vivo in disease propagation, e.g., wound healing or cancer progression \cite{bromberek2002,conte2022}. Fibroblasts migrate efficiently along collagen or fibronectin fibers in connective tissues during wound healing \cite{provenzano2009}. Cancer cells follow the aligned fibers at the tumor-stroma interface in the case of tumour spread. The nonlocal model (\ref{MMSL}) for cell distributions can capture cell migration at the mesoscopic level. The nonlocal single equation model resulting from the full chemotaxis system is often obtained by simplifying the complex system of PDEs that explain chemotaxis, which is the movement of cells or organisms in response to chemical gradients \cite{kavallaris2007b}. This reduction frequently occurs in research on chemotaxis systems' pattern formation, aggregation, and blow-up processes.

Different types of self-organization biological aggregations exist in nature, e.g., insect swarms, ungulate herds, fish schools, and bacterial colonies \cite{ben2000, camazine2020, bellomo2024, lee2001, jewell2023, painter2024}. These aggregations arise from different social phenomena without the influence of external stimuli such as attraction, group cohesion, repulsion, and collision avoidance \cite{mogilner1999}. Sometimes, biological aggregation forms distinct groups with a sharp edge, called clumping \cite{parrish1997, parrish1999}. This clumping behaviour can be observed analytically \cite{levine2000} and numerically \cite{topaz2004} for two or higher-dimensional models but not for one-dimensional models built on realistic assumptions about social interactions on swarming behaviours. On the other hand, the clumping behaviour can arise in a simple nonlocal biological aggregation model (\ref{FCSE}) for one or higher-dimensional models. The model (\ref{FCSE}) preserves integro-differential conservation law with two movement terms: nonlinear degenerate diffusion arising from anti-crowding behaviour, and the other describes attractive nonlocal social interactions \cite{topaz2006}. Further research into nonlocal interactions in cell biology will help to a greater understanding of cellular processes and their significance to development, illness, and tissue engineering as experimental tools and computational methodologies evolve.

Energy-harvesting electronics and gadgets have gained popularity in biomedical applications, having the potential to power medical equipment and sensors utilizing ambient energy sources. For example, electrical energy may be produced from mechanical vibrations or bodily motions. In addition, the temperature gradients within the body or the surrounding environment can be used to generate electricity. Energy-harvesting technologies provide implanted medical devices with a long-term power source, such as triboelectric or piezoelectric generators, which can eliminate the need for battery replacement \cite{krishnaswamy2020c, wang2017, buroni2020, krishnaswamy2019, krishnaswamy2019a, krishnaswamy2019b, krishnaswamy2020, wang2018, krishnaswamy2020a, krishnaswamy2020b, krishnaswamy2021, krishnaswamy2023, buroni2024}. Implantable devices driven by harvested energy provide precise and regulated medication delivery, increasing the efficacy of medicinal therapies \cite{zou2021}. The development of self-powered sensors is essential in remote patient monitoring systems and ensures continuous data collection without the need for frequent maintenance \cite{yang2013a}.

\subsection{Physics, chemistry, and mechanics helping life science applications}

The RD systems have also been applied to high energy physics (HEP) and statistical physics. RD models proposed by Balitsky and by Jalilian-Marian, Iancu, McLerran, Leonidov, Kovner, and Weigert (JIMLKW) have been used to understand the high energy hard scattering near or at the unitarity limit \cite{balitsky1996, fukushima2006}. Ikeda, Nagasawa, and Watanabe have introduced the branching Brownian motion in statistical physics \cite{ikeda1965, ikeda1968}, which can be associated with the classical Fisher-KPP equation, as shown by McKean in \cite{mckean1975}. Nonlocal interactions can occur due to long-range forces such as gravitational or electromagnetic interactions and can considerably impact the behaviour and dynamics of multi-body systems. These interactions are critical in astrophysics for analyzing the dynamics of galaxies because gravitational forces between stars and other celestial entities reach over large distances \cite{binney1987}. In galaxies, nonlocal interactions between stars and dark matter particles regulate galactic systems' overall structure, rotation, and stability. Furthermore, the authors of \cite{spergel2000} examined the self-interacting dark matter in cosmology, emphasizing the importance of nonlocal interactions in the dynamics of multi-body systems on cosmological scales.

Nonlocal variational formulations have been used to derive and study nonlocal PDEs, which appear in diverse physical and biological systems. For example, the nonlocal aggregation equation (\ref{SCE}) has been adapted to describe biological organisms \cite{bernoff2016, bertozzi2007, bertozzi2012}, and it minimizes energy
\begin{equation*}
E(u) = \frac{1}{2}\int\int u(\mathbf{x})\Psi (|\mathbf{x}-\mathbf{y}|)u(\mathbf{y})d\mathbf{y}d\mathbf{x},
\end{equation*}
and it can be used to analyze the existence and stability of equilibrium points \cite{bernoff2011,carrillo2014,simione2015}. Despite its popularity, it is not biological for a minimizer of mass $M$; doubling the mass yields a double-density minimizer. On the other hand, the dimensionless aggregation model (\ref{FCSE}) does form well-spaced groups. This model minimizes the energy \cite{topaz2006, bedrossian2011, burger2014}:
\begin{equation*}
E(u) = \frac{1}{2}\int\int u(\mathbf{x})\Psi (|\mathbf{x}-\mathbf{y}|)u(\mathbf{y})d\mathbf{y}d\mathbf{x}+\frac{1}{6}\int u^{3}d\mathbf{x}.
\end{equation*}
These nonlocal variational problems are also used in image denoising and restoration tasks. The nonlocal operators capture the similarities between different image patches and help preserve important structures while reducing noise \cite{gilboa2009}. Furthermore, these problems are employed to study material's deformation and fracture behaviour with long-range interactions and discontinuities, such as brittle materials and nanomaterials. 

Nonlocal models also describe Ohmic heat production in different industrial processes, e.g., food sterilization, thermistor devices \cite{kavallaris2018}, etc. Using Ohmic heating, the nonlocal model of food sterilization in the one-dimensional spatial domain $(0,1)$ is given as \cite{kavallaris2002}:
\begin{equation}{\label{NLOH}}
    \frac{\partial u}{\partial t} +\frac{\partial u}{\partial x} = \frac{\partial^{2} u}{\partial x^{2}}+\lambda f(u)\bigg{/}\bigg{(} \int_{0}^{1}f(u)dx\bigg{)}^{2},
\end{equation}
with the initial condition $u(x,0) = u_{0}(x), 0<x<1$ and the boundary condition $u(0,t)=u_{x}(1,t) = 0, t>0$. Here, the function $f>0$ is the dimensionless electrical reactivity, and $\lambda>0$ is the square of the applied potential difference. A nonlocal hyperbolic model arising in Ohmic heating is examined in \cite{kavallaris2004}, whilst a nonlocal model with nonlinear diffusion is introduced and examined in \cite{kavallaris2004a, kavallaris2007}.

Nonlocal models are used in many domains, including mechanics, materials science, and continuum physics. Nowadays, nanotechnology is one of the main technological interests in the research community, and nonlocal continuum mechanics could play a potential role in analyzing their applications \cite{peddieson2003, romano2018, barretta2023}. Piezoelectric materials are used to make ultrasound scanners, and these sensors use the converse piezoelectric effect to send out sound vibrations reflected by muscles, organs, etc. \cite{manjon2018}. These ultrasound devices also use the piezoelectric effect to detect the reflected sound. Then, they are displayed as images so that doctors can see those highly detailed 3D images for treatments. Other types of applications of these materials in the medical field are the making of sensors for electronic stethoscopes, hearing aids, flow meters to monitor vascular health and both external and internal sensors for health status monitoring. These materials are also used in the automobile industry, e.g., knock sensors, tyre pressure monitors, airbag and seatbelt sensors, gyroscopes, accelerometers, and engine fluid detection \cite{varghese2015}. Piezoelectric and other smart materials and devices are increasingly important in the digital and physical world and human-machine interactions \cite{wu2021}. In these areas, nonlocal models, including degenerate, have proven advantageous \cite{thaheld2003, de2009, kavallaris2015, melaibari2022}.

Quantum dots are examples of a low-dimensional nanostructure, and researchers have been focused on these due to their several nanoelectronic applications \cite{zhang2005, zhang2005a}. In the context of nano-inclusions and quantum dots, the size dependency of mechanical strain at the nanoscale has been brought to light in recent publications. Piezoelectricity is the growing importance of nonlocal models in the context of smart materials and structure \cite{yang2006}. Gradient theories are closer to microscopic theories like lattice dynamics than classical continuum theories. They are still applicable when the characteristic length of a problem is so small that classical continuum theories begin to fail \cite{yang2006}. The application of nonlocal continuum theory to nanotechnology was initially addressed by Peddieson et al. \cite{peddieson2003}. This model has been extended to the buckling and vibration analysis of multiwalled carbon nanotubes \cite{lu2006}, which have been used extensively in biology and medicine for cancer treatment, drug delivery, and other therapeutic applications. 

Peridynamics is a nonlocal theory and is one of the key methods for modelling biological material fracture \cite{madenci2014}. The main advantage of this nonlocal model is that it deals with continuous and discrete media \cite{lejeune2021}. It has been applied to find (i) the relationship between fracture and healing in cortical bone \cite{deng2009}; (ii) fracture in anisotropic cortical bone \cite{ghajari2014}; (iii) fracture in a complex wood microstructure \cite{madenci2014}; (iv) rupture in a micron-scale biological membrane \cite{taylor2016}; (v) fracture in a porous material \cite{chen2019}. Furthermore, peridynamics is used to model tissue growth and shrinkage. The distribution of division angle orientations influences tissue scale growth, and researchers have been trying to predict it in an individual cell \cite{gillies2011}. Authors in \cite{lejeune2021} proposed a mechanics-based model with the help of the peridynamics framework. In addition, the authors have shown cell death differences on cellular and larger length scales. In particular, cell death may create gaps between cells in some cases, leading to tissue shrinkage in other cases. 

The nonlocal effects in heat-mass transfer arise in a wide variety of modern practical applications in materials sciences, e.g., metallic nanowires \cite{sellitto2012}, high-power laser melting \cite{sobolev1994}, colloidal solidification \cite{sobolev2012}, frontal polymerization \cite{sobolev1998}, etc. Many nonlocal mathematical and computational models developed for these and other applications \cite{thaheld2003, kavallaris2015, kavallaris2007a, mcmahon2009, kavallaris2015a, abouelregal2022, abouelregal2025} can be used as a basis for their extensions to nonlocal diffusion phenomena in a more general setting. The classical diffusion theory based on Fick's law suggests the diffusion flux at a spatial point $x$, and a time moment $t$ depends on the concentration gradient at the same space-time point. These classical heat conduction laws apply at macro-temporal and spatial scales but not to thermal non-equilibrium situations, e.g., cryogenic engineering, laser-aided material processing, high-rate heat transfer in rarefied media, etc.  Nowadays, researchers believe that heat-mass transport is an inherently nonlocal phenomenon. For instance, in one spatial dimension, a particle can come to a spatial point $x$ at a time moment $t+\tau$ from a spatial point $(x-l)$ or $(x+l)$ where it was at a time moment $t$. It implies a time lag $\tau$ between the diffusion flux and the particle concentration. This time lag $\tau$ is called relaxation time, particularly the mean collision time for gas particles. In the one-dimensional case, the net flux at a spatial point $x$ and time $t+\tau$ is given by
\begin{equation*}
    J(t+\tau,x) = \frac{1}{2}(C(t,x-l)-C(t,x+l))\nu,
\end{equation*}
where $C$ is the particle concentration, and $\nu$ is the particle mean velocity. Now, expanding both sides as Taylor's series expansion for the small parameters $\tau$ and $l$, we obtain
\begin{equation}{\label{TSE}}
    \sum_{m=0}\frac{\tau^{m}}{m!}\frac{\partial^{m} J}{\partial t^{m}} = -\nu \sum_{n=0}\frac{l^{2n+1}}{(2n+1)!}\frac{\partial^{2n+1} C}{\partial x^{2n+1}}.
\end{equation}
After applying the zero-order approximation and Fick's law, one can obtain the classical diffusion equation of the parabolic type. Furthermore, this approximation is valid for slow processes, i.e., $\tau\ll t_{0}$ and $l\ll h$, where $t_{0}$ and $h$ are the characteristic time and distances, respectively. In such cases, nonlocal effects are minimal, and the classical local equilibrium approach can be employed. For the other cases, e.g., $\tau\varpropto t_{0}$ and/or $l\varpropto h$, higher order approximation should be used as the nonlocal effects play a significant role. If the diffusion-like condition connects the small parameters:
\begin{equation}{\label{PC}}
    \lim_{l\rightarrow 0, \tau\rightarrow 0}\frac{l^{2}}{\tau} = d~(\mbox{diffusion constant}),
\end{equation}
then, the first order approximation of (\ref{TSE}) along with the mass conservation law $(\partial C/\partial t = -\partial J/\partial x)$, one can obtain \cite{sobolev1993, sobolev1997, sobolev2014}:
\begin{equation}{\label{PTNL}}
    \frac{\partial C}{\partial t}+\tau\frac{\partial^{2} C}{\partial t^{2}} = d\frac{\partial^{2} C}{\partial x^{2}} + d\frac{l^{2}}{6}\frac{\partial^{4} C}{\partial x^{4}},
\end{equation}
where $d = \nu l = l^{2}/\tau$. This is a parabolic-type nonlocal equation, and for the same limiting condition (\ref{PC}), a higher-order approximation also gives the parabolic-type equation. For the wavelike dynamics, the limiting relation between the small parameters $\tau$ and $l$ is 
\begin{equation}{\label{HC}}
    \lim_{l\rightarrow 0, \tau\rightarrow 0}\frac{l}{\tau} = \nu < \infty.
\end{equation}
The first-order approximation of (\ref{TSE}) along with the condition (\ref{HC}) and the mass conservation law, one can find a hyperbolic type nonlocal equation:
\begin{equation}{\label{FOHE}}
    \frac{\partial C}{\partial t}+\tau\frac{\partial^{2} C}{\partial t^{2}} = d\frac{\partial^{2} C}{\partial x^{2}}.
\end{equation}

The limiting condition (\ref{HC}) also reduces to the hyperbolic type nonlocal equation for higher order approximation of (\ref{TSE}). Therefore, the limiting conditions (\ref{PC}) and (\ref{HC}) mainly characterize the types of PDEs, whether it is parabolic or hyperbolic, and it is true for higher order approximations of (\ref{TSE}). 

The local nonequilibrium processes are described using extended irreversible thermodynamics, which incorporates higher-order fluxes into a set of independent variables \cite{jou1996}. The most basic evolution equation for the diffusion flux that accounts for both nonlocal and relaxation effects is: 
\begin{equation*}
    J + \tau\frac{\partial J}{\partial t} = -D\frac{\partial C}{\partial x} + l^{2}\frac{\partial^{2} J}{\partial x^{2}}, 
\end{equation*}
where $l$ is the correlation length and $\tau$ is the relaxation time to local equilibrium \cite{sobolev2014}. Now, this equation with the mass conservation law, we obtain
\begin{equation}{\label{PTNL1}}
    \frac{\partial C}{\partial t}+\tau\frac{\partial^{2} C}{\partial t^{2}} = d\frac{\partial^{2} C}{\partial x^{2}} + l^{2}\frac{\partial^{3} C}{\partial t\partial x^{2}},
\end{equation}
which is of parabolic type and has an additional term (the mixed derivative) compared to the hyperbolic equation (\ref{FOHE}). Nevertheless, this equation coincides with the Jeffreys-type diffusion equation for liquids' stress and strain rate \cite{sobolev2014, jou1996, joseph1989}.

Nonlocal models have received a lot of interest in the life and health sciences and provided an improved and comprehensive framework for analyzing a wide range of applications, including tissue growth, disease spread, neural networks, and tumor dynamics. The table below provides a summary of the mathematical models that have been covered in this section, emphasizing both their advantages and limitations.

\begin{center}
\begin{longtable}{ | m{5.0em} | m{6cm}| m{6.5cm}|} 
\hline
Model & Advantages & Limitations \\ 
\hline
Extended Fisher-KPP model (\ref{SPNM}) & It applies to a wider variety of systems, including ecosystems where competition is influenced by both local and nonlocal factors & The exact analytical solutions are difficult to obtain, often requiring numerical methods and produce computational challenges \\
\hline
Age-structured models (\ref{NLSIS}) and (\ref{LSIS}) & These models incorporate the age-dependent infection, recovery, and mortality rates along with the long-range dispersal & Getting high-resolution data on both age distribution and interaction patterns between different age groups \\
\hline
Model with conservation law (\ref{FCSE}) & It conserves the zeroth and first moments of but not the higher moments & This model neglects the birth, death, immigration, and emigration of organisms \\
\hline
Transport model (\ref{MMSL}) & This model can integrate the chemotaxis for cell migration and follows the mass conservation law & This model is limited to the condition in which the cues solely alter cell polarization, with a uniform distribution of speeds \\
\hline
Diffusion-convection model (\ref{NLOH}) & This model can be applied to the food industry, illustrating the evolution of the temperature of the sterilized food &  Different solution approaches should be followed, though depending on the monotonicity of the nonlinearity because no maximum principle is available for such nonlocal parabolic problems\\
\hline
Nonlocal diffusion models (\ref{PTNL}), (\ref{FOHE}), and (\ref{PTNL1}) & These models are based on the ratio of the mean free path to the characteristic length of the system and can be used for the model of the processes far from equilibrium conditions &  These approaches can not be applied to a system if it does not follow such specific ratio \\
\hline
\end{longtable}
\end{center}

\section{Emerging Trends, Future Directions, and Outlook}{\label{SE4}}

The areas of mathematical modelling in biology are still evolving; various developing patterns in nonlocal mathematical models have been observed. As research progresses, these models are expected to play an increasingly significant role in addressing new challenges in biology and life sciences. These trends reflect advances in computational approaches, enhanced experimental data availability, and a rising acknowledgement of the importance of nonlocal interactions in understanding complex biological phenomena. Some of the emerging trends in nonlocal mathematical models are the following.

\textit{Multi-scale modelling:} Biological systems often span multiple spatial and temporal scales, and nonlocal models provide a robust and powerful framework to bridge these scales and capture the intricate interactions between different components of the system, from molecular to cellular to tissue and organism levels, however, traditional modelling approaches often focus only on specific scales \cite{nieminen2002, shaat2020}.

\textit{Human-environment systems, biodiversity, and ecological modelling:} Human-environment systems represent the intricate relationship between human societies and their surrounding environments \cite{millennium2005}. In addition, biodiversity is the variety of life on Earth, encompassing different species, ecosystems, and genetic diversity.  It is critical for maintaining ecological balance, delivering ecosystem services, and contributing to ecosystem resilience in the face of environmental change. Ecological modelling is developing and analyzing mathematical and computer models to replicate and comprehend the dynamics of ecosystems. Studying the spatio-temporal trends of human activities, such as deforestation rates, carbon emissions, and resource consumption, provides insights into the evolving nature of anthropogenic impacts on the environment. These models assist researchers in predicting the impact of human activities on biodiversity, studying ecosystem responses, and developing sustainable resource management techniques. 

\textit{Collective cell migration:} Collective cell migration is a phenomenon where groups of cells move together to achieve specific functions, such as embryonic development, wound healing, or cancer invasion \cite{sherratt1990, maini2004, chen2020, marchello2024}. These nonlocal interactions can influence cell behaviour, signalling, and collective dynamics in complex biological systems. These interactions play a significant role in coordinating the movements of cells within the migrating group, enabling efficient collective migration \cite{colombi2021}.

\textit{Proteomics, nucleic acids, and polymer modelling:} Proteins are composed of amino acid chains and are vital for numerous biological functions. Their stability is influenced by factors such as temperature, pH, and interactions with other molecules. Cooperative self-assembly is a ubiquitous phenomenon found in natural systems \cite{dankulov2019}. However, these proteins lose their native structure through a process called ``denaturation'' due to environmental changes, which cause either reversible or irreversible changes in function. On the other hand, nucleic acids (DNA and RNA) are responsible for storing and transmitting genetic information. Biological polymers are large molecules composed of repeating structural units, and they are versatile, exhibit unique properties, and have various applications. Polymers for gene delivery systems occur across different length scales, from the molecular level to macroscopic structures, and aim to safely and effectively transport genetic material, such as DNA or RNA, into target cells \cite{paliy2009, paliy2010, badu2014, badu2020, badu2020a, badu2020b}. Polymeric fluids refer to liquids containing long-chain polymers and can exhibit complex rheological behaviour due to the interplay of molecular weight, concentration, and temperature \cite{yang2007c, yang2008, melnik2002, melnik2003d, melnik2003e}. Furthermore, DNA damage within a biological network, such as a cellular or genomic network, can have profound implications for cellular function, integrity, and overall health.

\textit{Dissipative processes in biosystems and structures:} Dissipation is the conversion of energy from one form to another, which frequently results in the preservation of organized structures and the survival of life \cite{nicolis1971}. Living organisms are open systems that interchange matter and energy with their environment, allowing organized structures and functions to form. In addition, the non-equilibrium thermodynamic concepts regulate the dissipative processes in biosystems \cite{de2013}. The constant flow of energy maintains life by facilitating biochemical processes and sustaining dynamic equilibrium, such as the Belousov-Zhabotinsky reaction \cite{zhabotinsky2007}, from cellular metabolism to organismal homeostasis. These oscillations are examples of out-of-equilibrium dynamics highlighting biosystems' ability to maintain order through energy dissipation. 

\textit{Synthetic biology and nonlocal model development:} Synthetic biology is a multidisciplinary discipline that combines biological, engineering, and computer science ideas to produce synthetic creatures, genetic circuits, and biomolecular systems. Nonlocal models give a more complete picture of how biological components interact across distances inside a cell or a population of cells. This is especially important when creating synthetic biological systems with complex spatial and temporal relationships. These models are used in constructing gene regulatory networks, allowing long-range interactions to be considered, which improves the accuracy of synthetic biology techniques for regulating gene expression \cite{hasty2000}. Furthermore, similar principles are extended to the study of population dynamics to include interactions between cells or animals that occur at a distance, which aids in designing synthetic ecosystems and understanding emergent behaviours \cite{mee2012}. Nevertheless, wine fermentation involves the complex biochemical and physical processes transforming grapes into wine. The process is often characterized by the activities of numerous microorganisms, the consumption of sugars, and the creation of alcohol, carbon dioxide, and other byproducts.  Fermentation dynamics can be better understood using numerical modelling, particularly when it is based on integro-differential equations \cite{schenk2022}. As research in both domains advances, incorporating nonlocal models into synthetic biology adds to developing novel biological tools, systems, and creatures with better spatial and temporal functions.

\textit{Nonlocal models in systems biology:} Systems biology seeks to understand the complex processes inside biological systems by combining experimental data, computational modelling, and mathematical investigations. Nonlocal models are very useful in systems biology when dealing with distributed networks, such as signalling pathways, in which molecules can impact each other across distances inside a cell or tissue. It is also important to find the parameter values associated with the nonlocal model, called the inverse problem in systems biology \cite{oguz2013}. As collecting data and mathematical models advance, resolving challenges associated with inverse problems will lead to more precise and customized techniques in medicine, genetics, drug development, and other life science areas. 

\textit{Peridynamics:} The application of peridynamics in biology is still an emerging area, and its adoption and development will rely on further advancements in computational techniques, experimental validation, and interdisciplinary collaborations between researchers in engineering, materials science, and biology \cite{taylor2016, zhou2022}. As the field progresses, peridynamics holds the potential to offer valuable insights into the mechanics and behaviour of biological systems under various conditions \cite{lejeune2021}. 

\textit{Biomimetics, bionics, and bio-inspired information processing:} Genetic circuits and neuromorphic computation work in distinct biological environments, but their analogies emphasize that information processing, adaptability, and resilience principles are critical to operating both systems. In addition, intercellular coupling is essential for multicellular organisms to function properly and maintain homeostasis. Disruptions in these communication pathways can lead to various illnesses and developmental problems. The most significant type of natural computers may be found in biological systems that do computing on numerous levels \cite{chu2018}. Understanding these concepts can lead to breakthroughs in synthetic biology, biotechnology, and the creation of neuromorphic technological devices \cite{thieu2022, thieu2022a}. Integration of signal and image processing with biology not only aids in analysing biological data but also in developing computational models, diagnostic tools, and treatment methods. As bionic models evolve, driven by advances in materials, sensors, and robotics, the quality of life for people with disabilities will improve, human capacities will be enhanced, and numerous health concerns will be addressed \cite{ying2022, ying2022a, ying2023}.

\textit{Coupled effects and phenomena, anisotropy, and auxeticity in biosystems:} Many biological tissues display features dependent on direction, making the appropriate representation of anisotropy in biological tissues essential to understanding their mechanical behaviour. For instance, the human cornea involves capturing its complex structure and mechanical behaviour for various applications, such as understanding biomechanics, predicting responses to surgical procedures, and designing contact lenses. In addition, the modelling of scleral collagen requires the overall tissue architecture, anisotropy, and interactions with other ocular structures. Incorporating experimental data and imaging investigations into computational nonlocal models can improve their accuracy and applicability \cite{mahapatra2006}. Flexoelectricity, piezoelectricity, and auxeticity are coupled effects that can have a major impact on the mechanical and electrical behaviour of materials, particularly in the setting of biosystems \cite{singh2022}. Furthermore, the study of piezoelectricity and other coupled effects in biological tissues has gained significant attention due to their potential applications in various biomedical fields \cite{melnik1998, melnik2000a, melnik2000b, melnik2001, melnik2003, melnik2003a}. For instance, smart materials can exhibit ferroelectric and piezoelectric properties (ferroelectric ceramics). Additionally, multiferroic materials show the coexistence of ferroelectricity and ferromagnetism, and they have been used for different bio applications \cite{melnik2000f, melnik2004, matus2004, wang2004, melnik2004a, wang2006c, wang2006d, wang2007, wang2007a, mahapatra2007a, wang2007c, xiao2008, wang2009, wang2010, wang2011, he2018, wang2016, he2020, du2020, du2020b, he2020a, wang2017b, wang2015}. Shape memory alloys (SMAs) have distinct thermomechanical behaviour, and the intricate interaction of mechanical and thermal factors impacts their hysteresis loops. These SMAs have a wide range of biological applications, including thermoelasticity of red blood cell membranes and other biostructures \cite{melnik2001a, melnik2003b, mahapatra2006a, wang2006, wang2007b, mahapatra2007, wang2008, wang2008a, dhote2012, wang2012, dhote2013, dhote2015, dhote2015a, dhote2016, du2021, han2022, han2022a, melnik2001}. Nonlocal operations allow each pixel or area in an image to evaluate information from all other places, allowing the model to grasp the broader context \cite{fu2019}. This is useful for spotting small structures, deformities, or subtle patterns in biological pictures. These models also help to understand cellular processes, interactions, and behaviours at the nanoscale \cite{conte2023}. Fractional calculus presents the notion of fractional order derivatives, which enables the modelling of processes with memory effects, long-range interactions, and nonlocal behaviour. This allows us to more accurately represent biological and nanoscale processes with complex and nonlocal dynamics \cite{metzler1994}. 

\textit{Nonlocal models for advancing the frontiers of medicine, nanoscience and nanotechnology:} Nonlocal models provide a more accurate and comprehensive understanding of the behaviour of materials at the nanoscale. Nanocomposites are composed materials, a combination of nanoparticles and matrix material, which enhances their functionalities and opens up new possibilities in fields such as medicine, diagnostics, and biotechnology \cite{fallahpour2022}. Self-assembled bio-nanostructures, like RNA-nanostructures, have a large number of applications across various fields \cite{mahapatra2012, bahrami2010, prabhakar2010, patil2009, melnik2007}. These can be designed to encapsulate and deliver therapeutic agents like drugs or gene therapies to specific target cells or even in the development of RNA-based vaccines \cite{paliy2010, paliy2009, badu2014, badu2020, badu2020a, badu2020b, badu2020c, badu2020d, badu2015, badu2015a, badu2015b, melnik2002b}. Different biotechnologies based on RNA and DNA collectively contribute to advancements in medicine, basic research, and various industries \cite{yang2007, yang2007b}. The nano-template viruses can be modified to carry therapeutic agents to specific cells or tissues, enhancing drug delivery precision and reducing off-target effects. In addition, carbon-based nanomaterials exhibit remarkable properties and have been used in emerging technology with potential applications in medical imaging and other fields \cite{sinha2007, sinha2007a, mahapatra2008, roy2009, mahapatra2011, yavarian2023}. Biosensors detect specific analytes and convert this information into a measurable signal \cite{singh2022a}. Nevertheless, environmentally-friendly technologies (e.g., biomedical and chemical sensors) for life sciences play crucial roles in various applications, providing valuable information for monitoring, diagnosis, and research \cite{singh2022a, fallahpour2021}. As materials are scaled down to the nanoscale, size-dependent effects and phenomena become more evident, e.g., flexoelectric effects are becoming more pronounced at the nanoscale. For instance, the nanoscale structures have many applications in biomedical and bioengineering, such as nanowires, nanorods, and nanowire superlattices \cite{lew2004, lassen2004, willatzen2004, galeriu2004, lew2004, radulovic2006, lassen2006a, patil2009a, patil2009b, alvaro2013, prabhakar2013} and quantum dots \cite{melnik2004, melnik2003f, melnik2005, lew2005, prabhakar2012, prabhakar2012a, prabhakar2013a}. Furthermore, including the strain and the coupled effects together \cite{lassen2006, willatzen2006, lassen2005, lassen2005a}. The utilization of 2D materials such as graphene with pronounced coupled effects in biomedical engineering, e.g., graphene nanoribbons \cite{prabhakar2013a, prabhakar2015, prabhakar2016, prabhakar2017, prabhakar2019, leon2022}, and other graphene-based nanostructures \cite{prabhakar2014, prabhakar2015, prabhakar2015a, prabhakar2016, prabhakar2013b, prabhakar2022}.

\textit{Nonlocal neural models:} Neurons in the brain communicate and interact through synapses, which can involve long-range connections \cite{lovinger2008, seguin2023, papo2024}. Nonlocal neural models have been employed to describe neurons' collective activity and synchronization in neural networks. These models are gaining attention to represent neural networks' long-range interactions and memory effects \cite{heffernan2017}. They can provide insights into brain function, learning, and cognition, e.g., how the information is processed and transmitted between neurons in neural networks, and they can benefit from considering nonlocal interactions.

\textit{Fractional model:} Ecological systems often span multiple scales, from individual organisms to entire ecosystems. Fractional reaction-diffusion models can account for nonlocal interactions and dispersal across different spatial scales, making them valuable tools for studying population dynamics and species interactions in diverse ecological settings \cite{owolabi2023}. In addition, the study of anomalous diffusion, which departs from classical diffusion behaviour, is becoming more prevalent in nonlocal models. Fractional diffusion and subdiffusion models are being used to describe transport processes in complex biological media with irregular geometries or barriers \cite{obembe2020}. For instance, the drug delivery within the tumor often exhibits anomalous transport behaviour.

\textit{Nonlocal models in ecology:} Ecological systems are defined by complex interactions between species at various spatial scales \cite{lang2013, berrio2021}. Nonlocal models are used to investigate species dispersal, habitat connectivity, and the effect of environmental changes on population dynamics. Furthermore, depending on the nature of the interactions, several types of nonlocal interaction are considered in the reaction terms, as well as self- or cross-diffusion \cite{jdmurray, volpert2009, pal2018bmb, jungel2022, doumic2024, colombo2024, venturino2024, berestycki2024}. Nonlocal dispersal is an essential aspect of ecological processes, and understanding its implications is vital for predicting species distributions, ecosystem dynamics, and responses to environmental changes \cite{sherratt2016}. As ecological research advances, the study of nonlocal dispersal in biological systems will likely remain a critical investigation area.

\textit{Treating diseases with innovative procedures:} It involves exploring cutting-edge techniques, technologies, and approaches to improve patient outcomes. Some examples of innovative processes in disease treatment include cryosurgery - cell destruction, thermal ablation, and laser ablation \cite{singh2020, zhang2008, wang2006a, wang2006b}. Phase-lag models are mathematical frameworks which describe the heat transfer processes in biological tissues. These models are useful for understanding thermal processes in live beings and are widely used in thermal therapy, hyperthermia therapies, and thermal imaging \cite{singh2021, singh2022b, singh2020d, singh2020e}. Cardiac ablation, in particular, is a medical technique used to treat irregular heart rhythms by selectively eliminating or isolating portions of the heart tissue that create or conduct irregular electrical impulses \cite{singh2019, singh2019a, singh2019b, singh2020f}. 

\textit{Brain dynamics and modelling in neuroscience:} It is an interdisciplinary proposal to discover how the brain creates cognition and behaviour. Neuroimaging methods, such as fMRI and EEG, offer data for developing data-driven models to capture the spatio-temporal dynamics of brain activity and connection. These models are critical for creating brain-machine interfaces, which transform cerebral activity into commands for external devices, allowing persons with motor disabilities to communicate and operate technology \cite{dayan2005}. Nonlocal models provide a framework for representing the long-distance connections that contribute to information integration and processing, as neural interactions frequently extend beyond immediate neighbours. Furthermore, brain dynamic models aid in the understanding of aberrant brain dynamics associated with illnesses such as Alzheimer's disease and Parkinson's disease \cite{pal2022sr, shaheen2022, shaheen2023}. 

\textit{Nonlocal models on network:} Nonlocal interactions on networks have broader applications in systems biology, including protein-protein interaction networks, cell signalling networks, and ecological food webs \cite{autry2018, sevimoglu2014, lee2022}. Understanding and incorporating nonlocal interactions on networks are essential for capturing biological systems' emergent properties and collective behaviours. As research in network biology progresses, the significance of nonlocal interactions in shaping the dynamics and functions of biological networks will continue to be a topic of interest and investigation. 

\textit{Nonlocal models in personalized medicine:} Nonlocal models are being investigated in medicine to explain better individual variability and tailored responses to treatments \cite{falco2021}. Individual-specific dynamics in diseases are being characterized using fractional calculus, and individualized therapeutic actions are being predicted.

\textit{Nonlocal models in epidemiology:} Nonlocal epidemiological models are valuable in understanding disease dynamics in real-world scenarios, where the assumption of local interactions is often inadequate \cite{ruan2007}. They have applications in public health planning, disease control strategies, and outbreak predictions. As our understanding of infectious diseases and computational methods continue to advance, nonlocal epidemiological models will play an increasingly crucial role in addressing emerging challenges in biology and epidemiology.

\textit{Nonlocal interactions in social networks and biosocial dynamics:} Complex networks provide a powerful lens for understanding the structure and behaviour of interconnected systems in diverse domains \cite{melnik2008, dankulov2019}. For instance, social networks exhibit nonlocal interactions which extend connections among individuals. The structure of social networks is characterized by ties that go beyond direct acquaintances and influence the spread of information, behaviours, and opinions \cite{tadic2017, newman2018, dankulov2015, andjelkovic2016}. On the other hand, biosocial dynamics incorporate the spread of health behaviours, such as smoking cessation, exercise adoption, and vaccination uptake \cite{christakis2007}. The connectivity patterns within populations influence the dynamics of disease transmission, and network models assist in designing effective control strategies \cite{ tadic2021, tadic2020, mitrovic2023, mitrovic2022}. Human connectomes are highly connected hubs involving complex interactions between regions and their relevance for different processes \cite{tadic2019, andjelkovic2020}. Furthermore, nonlocal interactions contribute to the transmission of emotional experiences, influencing mental health and well-being \cite{kramer2014}. 

\textit{Conservation laws in biological transport:} The principles of conservation laws, such as mass or energy conservation, are fundamental in understanding various biological transport processes \cite{sterner2011}. For instance, they are applied in modelling blood flow in vessels, nutrient transport in plants, and fluid dynamics in biological systems \cite{sharma2018, el2022}. By employing nonlocal problems and conservation laws in mathematical modelling, researchers can better understand complex biological phenomena and develop more accurate and realistic representations of biological processes involving long-range interactions and mass conservation.

\textit{Nonlocal variational problems:} Nonlocal functionals arise in different contexts, such as fractional calculus, peridynamics, and nonlocal diffusion. The mathematical formulations of variational principles for nonlocal problems are a rapidly growing area in the interdisciplinary field.

\textit{Nonlocal models in integrating multiscale modelling and AI techniques:} Nonlocal models play a crucial role in integrating multiscale modelling and AI techniques, providing a framework that captures interactions and influences beyond immediate neighbours or specific scales. For instance, AI techniques in neural networks can learn and represent long-range interactions, which allows for more accurate predictions and simulations across different scales. In addition, AI techniques can assist in optimizing parameters across different scales based on nonlocal information, improving the overall fidelity of the multiscale model \cite{zhang2023}. There are mathematical and computational challenges in the modelling of complex systems that involve multiphysics and multiscale processes \cite{du2020}. But, overall, nonlocal models serve as a bridge in integrating multiscale modelling and AI techniques by providing a comprehensive framework for capturing new insights into disease mechanisms on different scales, which helps in identifying new targets and treatment strategies for the benefit of human health \cite{alber2019}.

\textit{Active matter and nonlocal models:} Active matter refers to a class of materials or systems that can convert energy into mechanical work to generate self-driven motion. These systems are distinguished by their individual components' capacity to drive themselves actively, showing aggregate behaviours that result from interactions between these active entities. Active matter is a large and multidisciplinary area that includes biological creatures, synthetic microswimmers, and even certain types of materials \cite{ghosh2020, gompper2020}. Nonlocality is essential for capturing the long-range correlations and collective behaviours that appear often in active matter systems. For example, the Vicsek model depicts the collective motion of self-propelled particles aligning with their neighbours \cite{vicsek1995}. It considers the average alignment of particles in a certain neighbourhood to add nonlocal interactions. Several theoretical frameworks and mathematical models, generally taken from statistical mechanics, fluid dynamics, and soft matter physics, have been constructed to characterize active matter and integrate nonlocal interactions. This method is critical for researching phenomena like flocking, swarming, and pattern development.

\textit{Nonlocal models in bio-inspired technologies:} Nonlocal models help capture the nonlocal interactions and behaviours observed in biological systems, allowing for the development of more efficient, adaptive, and biomimetic technologies. These models are employed in simulating the adaptive properties of biological materials and structures, e.g., self-healing, self-repair, and responsiveness to environmental changes. Furthermore, these models allow for the simulation of how bio-inspired technology often involves the development of sensors for enhancing sensory systems in areas including environmental monitoring and surveillance. These emerging developments emphasize the growing importance of nonlocal mathematical models in solving complicated biological challenges and exploring new mathematical biology frontiers. In addition, electron transfer processes are vital in many chemical reactions and biological systems. Nonlocal models, such as the Marcus theory, consider the nonlocal electron transfer interactions between donor and acceptor molecules. These models account for the influence of solvent polarization and nuclear motions on electron transfer rates, leading to a more accurate description of electron transfer phenomena. Nonlocal models are likely to play a critical role in enhancing our understanding of varied biological systems as interdisciplinary research expands. 

\textit{Complex adaptive systems:} There is a growing interest in complex adaptive systems, and a new framework is being created to combine classical and quantum systems using ``emergent'' quantum theory in complex adaptive systems \cite{paperin2011, hubsch2023}. Some hidden variables in classical nonlocal models impact the outcomes of measurements and can reveal correlations that local, realistic theories cannot explain. Quantum mechanics, on the other hand, introduces nonlocality through phenomena such as entanglement, in which particles become correlated. Nonlocality is a fundamental property of quantum physics, although it may also be investigated in some classical systems. Interestingly, the complex adaptive system is directly relevant to Lotka-Volterra systems, and researchers argue that the emergent stationary Schrodinger equation describes the emergent quantum theory of Lotka-Volterra dynamics. 

\textit{Coarse-grained models and computations with equation-free models:} Coarse-grained models are simplified representations of complicated systems that average or combine small components to capture the essential aspects. These models involve constructing a hierarchy of models that explain the system at various degrees of detail. On the other hand, equation-free models are computational tools that allow for the study and simulation of complex systems without explicit knowledge of governing equations. These strategies are especially effective when the system dynamics are unclear or computationally costly. Equation-free models are used in a wide range of disciplines, including neurology, climate science, and materials research. Coarse-grained models and equation-free models are both effective tools for modelling and evaluating complex systems in situations when standard methodologies are unfeasible, according to \cite{gear2003}. They enable researchers to get insights into the emergent behaviour and dynamics of multiscale systems without having to study the microscopic processes in depth.

\textit{Hyperbolic and thermal relaxation models, other sources of nonlocality:} Hyperbolic thermoelasticity is the coupled of heat conduction and elastic wave equations and describes the evolution of temperature and displacement fields in the material. It has many applications in various fields, including biology, the design of thermal devices, and the understanding of heat transfer mechanisms at the microscopic level \cite{strunin2000, strunin2001, hillen2002, melnik2002a, melnik2005}. In addition, the heat conduction equation with appropriate relaxation terms captures the time-dependent behaviour of the material during thermal ablation \cite{singh2019d}, which is typically used to destroy or damage tissues for the treatment of tumors. Furthermore, nonlocal PDEs play a significant role in modelling various phenomena, e.g., swarming behaviour, nonlinear diffusion, models in fluid dynamics and fluid-structure interaction for biological and life science problems \cite{liang2019, kamath2006}. Nonlinear nonlocal models also account for competing effects (e.g., attractive and repulsive forces). Biological tissues, such as tendons, ligaments, and bone, have intricate microstructures that can lead to non-uniform stress and strain distributions. Nonlocal plasticity models are motivated by the need to capture size effects, strain localization, and the influence of remote material points on deformation patterns in these tissues, enhancing our understanding of tissue mechanics and guiding advancements in bioengineering \cite{nilsson1998, shizawa1999, lion2000}. In addition, phase-field models in materials are used to study the evolution of phase boundaries, such as solidification, melting, and phase separation \cite{provatas2011, dhote2014, dhote2014a, dhote2014b, wang2017a, wang2018a}. The nonlocal effects in the phase-field models are especially relevant in materials where the interface width is not negligible compared to the system size. Contact mechanics also considers the nonlocal model, which deals with the deformation and interaction of solid surfaces in contact. Nevertheless, the spatial organization and arrangement of certain features or structures happens within a semiconductor material \cite{scholl2001}. This phenomenon is significant in semiconductor technology as it directly affects the operation and performance of electronic devices, including those critical in healthcare bioengineering and biomedicine.

\subsection{Future directions}

Nonlocal models have shown promise in comprehending phenomena in biological systems such as collective behaviour, pattern formation, and information transfer. To acquire insights into emergent behaviours, future studies could look into how to integrate nonlocal interactions into more complex biological systems, such as multi-cellular organisms or ecosystems. These developments will further enhance our understanding of complex biological phenomena and pave the way for new applications and discoveries. Some future directions of nonlocal mathematical models are as follows.

\textit{Coupling of multiple scales and multi-body systems:} Integrating multiple scales into nonlocal models allows researchers to bridge the gap between molecular-level interactions and tissue or organ-level behaviours. Future studies may focus on refining methods to couple various scales effectively, enabling more accurate predictions of biological processes at different hierarchical levels \cite{konieczny2023}. In the context of multi-scale porous media, such as biological tissues or cellular membranes, fractional diffusion models are used to describe how substances diffuse within the intricate network of interconnected pores. Traditional diffusion models are inadequate for accurately representing such complex structures, but fractional diffusion equations can effectively capture the anomalous transport observed in multiscale porous media \cite{sun2017, scheidweiler2020}. Furthermore, nonlocal interactions in multi-body systems are a fascinating research area, and their study has applications in diverse fields, including physics, materials science, and geophysics \cite{shaat2020}. As experimental techniques and computational methods continue to advance, our understanding of the role of nonlocal interactions in multi-body systems is expected to deepen, leading to discoveries and applications in various disciplines.

\textit{Cell migration:} Nonlocal interactions in cell biology are a fascinating area of research, and they are essential in understanding how cells coordinate their activities in complex biological processes. Nonlocal models can be used to study the spatial and temporal aspects of immune responses, such as immune cell migration, pathogen spreading, and tissue repair. As experimental techniques and computational methods advance, the study of nonlocal interactions in cell biology will continue to provide valuable insights into the emergent behaviours and collective dynamics of cells in tissues and organisms \cite{buttenschon2020}. In addition, nonlocal models using fractional calculus could be used to capture the long-range communication and coordination among migrating cells, shedding light on the mechanisms governing collective cell behaviours.

\textit{Hybrid models:} Hybrid models can combine PDEs that describe spatial processes with ODEs that represent temporal dynamics. In this type of hybrid model, the local interactions represent the immediate influence of neighbouring elements, while nonlocal interactions account for long-range dependencies. These models could be used to study systems with spatial heterogeneity and temporal dynamics, such as cancer growth, morphogenesis, pattern formation, metastasis, competition, and predation in predator-prey interactions \cite{sherratt2016, volpert2009, bitsouni2018}. These models could be used further to model the spreading of signals, morphogens, or chemicals over extended distances in finding spatial patterns.

\textit{Peridynamics:} Peridynamics is a mathematical framework that extends classical continuum mechanics to describe the behaviour of materials with discontinuities and fractures. While peridynamics has been widely used in engineering and materials science, its applications in biology are still relatively new and evolving. It can be applied to model the mechanics of biological tissues, such as soft tissues (e.g., muscle, cartilage), hard tissues (e.g., bone), and invasion of cancer cells into surrounding tissues \cite{lejeune2021}. Furthermore, it can be employed to study the mechanics of biological membranes (e.g., lipid bilayers), cell migration, and the interactions between migrating cells and their surroundings.

\textit{Epidemiological models:} Nonlocal epidemiological models continue to be an important research area, especially in the context of emerging infectious diseases and the influence of spatial factors on disease transmission dynamics \cite{ruan2007, li2021}. Different types of physiological age structure characteristics of the population can be considered in the model to improve the predictions for infectious disease transmissions. As research in this field progresses, nonlocal models are expected to be increasingly important to guide public health interventions and inform disease control strategies in understanding infectious diseases, and computational capabilities also have to improve in parallel. 

\textit{Neural networks:} In computational neuroscience, hybrid neural network models combine local, point-to-point synaptic connections with nonlocal, spatially extended interactions. This allows for the integration of both short-range and long-range interactions observed in real neural networks \cite{senk2022}. These models are employed to study brain function, learning, and memory processes. In addition, neurons in the brain have a multi-compartmental structure with dendrites, soma, and axons. Fractional-order differential equations can be used to model the electrical activity within neurons, accounting for the nonlocal interactions between different compartments \cite{magin2010, drapaca2016}. These models are essential for understanding how information is integrated and processed in complex neural networks.

\textit{Fractional derivative models:} Extending nonlocal models to the time domain using fractional calculus will enable modelling memory effects and history-dependent processes in biological systems. In the context of emerging infectious diseases, traditional local epidemiological models may fail to account for the impact of long-range dispersal and spatial heterogeneity. However, fractional epidemic models with nonlocal interactions can be used to study the spread of diseases in heterogeneous environments \cite{chen2021}. In addition, the folding and aggregation of proteins are complex processes influenced by long-range interactions between amino acids \cite{vascon2020}. In this case, the fractional models can be applied to study protein folding kinetics and aggregation dynamics to understand the underlying mechanisms of protein misfolding diseases, such as Alzheimer's and Parkinson's.

\textit{Variational problems:} There are many challenges and unique features associated with variational problems in understanding complex systems in biology, materials science, and social dynamics \cite{li2004, ramstead2019}. For instance, the analytical and numerical methods employed to tackle nonlocal variational problems, including the use of nonlocal differential operators and integral equations.

\textit{Stochastic nonlocal models:} Future nonlocal models may incorporate stochasticity to account for inherent randomness and uncertainty in biological systems. Stochastic nonlocal models will allow for a more realistic representation of noise-driven processes and enable the exploration of system robustness and variability.

\textit{Biosocial dynamics:} The collective behaviour of animal groups, such as flocking birds or schooling fish, relies on nonlocal interactions that enable coordinated movements over large distances \cite{giardina2008}. Nonlocal models have been used to investigate these animal groups' emergent patterns and self-organization, offering insights into their decision-making processes and survival strategies. Nevertheless, nonlocal models can offer a powerful framework for studying human social dynamics, allowing researchers to capture the complexity of interactions that extend beyond local connections, e.g., information spread, the epidemiological spread of behaviours, migration, long-term social evolution, etc.

\textit{Data-driven approaches:} As biological systems are often complex and high-dimensional, future research might explore the development of efficient numerical methods and simulations tailored to handle nonlocal interactions \cite{du2019, you2022}. This includes optimization techniques for parameter estimation and sensitivity analysis to gain deeper insights into model behaviour. The integration of machine learning and artificial intelligence techniques with nonlocal models could enhance the predictive capabilities and understanding of complex biological phenomena \cite{boussange2023, anstine2023, farajpour2024}. Future research may explore how artificial intelligence can assist in pattern recognition, model calibration, and model selection for nonlocal models. Developing efficient and scalable numerical methods for nonlocal models will be crucial to handling the computational demands of large-scale biological systems and high-dimensional data.

\textit{Brain:} The brain is a highly complex organ with complicated networks and connections, and nonlocal effects can play a crucial role in understanding various aspects of brain function \cite{papo2024}. Neuronal networks often exhibit long-range connections, where the activity in one region can influence distant regions. Modelling with these nonlocal models helps to understand how information is processed and integrated across different brain regions. In addition, nonlocal diffusion processes can be used to model the spread of molecules, neurotransmitters, or electrical signals across the brain tissue.

\textit{Genetics:} Nonlocal problems in genetics refer to scenarios where genetic influences extend beyond the immediate vicinity, challenging the traditional assumptions of local interactions. Studying these problems helps to understand how genetic information spreads over space and time, which is crucial for deciphering the dynamics of evolution. It has different challenges, e.g., developing appropriate mathematical models, computational methods, and experimental techniques to observe and validate nonlocal genetic interactions.

\subsection{Outlook}

There are many advantages to considering the nonlocal models, such as capturing better spatially extended interactions, more accurate representation of long-range dependencies and enhanced predictive capabilities. Advances in experimental techniques and data collection have opened up new opportunities for data-driven modelling in biology. Future nonlocal models may better incorporate experimental data to parameterize and validate these models, leading to more accurate and realistic predictions. However, several problems must be solved, such as numerical approaches for nonlocal models, parameter estimates in conjunction with model validation, and adding noise and uncertainty. Overall, the outlook for nonlocal models in biology is optimistic, with ongoing research aimed at addressing challenges and unlocking the full potential of these models. As our ability to gather and analyze biological data improves, nonlocal models will likely play an increasingly important role in advancing our understanding of complex biological systems.

\section{Conclusions}{\label{SE5}}

Nonlocal problems in biology and life sciences encompass a broad spectrum of applications and involve a wide range of mathematical tools, including differential equations with nonlocal operators, such as fractional derivatives or integral operators. These operators account for nonlocal interactions and long-range dependencies, making them suitable for modelling biological processes that exhibit spatial memory effects and anomalous transport. As we have seen in this review, examples of nonlocal problems in biology and life sciences include fractional diffusion models, fractional reaction-diffusion equations, and integro-differential equations representing nonlocal transport processes in biological tissues, cellular dynamics, and ecological systems. Nonlocal models have demonstrated their significance in bridging the gap between local models and the complexities of biological systems. Furthermore, these models provide a more realistic and comprehensive representation of biological processes across different scales (multiscale and coupled) by incorporating long-range interactions and spatial dependencies. This review has explored the theoretical foundations, applications, advantages, and challenges of nonlocal models in mathematical biology. As interdisciplinary research continues to flourish, nonlocal mathematical models will more decisively play a pivotal role in advancing our understanding of complex biological systems. Future research in this area holds great potential for further enhancing our understanding of biological processes. In particular, increasing collaboration between mathematicians, modellers, and experimental biologists will facilitate validating and refining nonlocal models using experimental data. This will enhance the models’ predictive capabilities and will better reflect relevance to real-world biological processes.

\section*{Acknowledgements}
The authors are grateful to the NSERC and the CRC Program for their support. RM also acknowledges the support of the BERC 2022–2025 program and the Spanish Ministry of Science, Innovation and Universities through the Agencia Estatal de Investigacion (AEI) BCAM Severo Ochoa excellence accreditation SEV-2017–0718 and the Basque Government fund AI in BCAM EXP. 2019/00432. This research was enabled in part by support provided by SHARCNET and the Digital Research Alliance of Canada.

\bibliography{References}

\begin{thebibliography}{100}
\expandafter\ifx\csname url\endcsname\relax
  \def\url#1{\texttt{#1}}\fi
\expandafter\ifx\csname urlprefix\endcsname\relax\def\urlprefix{URL }\fi
\expandafter\ifx\csname href\endcsname\relax
  \def\href#1#2{#2} \def\path#1{#1}\fi

\bibitem{feldman1923}
W.~M. Feldman, Biomathematics: being the principles of mathematics for students
  of biological science, C. Griffin \& Company, Limited, 1923.

\bibitem{rashevsky1938}
N.~Rashevsky, Mathematical Biophysics: Physico-Mathematical Foundations of
  Biology, Dover Publications, Inc., 1938.

\bibitem{amturing}
A.~M. Turing, The chemical basis of morphogenesis, Bulletin of Mathematical
  Biology 52~(1) (1952) 153--197.

\bibitem{jdmurray}
J.~D. Murray, Mathematical Biology \mbox{I}. \mbox{An Introduction}, Springer,
  2002.

\bibitem{edelstein2005}
L.~Edelstein-Keshet, Mathematical models in biology, SIAM, 2005.

\bibitem{tomlin2007}
C.~J. Tomlin, J.~D. Axelrod, Biology by numbers: mathematical modelling in
  developmental biology, Nature Reviews Genetics 8~(5) (2007) 331--340.

\bibitem{ellis2019}
G.~F. Ellis, J.~Kopel, The dynamical emergence of biology from physics:
  branching causation via biomolecules, Frontiers in physiology 9 (2019) 1966.

\bibitem{vandermeer2010}
J.~Vandermeer, How populations grow: the exponential and logistic equations,
  Nature Education Knowledge 3~(10) (2010) 15.

\bibitem{mittelbach2019}
G.~G. Mittelbach, B.~J. McGill, The fundamentals of predator–prey
  interactions, Oxford University Press, 2019.

\bibitem{azuaje2011}
F.~Azuaje, Computational discrete models of tissue growth and regeneration,
  Briefings in Bioinformatics 12~(1) (2011) 64--77.

\bibitem{azizi2021}
T.~Azizi, B.~Alali, G.~Kerr, Discrete dynamical systems: With applications in
  biology, Discrete Dynamical Systems: With Applications in Biology-2nd Edition
  (2021) 1--103.

\bibitem{cappuccino1995}
N.~Cappuccino, P.~W. Price, Population dynamics: new approaches and synthesis,
  Elsevier, 1995.

\bibitem{ross1916}
R.~Ross, An application of the theory of probabilities to the study of a priori
  pathometry.-\mbox{Part I}, Proceedings of the Royal Society of London. Series
  A, Containing papers of a mathematical and physical character 92~(638) (1916)
  204--230.

\bibitem{ross1917}
R.~Ross, H.~P. Hudson, An application of the theory of probabilities to the
  study of a priori pathometry.—\mbox{Part III}, Proceedings of the Royal
  Society of London. Series A, Containing papers of a mathematical and physical
  character 93~(650) (1917) 225--240.

\bibitem{kermack1927}
W.~O. Kermack, A.~G. McKendrick, A contribution to the mathematical theory of
  epidemics, Proceedings of the Royal Society of London. Series A, Containing
  papers of a mathematical and physical character 115~(772) (1927) 700--721.

\bibitem{jones2009}
D.~S. Jones, M.~Plank, B.~D. Sleeman, Differential equations and mathematical
  biology, Chapman and Hall/CRC, 2009.

\bibitem{perko2013}
L.~Perko, Differential equations and dynamical systems, Vol.~7, Springer
  Science \& Business Media, 2013.

\bibitem{lorenz1963}
E.~N. Lorenz, Deterministic nonperiodic flow, Journal of Atmospheric Sciences
  20~(2) (1963) 130--141.

\bibitem{miller1993}
K.~S. Miller, B.~Ross, An introduction to the fractional calculus and
  fractional differential equations, Wiley, 1993.

\bibitem{caputo2015}
M.~Caputo, M.~Fabrizio, A new definition of fractional derivative without
  singular kernel, Progress in Fractional Differentiation \& Applications 1~(2)
  (2015) 73--85.

\bibitem{cheng2019}
X.~Cheng, J.~Duan, D.~Li, A novel compact \mbox{ADI} scheme for two-dimensional
  \mbox{Riesz} space fractional nonlinear reaction-diffusion equations, Applied
  Mathematics and Computation 346 (2019) 452--464.

\bibitem{cole1933}
K.~S. Cole, Electric conductance of biological systems, in: Cold Spring Harbor
  symposia on quantitative biology, Vol.~1, Cold Spring Harbor Laboratory
  Press, 1933, pp. 107--116.

\bibitem{djordjevic2003}
V.~D. Djordjevi{\'c}, J.~Jari{\'c}, B.~Fabry, J.~J. Fredberg,
  D.~Stamenovi{\'c}, Fractional derivatives embody essential features of cell
  rheological behavior, Annals of Biomedical Engineering 31~(6) (2003)
  692--699.

\bibitem{saeedian2017}
M.~Saeedian, M.~Khalighi, N.~Azimi-Tafreshi, G.~Jafari, M.~Ausloos, Memory
  effects on epidemic evolution: The susceptible-infected-recovered epidemic
  model, Physical Review E 95~(2) (2017) 022409.

\bibitem{auchincloss2012}
A.~H. Auchincloss, S.~Y. Gebreab, C.~Mair, A.~V. Diez~Roux, A review of spatial
  methods in epidemiology, 2000--2010, Annual Review of Public Health 33 (2012)
  107--122.

\bibitem{yu2018}
G.~Yu, R.~Yang, Y.~Wei, D.~Yu, W.~Zhai, J.~Cai, B.~Long, S.~Chen, J.~Tang,
  G.~Zhong, et~al., Spatial, temporal, and spatiotemporal analysis of mumps in
  \mbox{Guangxi Province, China}, 2005--2016, BMC Infectious Diseases 18~(1)
  (2018) 1--13.

\bibitem{byun2021}
H.~G. Byun, N.~Lee, S.-s. Hwang, A systematic review of spatial and
  spatio-temporal analyses in public health research in korea, Journal of
  Preventive Medicine and Public Health 54~(5) (2021) 301.

\bibitem{fryxell2014}
J.~M. Fryxell, A.~R. Sinclair, G.~Caughley, Wildlife ecology, conservation, and
  management, John Wiley \& Sons, 2014.

\bibitem{salevin}
S.~A. Levin, The problem of pattern and scale in ecology: the \mbox{Robert H.
  MacArthur} award lecture, Ecology 73~(6) (1992) 1943--1967.

\bibitem{page2003pattern}
K.~Page, P.~K. Maini, N.~A. Monk, Pattern formation in spatially heterogeneous
  \mbox{Turing} reaction--diffusion models, Physica D: Nonlinear Phenomena
  181~(1-2) (2003) 80--101.

\bibitem{colizza2007reaction}
V.~Colizza, R.~Pastor-Satorras, A.~Vespignani, Reaction--diffusion processes
  and metapopulation models in heterogeneous networks, Nature Physics 3~(4)
  (2007) 276--282.

\bibitem{song2019spatial}
P.~Song, Y.~Lou, Y.~Xiao, A spatial \mbox{SEIRS} reaction-diffusion model in
  heterogeneous environment, Journal of Differential Equations 267~(9) (2019)
  5084--5114.

\bibitem{wuyts2019tropical}
B.~Wuyts, A.~R. Champneys, N.~Verschueren, J.~I. House, Tropical tree cover in
  a heterogeneous environment: A reaction-diffusion model, PloS One 14~(6)
  (2019) e0218151.

\bibitem{kholodenko2003}
B.~N. Kholodenko, Four-dimensional organization of protein kinase signaling
  cascades: the roles of diffusion, endocytosis and molecular motors, Journal
  of Experimental Biology 206~(12) (2003) 2073--2082.

\bibitem{meinhardt1982}
H.~Meinhardt, Models of biological pattern formation, New York 118 (1982).

\bibitem{meinhardt2009}
H.~Meinhardt, The algorithmic beauty of sea shells, Springer Science \&
  Business Media, 2009.

\bibitem{castets1990}
V.~Castets, E.~Dulos, J.~Boissonade, P.~De~Kepper, Experimental evidence of a
  sustained standing \mbox{Turing-type} nonequilibrium chemical pattern,
  Physical Review Letters 64~(24) (1990) 2953.

\bibitem{de1991}
P.~De~Kepper, V.~Castets, E.~Dulos, J.~Boissonade, Turing-type chemical
  patterns in the chlorite-iodide-malonic acid reaction, Physica D: Nonlinear
  Phenomena 49~(1-2) (1991) 161--169.

\bibitem{lengyel1991}
I.~Lengyel, I.~R. Epstein, Modeling of \mbox{Turing} structures in the
  chlorite-iodide-malonic acid-starch reaction system, Science 251~(4994)
  (1991) 650--652.

\bibitem{diggle2019}
P.~Diggle, E.~Giorgi, M.~Chipeta, S.~B. Macfarlane, Tracking health outcomes in
  space and time: Spatial and spatio-temporal methods, in: The Palgrave
  Handbook of Global Health Data Methods for Policy and Practice, Springer,
  2019, pp. 383--401.

\bibitem{silling2019}
S.~A. Silling, E.~Madenci, The world is nonlocal, Journal of Peridynamics and
  Nonlocal Modeling 1~(1) (2019) 1--2.

\bibitem{lohrey2020}
A.~Lohrey, B.~Boreham, The nonlocal universe, Communicative \& Integrative
  Biology 13~(1) (2020) 147--159.

\bibitem{krause2021}
A.~L. Krause, V.~Klika, P.~K. Maini, D.~Headon, E.~A. Gaffney, Isolating
  patterns in open reaction--diffusion systems, Bulletin of Mathematical
  Biology 83~(7) (2021) 82.

\bibitem{chen2021}
Y.~Chen, F.~Liu, Q.~Yu, T.~Li, Review of fractional epidemic models, Applied
  Mathematical Modelling 97 (2021) 281--307.

\bibitem{thompson2020}
T.~B. Thompson, P.~Chaggar, E.~Kuhl, A.~Goriely, A.~D.~N. Initiative,
  Protein-protein interactions in neurodegenerative diseases: A conspiracy
  theory, PLoS Computational Biology 16~(10) (2020) e1008267.

\bibitem{pal2022sr}
S.~Pal, R.~Melnik, Nonlocal models in the analysis of brain neurodegenerative
  protein dynamics with application to \mbox{Alzheimer’s} disease, Scientific
  Reports 12~(1) (2022) 1--13.

\bibitem{furter1989}
J.~Furter, M.~Grinfeld, Local vs. non-local interactions in population
  dynamics, Journal of Mathematical Biology 27~(1) (1989) 65--80.

\bibitem{eftimie2007}
R.~Eftimie, G.~de~Vries, M.~A. Lewis, Complex spatial group patterns result
  from different animal communication mechanisms, Proceedings of the National
  Academy of Sciences 104~(17) (2007) 6974--6979.

\bibitem{levine2000}
H.~Levine, W.-J. Rappel, I.~Cohen, Self-organization in systems of
  self-propelled particles, Physical Review E 63~(1) (2000) 017101.

\bibitem{saha2024a}
S.~Saha, S.~Pal, R.~Melnik, The role of inducible defence in ecological models:
  Effects of nonlocal intraspecific competitions, arXiv preprint
  arXiv:2411.10551 (2024).

\bibitem{saha2025}
S.~Saha, S.~Pal, R.~Melnik, Innate behavioural mechanisms and defensive traits
  in ecological models of predator-prey types, arXiv preprint arXiv:2501.01687
  (2025).

\bibitem{chen2020}
L.~Chen, K.~Painter, C.~Surulescu, A.~Zhigun, Mathematical models for cell
  migration: a non-local perspective, Philosophical Transactions of the Royal
  Society B 375~(1807) (2020) 20190379.

\bibitem{kornberg2014}
T.~B. Kornberg, S.~Roy, Cytonemes as specialized signaling filopodia,
  Development 141~(4) (2014) 729--736.

\bibitem{sherratt2016}
J.~A. Sherratt, Invasion generates periodic traveling waves (wavetrains) in
  predator-prey models with nonlocal dispersal, SIAM Journal on Applied
  Mathematics 76~(1) (2016) 293--313.

\bibitem{nathan2012}
R.~Nathan, E.~Klein, J.~Robledo-Arnuncio, Dispersal kernels: Review, in
  dispersal ecology and evolution, edited by j. clobert, m. baguette, tg
  benton, and jm bullock (2012).

\bibitem{painter2024}
K.~J. Painter, T.~Hillen, J.~R. Potts, Biological modelling with nonlocal
  advection diffusion equations, Mathematical Models and Methods in Applied
  Sciences 34~(01) (2024) 57--107.

\bibitem{kendall1965}
D.~G. Kendall, Mathematical models of the spread of infection, Mathematics and
  Computer Science in Biology and Medicine (1965) 213--225.

\bibitem{kang2021}
H.~Kang, S.~Ruan, Nonlinear age-structured population models with nonlocal
  diffusion and nonlocal boundary conditions, Journal of Differential Equations
  278 (2021) 430--462.

\bibitem{eringen2003}
A.~C. Eringen, J.~Wegner, Nonlocal continuum field theories, Appl. Mech. Rev.
  56~(2) (2003) B20--B22.

\bibitem{lavrik2004}
N.~V. Lavrik, M.~J. Sepaniak, P.~G. Datskos, Cantilever transducers as a
  platform for chemical and biological sensors, Review of Scientific
  Instruments 75~(7) (2004) 2229--2253.

\bibitem{ekinci2005}
K.~Ekinci, M.~Roukes, Nanoelectromechanical systems, Review of Scientific
  Instruments 76~(6) (2005) 061101.

\bibitem{lu2006}
P.~Lu, H.~Lee, C.~Lu, P.~Zhang, Dynamic properties of flexural beams using a
  nonlocal elasticity model, Journal of Applied Physics 99~(7) (2006) 073510.

\bibitem{hutson2003}
V.~Hutson, S.~Martinez, K.~Mischaikow, G.~T. Vickers, The evolution of
  dispersal, Journal of Mathematical Biology 47~(6) (2003) 483--517.

\bibitem{nieminen2002}
R.~M. Nieminen, From atomistic simulation towards multiscale modelling of
  materials, Journal of Physics: Condensed Matter 14~(11) (2002) 2859.

\bibitem{shaat2020}
M.~Shaat, E.~Ghavanloo, S.~A. Fazelzadeh, Review on nonlocal continuum
  mechanics: physics, material applicability, and mathematics, Mechanics of
  Materials 150 (2020) 103587.

\bibitem{deng2009}
Q.~Deng, Y.~Chen, J.~Lee, An investigation of the microscopic mechanism of
  fracture and healing processes in cortical bone, International Journal of
  Damage Mechanics 18~(5) (2009) 491--502.

\bibitem{taylor2016}
M.~Taylor, I.~G{\"o}zen, S.~Patel, A.~Jesorka, K.~Bertoldi, Peridynamic
  modeling of ruptures in biomembranes, PloS One 11~(11) (2016) e0165947.

\bibitem{lejeune2021}
E.~Lejeune, C.~Linder, Modeling biological materials with peridynamics, in:
  Peridynamic Modeling, Numerical Techniques, and Applications, Elsevier, 2021,
  pp. 249--273.

\bibitem{silling2000}
S.~A. Silling, Reformulation of elasticity theory for discontinuities and
  long-range forces, Journal of the Mechanics and Physics of Solids 48~(1)
  (2000) 175--209.

\bibitem{manjon2018}
A.~M. Manj{\'o}n-Sanz, M.~R. Dolgos, Applications of piezoelectrics: Old and
  new, Chemistry of Materials 30~(24) (2018) 8718--8726.

\bibitem{singh2024}
S.~Singh, L.~Bianchi, S.~Korganbayev, P.~Namakshenas, R.~Melnik, P.~Saccomandi,
  Non-fourier bioheat transfer analysis in brain tissue during interstitial
  laser ablation: Analysis of multiple influential factors, Annals of
  Biomedical Engineering 52~(4) (2024) 967--981.

\bibitem{volpert2009}
V.~Volpert, S.~Petrovskii, Reaction--diffusion waves in biology, Physics of
  Life Reviews 6~(4) (2009) 267--310.

\bibitem{kolmogorov1937}
A.~Kolmogorov, I.~Petrovsky, N.~Piskunov, Investigation of the equation of
  diffusion combined with increasing of the substance and its application to a
  biology problem, Bull. Moscow State Univ. Ser. A: Math. Mech 1~(6) (1937)
  1--25.

\bibitem{simpson2013}
M.~J. Simpson, K.~K. Treloar, B.~J. Binder, P.~Haridas, K.~J. Manton, D.~I.
  Leavesley, D.~S. McElwain, R.~E. Baker, Quantifying the roles of cell
  motility and cell proliferation in a circular barrier assay, Journal of the
  Royal Society Interface 10~(82) (2013) 20130007.

\bibitem{sherratt1990}
J.~A. Sherratt, J.~D. Murray, Models of epidermal wound healing, Proceedings of
  the Royal Society of London. Series B: Biological Sciences 241~(1300) (1990)
  29--36.

\bibitem{maini2004}
P.~K. Maini, D.~S. McElwain, D.~I. Leavesley, Traveling wave model to interpret
  a wound-healing cell migration assay for human peritoneal mesothelial cells,
  Tissue Engineering 10~(3-4) (2004) 475--482.

\bibitem{fornari2019}
S.~Fornari, A.~Sch{\"a}fer, M.~Jucker, A.~Goriely, E.~Kuhl, Prion-like
  spreading of \mbox{Alzheimer’s} disease within the brain’s connectome,
  Journal of the Royal Society Interface 16~(159) (2019) 20190356.

\bibitem{schafer2021}
A.~Sch{\"a}fer, M.~Peirlinck, K.~Linka, E.~Kuhl, A.~D. N.~I. (ADNI, et~al.,
  Bayesian physics-based modeling of tau propagation in \mbox{Alzheimer's}
  disease, Frontiers in Physiology (2021) 1081.

\bibitem{petrovskii2005}
S.~V. Petrovskii, B.-L. Li, Exactly solvable models of biological invasion,
  Chapman and Hall/CRC, 2005.

\bibitem{skellam1951}
J.~G. Skellam, Random dispersal in theoretical populations, Biometrika 38~(1/2)
  (1951) 196--218.

\bibitem{levin2003}
S.~A. Levin, H.~C. Muller-Landau, R.~Nathan, J.~Chave, The ecology and
  evolution of seed dispersal: a theoretical perspective, Annual Review of
  Ecology, Evolution, and Systematics (2003) 575--604.

\bibitem{vasilopoulos2012}
G.~Vasilopoulos, Local and non-local mathematical modelling of signalling
  during embryonic development, Ph.D. thesis, Heriot-Watt University (2012).

\bibitem{viana2016}
D.~S. Viana, L.~Gangoso, W.~Bouten, J.~Figuerola, Overseas seed dispersal by
  migratory birds, Proceedings of the Royal Society B: Biological Sciences
  283~(1822) (2016) 20152406.

\bibitem{li2024a}
J.~Li, G.~Guo, H.~Yuan, Nonlocal delay gives rise to vegetation patterns in a
  vegetation-sand model, Mathematical Biosciences and Engineering 21~(3) (2024)
  4521--4553.

\bibitem{simoy2023}
M.~I. Simoy, M.~N. Kuperman, Non-local interaction effects in models of
  interacting populations, Chaos, Solitons \& Fractals 167 (2023) 112993.

\bibitem{banerjee2022review}
M.~Banerjee, M.~Kuznetsov, O.~Udovenko, V.~Volpert, Nonlocal
  reaction--diffusion equations in biomedical applications, Acta Biotheoretica
  70~(2) (2022) 1--28.

\bibitem{billingham2020}
J.~Billingham, Slow travelling wave solutions of the nonlocal \mbox{Fisher-KPP}
  equation, Nonlinearity 33~(5) (2020) 2106.

\bibitem{pal2018bmb}
S.~Pal, S.~Ghorai, M.~Banerjee, Analysis of a prey--predator model with
  non-local interaction in the prey population, Bulletin of Mathematical
  Biology 80~(4) (2018) 906--925.

\bibitem{tian2021}
G.~Tian, Z.-C. Wang, G.-B. Zhang, Stability of traveling waves of the nonlocal
  \mbox{Fisher-KPP} equation, Nonlinear Analysis 211 (2021) 112399.

\bibitem{banerjee2017ec}
M.~Banerjee, V.~Volpert, Spatio-temporal pattern formation in
  \mbox{Rosenzweig-Macarthur} model: effect of nonlocal interactions,
  Ecological Complexity 30 (2017) 2--10.

\bibitem{dornelas2021}
V.~Dornelas, E.~H. Colombo, C.~L{\'o}pez, E.~Hern{\'a}ndez-Garc{\'\i}a,
  C.~Anteneodo, Landscape-induced spatial oscillations in population dynamics,
  Scientific Reports 11~(1) (2021) 1--11.

\bibitem{pal2019mbe}
S.~Pal, S.~Ghorai, M.~Banerjee, Effect of kernels on spatio-temporal patterns
  of a non-local prey-predator model, Mathematical Biosciences 310 (2019)
  96--107.

\bibitem{segal2013}
B.~Segal, V.~Volpert, A.~Bayliss, Pattern formation in a model of competing
  populations with nonlocal interactions, Physica D: Nonlinear Phenomena 253
  (2013) 12--22.

\bibitem{pal2020amm}
S.~Pal, M.~Banerjee, S.~Ghorai, Effects of boundary conditions on pattern
  formation in a nonlocal prey--predator model, Applied Mathematical Modelling
  79 (2020) 809--823.

\bibitem{merchant2011}
S.~M. Merchant, W.~Nagata, Instabilities and spatiotemporal patterns behind
  predator invasions with nonlocal prey competition, Theoretical Population
  Biology 80~(4) (2011) 289--297.

\bibitem{pal2019ijbc}
S.~Pal, M.~Banerjee, S.~Ghorai, Spatio-temporal pattern formation in
  \mbox{Holling--Tanner} type model with nonlocal consumption of resources,
  International Journal of Bifurcation and Chaos 29~(01) (2019) 1930002.

\bibitem{bian2017}
S.~Bian, L.~Chen, E.~A. Latos, Global existence and asymptotic behavior of
  solutions to a nonlocal \mbox{Fisher--KPP} type problem, Nonlinear Analysis:
  Theory, Methods \& Applications 149 (2017) 165--176.

\bibitem{kavallaris2023}
N.~I. Kavallaris, E.~Latos, T.~Suzuki, Diffusion-driven blow-up for a nonlocal
  \mbox{Fisher-KPP} type model, SIAM Journal on Mathematical Analysis 55~(3)
  (2023) 2411--2433.

\bibitem{britton1989}
N.~Britton, Aggregation and the competitive exclusion principle, Journal of
  Theoretical Biology 136~(1) (1989) 57--66.

\bibitem{britton1990}
N.~F. Britton, Spatial structures and periodic travelling waves in an
  integro-differential reaction-diffusion population model, SIAM Journal on
  Applied Mathematics 50~(6) (1990) 1663--1688.

\bibitem{pal2020mbe}
S.~Pal, M.~Banerjee, V.~Volpert, Spatio-temporal \mbox{Bazykin’s} model with
  space-time nonlocality, Mathematical Biosciences and Engineering 17~(5)
  (2020) 4801--4824.

\bibitem{laine2012}
E.-M. Laine, H.-P. Breuer, J.~Piilo, C.-F. Li, G.-C. Guo, Nonlocal memory
  effects in the dynamics of open quantum systems, Physical review letters
  108~(21) (2012) 210402.

\bibitem{tarasov2021}
V.~E. Tarasov, Non-markovian dynamics of open quantum system with memory,
  Annals of Physics 434 (2021) 168667.

\bibitem{li2006}
Y.~Li, D.~X. Dai, Biomechanical engineering of textiles and clothing, Woodhead
  Publishing, 2006.

\bibitem{chanet2014}
S.~Chanet, A.~C. Martin, Mechanical force sensing in tissues, Progress in
  molecular biology and translational science 126 (2014) 317--352.

\bibitem{al2003}
M.~Al-Hajj, M.~S. Wicha, A.~Benito-Hernandez, S.~J. Morrison, M.~F. Clarke,
  Prospective identification of tumorigenic breast cancer cells, Proceedings of
  the National Academy of Sciences 100~(7) (2003) 3983--3988.

\bibitem{bonnet1997}
D.~Bonnet, J.~E. Dick, Human acute myeloid leukemia is organized as a hierarchy
  that originates from a primitive hematopoietic cell, Nature Medicine 3~(7)
  (1997) 730--737.

\bibitem{dick2003}
J.~E. Dick, Breast cancer stem cells revealed, Proceedings of the National
  Academy of Sciences 100~(7) (2003) 3547--3549.

\bibitem{enderling2009}
H.~Enderling, A.~R. Anderson, M.~A. Chaplain, A.~Beheshti, L.~Hlatky,
  P.~Hahnfeldt, Paradoxical dependencies of tumor dormancy and progression on
  basic cell kinetics tumor dormancy and progression, Cancer Research 69~(22)
  (2009) 8814--8821.

\bibitem{hillen2013}
T.~Hillen, H.~Enderling, P.~Hahnfeldt, The tumor growth paradox and immune
  system-mediated selection for cancer stem cells, Bulletin of Mathematical
  Biology 75~(1) (2013) 161--184.

\bibitem{yang2007}
X.-D. Yang, D.~R. Mahapatra, R.~V. Melnik, Simulation of \mbox{RNA} silencing
  pathway for time-dependent transgene transcription rate, in: AIP Conference
  Proceedings, Vol. 952(1), American Institute of Physics, 2007, pp. 229--237.

\bibitem{lambeth2013}
L.~S. Lambeth, C.~A. Smith, Short hairpin \mbox{RNA-mediated} gene silencing,
  siRNA Design: Methods and Protocols (2013) 205--232.

\bibitem{macfarlane2010}
L.-A. MacFarlane, P.~R~Murphy, Microrna: biogenesis, function and role in
  cancer, Current Genomics 11~(7) (2010) 537--561.

\bibitem{redman2016}
M.~Redman, A.~King, C.~Watson, D.~King, What is crispr/cas9?, Archives of
  Disease in Childhood-Education and Practice 101~(4) (2016) 213--215.

\bibitem{rinaldi2018}
C.~Rinaldi, M.~J. Wood, Antisense oligonucleotides: the next frontier for
  treatment of neurological disorders, Nature Reviews Neurology 14~(1) (2018)
  9--21.

\bibitem{guo2020}
C.-h. Guo, T.~Cao, L.-t. Zheng, J.~L. Waddington, X.-c. Zhen, Development and
  characterization of an inducible dicer conditional knockout mouse model of
  parkinson’s disease: validation of the antiparkinsonian effects of a
  sigma-1 receptor agonist and dihydromyricetin, Acta Pharmacologica Sinica
  41~(4) (2020) 499--507.

\bibitem{shomar2020}
A.~Shomar, O.~Barak, N.~Brenner, Local and global features of genetic networks
  supporting a phenotypic switch, Plos one 15~(9) (2020) e0238433.

\bibitem{singh2020}
S.~Singh, J.~A. Krishnaswamy, R.~Melnik, Biological cells and coupled
  electro-mechanical effects: The role of organelles, microtubules, and
  nonlocal contributions, journal of the Mechanical Behavior of Biomedical
  Materials 110 (2020) 103859.

\bibitem{singh2020iw}
S.~Singh, R.~Melnik, Coupled electro-mechanical behavior of microtubules, in:
  Bioinformatics and Biomedical Engineering: 8th International Work-Conference,
  IWBBIO 2020, Granada, Spain, May 6--8, 2020, Proceedings 8, Springer, 2020,
  pp. 75--86.

\bibitem{singh2020ic}
S.~Singh, R.~Melnik, Microtubule biomechanics and the effect of degradation of
  elastic moduli, in: Computational Science--ICCS 2020: 20th International
  Conference, Amsterdam, The Netherlands, June 3--5, 2020, Proceedings, Part VI
  20, Springer, 2020, pp. 348--358.

\bibitem{melnik2009}
R.~V. Melnik, X.~Wei, G.~MORENO-HAGELSIEB, Nonlinear dynamics of cell cycles
  with stochastic mathematical models, Journal of Biological Systems 17~(03)
  (2009) 425--460.

\bibitem{fazelpour2022}
F.~Fazelpour, Z.~Tang, K.~E. Daniels, The effect of grain shape and material on
  the nonlocal rheology of dense granular flows, Soft Matter 18~(7) (2022)
  1435--1442.

\bibitem{carmeliet2005}
P.~Carmeliet, Angiogenesis in life, disease and medicine, Nature 438~(7070)
  (2005) 932--936.

\bibitem{li2014}
F.~Li, Y.~Lou, Y.~Wang, Global dynamics of a competition model with non-local
  dispersal \mbox{I}: The shadow system, Journal of Mathematical Analysis and
  Applications 412~(1) (2014) 485--497.

\bibitem{khodabakhshi2021}
P.~Khodabakhshi, K.~E. Willcox, M.~Gunzburger, A multifidelity method for a
  nonlocal diffusion model, Applied Mathematics Letters 121 (2021) 107361.

\bibitem{garcia2012}
V.~Garc{\'\i}a-Morales, K.~Krischer, The complex \mbox{Ginzburg--Landau}
  equation: an introduction, Contemporary Physics 53~(2) (2012) 79--95.

\bibitem{kao2010}
C.-Y. Kao, Y.~Lou, W.~Shen, Random dispersal vs. non-local dispersal, Discrete
  Contin. Dyn. Syst 26~(2) (2010) 551--596.

\bibitem{zhan2022}
H.~Zhan, F.~Gao, L.~Guo, Global boundedness and \mbox{Allee} effect for a
  nonlocal time fractional \mbox{p-Laplacian} reaction-diffusion equation,
  arXiv preprint arXiv:2202.04928 (2022).

\bibitem{alarcon2003}
T.~Alarc{\'o}n, H.~M. Byrne, P.~K. Maini, A cellular automaton model for tumour
  growth in inhomogeneous environment, Journal of Theoretical Biology 225~(2)
  (2003) 257--274.

\bibitem{pal2024a}
S.~Pal, R.~Melnik, Parkinson’s disease progression and treatment dynamics
  accounting for nonlocality of bioneurological processes, in: International
  Work-Conference on Bioinformatics and Biomedical Engineering, Springer, 2024,
  pp. 42--54.

\bibitem{zhou2025}
X.~Zhou, X.~Wang, G.~Zhang, Dynamics of a non-local intraspecific competition
  predator--prey model with memory effect, Applied Mathematics Letters 160
  (2025) 109334.

\bibitem{du2023}
Q.~Du, X.~Tian, Z.~Zhou, Nonlocal diffusion models with consistent local and
  fractional limits, in: A$^3$N$^2$M: Approximation, Applications, and Analysis
  of Nonlocal, Nonlinear Models: Proceedings of the 50th John H. Barrett
  Memorial Lectures, Springer, 2023, pp. 175--213.

\bibitem{li2019}
G.~Li, H.~Zhang, B.~Zhang, Continuous time random walk with \mbox{A}
  $\rightarrow$ \mbox{B} reaction in flows, Physica A: Statistical Mechanics
  and its Applications 532 (2019) 121917.

\bibitem{ruan2007}
S.~Ruan, Spatial-temporal dynamics in nonlocal epidemiological models, in:
  Mathematics for life science and medicine, Springer, 2007, pp. 97--122.

\bibitem{chang2023}
K.~Chang, Z.~Zhang, G.~Liang, Dynamics analysis of a nonlocal diffusion dengue
  model, Scientific Reports 13~(1) (2023) 15239.

\bibitem{wang2024}
F.-B. Wang, R.~Wu, X.-Q. Zhao, A nonlocal reaction-diffusion model of west nile
  virus with vertical transmission, Journal of Nonlinear Science 34~(1) (2024)
  13.

\bibitem{li2024b}
Z.~Li, X.-Q. Zhao, Global dynamics of a time-delayed nonlocal
  reaction-diffusion model of within-host viral infections, Journal of
  Mathematical Biology 88~(3) (2024) 38.

\bibitem{sohrabi2020}
C.~Sohrabi, Z.~Alsafi, N.~O'neill, M.~Khan, A.~Kerwan, A.~Al-Jabir,
  C.~Iosifidis, R.~Agha, World health organization declares global emergency: A
  review of the 2019 novel coronavirus \mbox{(COVID-19)}, International Journal
  of Surgery 76 (2020) 71--76.

\bibitem{biswas2020}
S.~K. Biswas, J.~K. Ghosh, S.~Sarkar, U.~Ghosh, \mbox{COVID-19 pandemic in
  India:} a mathematical model study, Nonlinear Dynamics 102 (2020) 537--553.

\bibitem{shi2023}
L.~Shi, Z.~Chen, P.~Wu, Spatial and temporal dynamics of \mbox{COVID-19} with
  nonlocal dispersal in heterogeneous environment: Modeling, analysis and
  simulation, Chaos, Solitons \& Fractals 174 (2023) 113891.

\bibitem{xu2023}
C.~Xu, Z.~Liu, Y.~Pang, A.~Akg{\"u}l, Stochastic analysis of a \mbox{COVID-19}
  model with effects of vaccination and different transition rates: Real data
  approach, Chaos, Solitons \& Fractals 170 (2023) 113395.

\bibitem{smith2009}
J.~E. Smith-Garvin, G.~A. Koretzky, M.~S. Jordan, T cell activation, Annual
  Review of Immunology 27 (2009) 591--619.

\bibitem{bachmann2010}
M.~F. Bachmann, G.~T. Jennings, Vaccine delivery: a matter of size, geometry,
  kinetics and molecular patterns, Nature Reviews Immunology 10~(11) (2010)
  787--796.

\bibitem{rosenberg2014}
S.~A. Rosenberg, \mbox{IL-2:} the first effective immunotherapy for human
  cancer, The Journal of Immunology 192~(12) (2014) 5451--5458.

\bibitem{tokoyoda2009}
K.~Tokoyoda, S.~Zehentmeier, A.~N. Hegazy, I.~Albrecht, J.~R. Gr{\"u}n,
  M.~L{\"o}hning, A.~Radbruch, Professional memory \mbox{CD4+ T} lymphocytes
  preferentially reside and rest in the bone marrow, Immunity 30~(5) (2009)
  721--730.

\bibitem{boyman2012}
O.~Boyman, J.~Sprent, The role of interleukin-2 during homeostasis and
  activation of the immune system, Nature Reviews Immunology 12~(3) (2012)
  180--190.

\bibitem{de1979}
P.~De~Mottoni, E.~Orlandi, A.~Tesei, Asymptotic behavior for a system
  describing epidemics with migration and spatial spread of infection,
  Nonlinear Analysis: Theory, Methods \& Applications 3~(5) (1979) 663--675.

\bibitem{liang2003}
Liang, Travelling waves and numerical approximations in a reaction advection
  diffusion equation with nonlocal delayed effects, Journal of Nonlinear
  Science 13 (2003) 289--310.

\bibitem{noftle2010}
E.~E. Noftle, W.~Fleeson, Age differences in big five behavior averages and
  variabilities across the adult life span: moving beyond retrospective, global
  summary accounts of personality., Psychology and Aging 25~(1) (2010) 95.

\bibitem{hoy2020}
S.~R. Hoy, D.~R. MacNulty, D.~W. Smith, D.~R. Stahler, X.~Lambin, R.~O.
  Peterson, J.~S. Ruprecht, J.~A. Vucetich, Fluctuations in age structure and
  their variable influence on population growth, Functional Ecology 34~(1)
  (2020) 203--216.

\bibitem{ripoll2023}
J.~Ripoll, J.~Font, Numerical approach to an age-structured lotka-volterra
  model, Mathematical Biosciences and Engineering 20~(9) (2023) 15603--15622.

\bibitem{chauvet1993}
G.~A. Chauvet, Non-locality in biological systems results from hierarchy:
  Application to the nervous system, Journal of mathematical biology 31 (1993)
  475--486.

\bibitem{buttenschon2018}
A.~Buttensch{\"o}n, T.~Hillen, A.~Gerisch, K.~J. Painter, A space-jump
  derivation for non-local models of cell-cell adhesion and non-local
  chemotaxis, Journal of Mathematical Biology 76~(1) (2018) 429--456.

\bibitem{armstrong2006}
N.~J. Armstrong, K.~J. Painter, J.~A. Sherratt, A continuum approach to
  modelling cell--cell adhesion, Journal of Theoretical Biology 243~(1) (2006)
  98--113.

\bibitem{sherratt2009}
J.~A. Sherratt, S.~A. Gourley, N.~J. Armstrong, K.~J. Painter, Boundedness of
  solutions of a non-local reaction--diffusion model for adhesion in cell
  aggregation and cancer invasion, European Journal of Applied Mathematics
  20~(1) (2009) 123--144.

\bibitem{colombo2018}
R.~M. Colombo, E.~Rossi, Nonlocal conservation laws in bounded domains, SIAM
  Journal on Mathematical Analysis 50~(4) (2018) 4041--4065.

\bibitem{vo2024}
A.~T. Vo, Nonlocal frameworks for nonlinear conservation laws and
  advection-diffusion processes, Ph.D. thesis, The University of
  Nebraska-Lincoln (2024).

\bibitem{hsu2014}
S.~Hsu, J.~L{\'o}pez-G{\'o}mez, L.~Mei, M.~Molina-Meyer, A nonlocal problem
  from conservation biology, SIAM Journal on Mathematical Analysis 46~(6)
  (2014) 4035--4059.

\bibitem{xu2022}
X.~Xu, B.~Qin, A variational approach for \mbox{Kirchhoff-Carrier} type
  non-local equation boundary value problems, Journal of Mathematical Analysis
  and Applications 508~(2) (2022) 125885.

\bibitem{leimkuhler2004}
B.~Leimkuhler, S.~Reich, Simulating hamiltonian dynamics, 14, Cambridge
  university press, 2004.

\bibitem{luenberger1971}
D.~Luenberger, An introduction to observers, IEEE Transactions on automatic
  control 16~(6) (1971) 596--602.

\bibitem{rietkerk2008}
M.~Rietkerk, J.~Van~de Koppel, Regular pattern formation in real ecosystems,
  Trends in ecology \& evolution 23~(3) (2008) 169--175.

\bibitem{sun2019}
Y.-J. Sun, L.~Zhang, W.-T. Li, Z.-C. Wang, Entire solutions in nonlocal
  monostable equations: Asymmetric case, Communications on Pure \& Applied
  Analysis 18~(3) (2019) 1049.

\bibitem{lutscher2005}
F.~Lutscher, E.~Pachepsky, M.~A. Lewis, The effect of dispersal patterns on
  stream populations, SIAM Review 47~(4) (2005) 749--772.

\bibitem{myerscough1996}
M.~Myerscough, M.~Darwen, W.~Hogarth, Stability, persistence and structural
  stability in a classical predator-prey model, Ecological Modelling 89~(1-3)
  (1996) 31--42.

\bibitem{pal2021cnsns}
S.~Pal, S.~Petrovskii, S.~Ghorai, M.~Banerjee, Spatiotemporal pattern formation
  in \mbox{2D} prey-predator system with nonlocal intraspecific competition,
  Communications in Nonlinear Science and Numerical Simulation 93 (2021)
  105478.

\bibitem{real1977}
L.~A. Real, The kinetics of functional response, The American Naturalist
  111~(978) (1977) 289--300.

\bibitem{pal2024}
S.~Pal, R.~Melnik, M.~Banerjee, Pattern alternations induced by nonlocal
  interactions, Communications in Mathematical Sciences 22~(7) (2024)
  1817--1838.

\bibitem{yadav2023}
R.~Yadav, S.~Pal, M.~Sen, The effect of nonlocal interaction on chaotic
  dynamics, turing patterns, and population invasion in a prey--predator model,
  Chaos: An Interdisciplinary Journal of Nonlinear Science 33~(10) (2023).

\bibitem{skalski2001}
G.~T. Skalski, J.~F. Gilliam, Functional responses with predator interference:
  viable alternatives to the \mbox{Holling type II} model, Ecology 82~(11)
  (2001) 3083--3092.

\bibitem{pal2025}
S.~Pal, M.~Banerjee, R.~Melnik, Nonequilibrium dynamics in a noise-induced
  predator--prey model, Chaos, Solitons \& Fractals 191 (2025) 115884.

\bibitem{berryman1992}
A.~A. Berryman, The orgins and evolution of predator-prey theory, Ecology
  73~(5) (1992) 1530--1535.

\bibitem{pal2024b}
S.~Pal, M.~Banerjee, R.~Melnik, The role of soil surface in a sustainable
  semiarid ecosystem, PloS one 19~(12) (2024) e0314910.

\bibitem{hsu2015}
S.-B. Hsu, S.~Ruan, T.-H. Yang, Analysis of three species
  \mbox{Lotka--Volterra} food web models with omnivory, Journal of Mathematical
  Analysis and Applications 426~(2) (2015) 659--687.

\bibitem{autry2018}
E.~A. Autry, A.~Bayliss, V.~A. Volpert, Biological control with nonlocal
  interactions, Mathematical Biosciences 301 (2018) 129--146.

\bibitem{m1925}
A.~M'kendrick, Applications of mathematics to medical problems, Proceedings of
  the Edinburgh Mathematical Society 44 (1925) 98--130.

\bibitem{von1959}
H.~Von~Foerster, Some remarks on changing populations, The Kinetics of Cellular
  Proliferation (1959) 382--407.

\bibitem{trucco1965}
E.~Trucco, Mathematical models for cellular systems the \mbox{von Foerster
  equation. Part I}, The Bulletin of Mathematical Biophysics 27 (1965)
  285--304.

\bibitem{kavallaris2021}
N.~I. Kavallaris, R.~Barreira, A.~Madzvamuse, Dynamics of shadow system of a
  singular \mbox{Gierer--Meinhardt} system on an evolving domain, Journal of
  Nonlinear Science 31~(1) (2021) 5.

\bibitem{marciniak2017}
A.~Marciniak-Czochra, G.~Karch, K.~Suzuki, Instability of turing patterns in
  reaction-diffusion-\mbox{ODE} systems, Journal of mathematical biology 74
  (2017) 583--618.

\bibitem{marciniak2013}
A.~Marciniak-Czochra, G.~Karch, K.~Suzuki, Unstable patterns in
  reaction--diffusion model of early carcinogenesis, Journal de
  Math{\'e}matiques Pures et Appliqu{\'e}es 99~(5) (2013) 509--543.

\bibitem{kavallaris2017}
N.~I. Kavallaris, T.~Suzuki, On the dynamics of a non-local parabolic equation
  arising from the \mbox{Gierer--Meinhardt} system, Nonlinearity 30~(5) (2017)
  1734.

\bibitem{duong2021}
G.~K. Duong, N.~I. Kavallaris, H.~Zaag, Diffusion-induced blowup solutions for
  the shadow limit model of a singular \mbox{Gierer--Meinhardt} system,
  Mathematical Models and Methods in Applied Sciences 31~(07) (2021)
  1469--1503.

\bibitem{gierer1972}
A.~Gierer, H.~Meinhardt, A theory of biological pattern formation, Kybernetik
  12 (1972) 30--39.

\bibitem{tadic2023}
B.~Tadi{\'c}, M.~M. Dankulov, R.~Melnik, Evolving cycles and self-organised
  criticality in social dynamics, Chaos, Solitons \& Fractals 171 (2023)
  113459.

\bibitem{mitrovic2023}
M.~Mitrovi{\'c}~Dankulov, B.~Tadi{\'c}, R.~Melnik, Robust global trends during
  pandemics: Analysing the interplay of biological and social processes,
  Dynamics 3~(4) (2023) 764--776.

\bibitem{elliot2000}
P.~Elliot, J.~C. Wakefield, N.~G. Best, D.~J. Briggs, et~al., Spatial
  epidemiology: methods and applications, Oxford University Press, 2000.

\bibitem{zhigun2024}
A.~Zhigun, M.~L. Rajendran, Modelling non-local cell-cell adhesion: a
  multiscale approach, Journal of Mathematical Biology 88~(5) (2024) 55.

\bibitem{miller2001}
M.~B. Miller, B.~L. Bassler, Quorum sensing in bacteria, Annual Reviews in
  Microbiology 55~(1) (2001) 165--199.

\bibitem{otten2003}
C.~Otten, For quantum confinement, size matters, but so does shape, Washington
  University, St. Louis (2003).

\bibitem{michalek2015}
G.~Micha{\l}ek, T.~Doma{\'n}ski, B.~Bu{\l}ka, K.~I. Wysoki{\'n}ski, Novel
  non-local effects in three-terminal hybrid devices with quantum dot,
  Scientific Reports 5~(1) (2015) 14572.

\bibitem{taghipour2017}
Y.~Taghipour, G.~H. Baradaran, Large deflection analysis of nanowires based on
  nonlocal theory using total lagrangian finite element method, Acta Mechanica
  228 (2017) 2429--2442.

\bibitem{schornbaum2015}
J.~Schornbaum, Y.~Zakharko, M.~Held, S.~Thiemann, F.~Gannott, J.~Zaumseil,
  Light-emitting quantum dot transistors: emission at high charge carrier
  densities, Nano letters 15~(3) (2015) 1822--1828.

\bibitem{kahmann2020}
S.~Kahmann, A.~Shulga, M.~A. Loi, Quantum dot light-emitting
  transistors—powerful research tools and their future applications, Advanced
  Functional Materials 30~(20) (2020) 1904174.

\bibitem{freeney2022}
S.~E. Freeney, M.~R. Slot, T.~S. Gardenier, I.~Swart, D.~Vanmaekelbergh,
  Electronic quantum materials simulated with artificial model lattices, ACS
  Nanoscience Au 2~(3) (2022) 198--224.

\bibitem{ishihara2018}
K.~Ishihara, E.~M. Tanaka, Spontaneous symmetry breaking and pattern formation
  of organoids, Current Opinion in Systems Biology 11 (2018) 123--128.

\bibitem{denk2020}
J.~Denk, E.~Frey, Pattern-induced local symmetry breaking in active-matter
  systems, Proceedings of the National Academy of Sciences 117~(50) (2020)
  31623--31630.

\bibitem{crawford1991}
J.~D. Crawford, E.~Knobloch, Symmetry and symmetry-breaking bifurcations in
  fluid dynamics, Annual Review of Fluid Mechanics 23~(1) (1991) 341--387.

\bibitem{islas2005}
J.~R. Islas, D.~Lavabre, J.-M. Grevy, R.~H. Lamoneda, H.~R. Cabrera, J.-C.
  Micheau, T.~Buhse, Mirror-symmetry breaking in the soai reaction: A kinetic
  understanding, Proceedings of the National Academy of Sciences 102~(39)
  (2005) 13743--13748.

\bibitem{li2010}
R.~Li, B.~Bowerman, Symmetry breaking in biology, Cold Spring Harbor
  perspectives in biology 2~(3) (2010) a003475.

\bibitem{tanzy2013}
M.~Tanzy, V.~Volpert, A.~Bayliss, M.~Nehrkorn, Stability and pattern formation
  for competing populations with asymmetric nonlocal coupling, Mathematical
  Biosciences 246~(1) (2013) 14--26.

\bibitem{mcglinchey2022}
M.~J. McGlinchey, Making and Breaking Symmetry in Chemistry: Syntheses,
  Mechanisms and Molecular Rearrangements, World Scientific, 2022.

\bibitem{hutchinson1961}
G.~E. Hutchinson, The paradox of the plankton, The American Naturalist 95~(882)
  (1961) 137--145.

\bibitem{levins1971}
R.~Levins, D.~Culver, Regional coexistence of species and competition between
  rare species, Proceedings of the National Academy of Sciences 68~(6) (1971)
  1246--1248.

\bibitem{chesson2000}
P.~Chesson, Mechanisms of maintenance of species diversity, Annual Review of
  Ecology and Systematics (2000) 343--366.

\bibitem{amarasekare2003}
P.~Amarasekare, Competitive coexistence in spatially structured environments: a
  synthesis, Ecology Letters 6~(12) (2003) 1109--1122.

\bibitem{kartzinel2015}
T.~R. Kartzinel, P.~A. Chen, T.~C. Coverdale, D.~L. Erickson, W.~J. Kress,
  M.~L. Kuzmina, D.~I. Rubenstein, W.~Wang, R.~M. Pringle, Dna metabarcoding
  illuminates dietary niche partitioning by african large herbivores,
  Proceedings of the National Academy of Sciences 112~(26) (2015) 8019--8024.

\bibitem{maciel2018}
G.~A. Maciel, F.~Lutscher, Movement behaviour determines competitive outcome
  and spread rates in strongly heterogeneous landscapes, Theoretical Ecology
  11~(3) (2018) 351--365.

\bibitem{maciel2021}
G.~A. Maciel, R.~Martinez-Garcia, Enhanced species coexistence in
  \mbox{Lotka-Volterra} competition models due to nonlocal interactions,
  Journal of Theoretical Biology 530 (2021) 110872.

\bibitem{tarnita2017}
C.~E. Tarnita, J.~A. Bonachela, E.~Sheffer, J.~A. Guyton, T.~C. Coverdale,
  R.~A. Long, R.~M. Pringle, A theoretical foundation for multi-scale regular
  vegetation patterns, Nature 541~(7637) (2017) 398--401.

\bibitem{bais2003}
H.~P. Bais, R.~Vepachedu, S.~Gilroy, R.~M. Callaway, J.~M. Vivanco, Allelopathy
  and exotic plant invasion: from molecules and genes to species interactions,
  Science 301~(5638) (2003) 1377--1380.

\bibitem{granato2019}
E.~T. Granato, T.~A. Meiller-Legrand, K.~R. Foster, The evolution and ecology
  of bacterial warfare, Current Biology 29~(11) (2019) R521--R537.

\bibitem{colombo2014}
R.~M. Colombo, E.~Rossi, Hyperbolic predators vs parabolic preys,
  Communications in Mathematical Sciences 13~(2) (2014) 369--400.

\bibitem{ichikawa2012}
K.~Ichikawa, M.~Rouzimaimaiti, T.~Suzuki, Reaction diffusion equation with
  non-local term arises as a mean field limit of the master equation, Discrete
  Contin. Dyn. Syst. S 5~(1) (2012) 115--126.

\bibitem{kavallaris2013}
N.~I. Kavallaris, T.~Suzuki, Non-local reaction--diffusion system involved by
  reaction radius \mbox{I}, The IMA Journal of Applied Mathematics 78~(3)
  (2013) 614--632.

\bibitem{kavallaris2014}
N.~I. Kavallaris, T.~Suzuki, Non-local reaction--diffusion system involving
  reaction radius \mbox{II}: rate of convergence, IMA Journal of Applied
  Mathematics 79~(1) (2014) 1--21.

\bibitem{golse2003}
F.~Golse, The mean-field limit for the dynamics of large particle systems,
  Journ{\'e}es {\'e}quations aux d{\'e}riv{\'e}es partielles (2003) 1--47.

\bibitem{kravvaritis2010}
D.~Kravvaritis, V.~Papanicolaou, A.~Xepapadeas, A.~Yannacopoulos, On a class of
  operator equations arising in infinite dimensional replicator dynamics,
  Nonlinear Analysis: Real World Applications 11~(4) (2010) 2537--2556.

\bibitem{kavallaris2017a}
N.~I. Kavallaris, J.~Lankeit, M.~Winkler, On a degenerate nonlocal parabolic
  problem describing infinite dimensional replicator dynamics, SIAM Journal on
  Mathematical Analysis 49~(2) (2017) 954--983.

\bibitem{alberts2017}
B.~Alberts, Molecular biology of the cell, Garland science, 2017.

\bibitem{orth2010}
J.~D. Orth, I.~Thiele, B.~{\O}. Palsson, What is flux balance analysis?, Nature
  Biotechnology 28~(3) (2010) 245--248.

\bibitem{wolpert2015}
L.~Wolpert, C.~Tickle, A.~M. Arias, Principles of development, Oxford
  University Press, USA, 2015.

\bibitem{madsen2015}
E.~L. Madsen, Environmental microbiology: from genomes to biogeochemistry, John
  Wiley \& Sons, 2015.

\bibitem{goldstein2002}
H.~Goldstein, C.~Poole, J.~Safko, Classical mechanics (2002).

\bibitem{feynman2018}
R.~P. Feynman, Statistical mechanics: a set of lectures, CRC press, 2018.

\bibitem{melnik2012}
R.~V. Melnik, I.~S. Kotsireas, Interconnected challenges and new perspectives
  in applied mathematical and computational sciences, in: Advances in Applied
  Mathematics, Modeling, and Computational Science, Springer, 2012, pp. 1--10.

\bibitem{antoniouk2013}
A.~V. Antoniouk, R.~Melnik, Mathematics and life sciences, De Gruyter Berlin,
  Germany, 2013.

\bibitem{melnik2015}
R.~Melnik, Universality of mathematical models in understanding nature,
  society, and man-made world, Mathematical and Computational Modeling: With
  Applications in Natural and Social Sciences, Engineering, and the Arts (2015)
  1--16.

\bibitem{jensen1998}
H.~J. Jensen, Self-organized criticality: emergent complex behavior in physical
  and biological systems, Vol.~10, Cambridge university press, 1998.

\bibitem{tadic2017}
B.~Tadi{\'c}, M.~M. Dankulov, R.~Melnik, Mechanisms of self-organized
  criticality in social processes of knowledge creation, Physical Review E
  96~(3) (2017) 032307.

\bibitem{tadic2021a}
B.~Tadi{\'c}, R.~Melnik, Self-organised critical dynamics as a key to
  fundamental features of complexity in physical, biological, and social
  networks, Dynamics 1~(2) (2021) 181--197.

\bibitem{grassberger1993}
P.~Grassberger, On a self-organized critical forest-fire model, Journal of
  Physics A: Mathematical and General 26~(9) (1993) 2081.

\bibitem{plenz2021}
D.~Plenz, T.~L. Ribeiro, S.~R. Miller, P.~A. Kells, A.~Vakili, E.~L. Capek,
  Self-organized criticality in the brain, Frontiers in Physics 9 (2021)
  639389.

\bibitem{rupe2022}
A.~Rupe, V.~V. Vesselinov, J.~P. Crutchfield, Nonequilibrium statistical
  mechanics and optimal prediction of partially-observed complex systems, New
  Journal of Physics 24~(10) (2022) 103033.

\bibitem{melnik1997}
V.~N. Melnik, On consistent regularities of control and value functions,
  Numerical Functional Analysis and Optimization 18~(3-4) (1997) 401--426.

\bibitem{melnik1998a}
R.~V.~N. Melnik, et~al., Dynamic system evolution and markov chain
  approximation, Discrete Dynamics in Nature and Society 2 (1998) 7--39.

\bibitem{russell2019}
S.~Russell, Human compatible: Artificial intelligence and the problem of
  control, Penguin, 2019.

\bibitem{burdet2013}
E.~Burdet, D.~W. Franklin, T.~E. Milner, Human robotics: neuromechanics and
  motor control, MIT press, 2013.

\bibitem{battiston2020}
F.~Battiston, G.~Cencetti, I.~Iacopini, V.~Latora, M.~Lucas, A.~Patania, J.-G.
  Young, G.~Petri, Networks beyond pairwise interactions: Structure and
  dynamics, Physics Reports 874 (2020) 1--92.

\bibitem{bick2023}
C.~Bick, T.~B{\"o}hle, C.~Kuehn, Phase oscillator networks with nonlocal
  higher-order interactions: Twisted states, stability, and bifurcations, SIAM
  Journal on Applied Dynamical Systems 22~(3) (2023) 1590--1638.

\bibitem{herzog2022}
R.~Herzog, F.~E. Rosas, R.~Whelan, S.~Fittipaldi, H.~Santamaria-Garcia,
  J.~Cruzat, A.~Birba, S.~Moguilner, E.~Tagliazucchi, P.~Prado, et~al., Genuine
  high-order interactions in brain networks and neurodegeneration, Neurobiology
  of Disease 175 (2022) 105918.

\bibitem{uzel2022}
K.~Uzel, S.~Kato, M.~Zimmer, A set of hub neurons and non-local connectivity
  features support global brain dynamics in c. elegans, Current Biology 32~(16)
  (2022) 3443--3459.

\bibitem{reddy2007}
J.~Reddy, Nonlocal theories for bending, buckling and vibration of beams,
  International Journal of Engineering Science 45~(2-8) (2007) 288--307.

\bibitem{park2005}
H.~S. Park, E.~G. Karpov, W.~K. Liu, P.~A. Klein, The bridging scale for
  two-dimensional atomistic/continuum coupling, Philosophical Magazine 85~(1)
  (2005) 79--113.

\bibitem{liu2006}
W.~K. Liu, H.~S. Park, D.~Qian, E.~G. Karpov, H.~Kadowaki, G.~J. Wagner,
  Bridging scale methods for nanomechanics and materials, Computer Methods in
  Applied Mechanics and Engineering 195~(13-16) (2006) 1407--1421.

\bibitem{guo2012}
J.-S. Guo, N.~I. Kavallaris, On a nonlocal parabolic problem arising in
  electrostatic \mbox{MEMS} control, Discrete Contin. Dyn. Syst 32~(5) (2012)
  1723--1746.

\bibitem{kavallaris2011}
N.~I. Kavallaris, A.~Lacey, C.~Nikolopoulos, D.~Tzanetis, A hyperbolic
  non-local problem modelling \mbox{MEMS} technology, The Rocky Mountain
  Journal of Mathematics (2011) 505--534.

\bibitem{kavallaris2016}
N.~I. Kavallaris, A.~A. Lacey, C.~V. Nikolopoulos, On the quenching of a
  nonlocal parabolic problem arising in electrostatic \mbox{MEMS} control,
  Nonlinear Analysis 138 (2016) 189--206.

\bibitem{drosinou2021}
O.~Drosinou, N.~I. Kavallaris, C.~V. Nikolopoulos, A study of a nonlocal
  problem with robin boundary conditions arising from technology, Mathematical
  Methods in the Applied Sciences 44~(13) (2021) 10084--10120.

\bibitem{kavallaris2023a}
N.~I. Kavallaris, C.~V. Nikolopoulos, A.~N. Yannacopoulos, On the impact of
  noise on quennching for a nonlocal diffusion model driven by a mixture of
  \mbox{Brownian and fractional Brownian motions}, Discrete and Continuous
  Dynamical Systems - S 17~(3) (2024) 1222--1268.

\bibitem{ei2021}
S.-I. Ei, H.~Ishii, S.~Kondo, T.~Miura, Y.~Tanaka, Effective nonlocal kernels
  on reaction--diffusion networks, Journal of theoretical biology 509 (2021)
  110496.

\bibitem{li2024}
X.~Li, H.~Shao, Investigation of bio-thermo-mechanical responses based on
  nonlocal elasticity theory and fractional pennes equation, Applied
  Mathematical Modelling 125 (2024) 390--401.

\bibitem{eringen1983}
A.~C. Eringen, On differential equations of nonlocal elasticity and solutions
  of screw dislocation and surface waves, Journal of applied physics 54~(9)
  (1983) 4703--4710.

\bibitem{lu2007}
P.~Lu, H.~Lee, C.~Lu, P.~Zhang, Application of nonlocal beam models for carbon
  nanotubes, International Journal of Solids and Structures 44~(16) (2007)
  5289--5300.

\bibitem{weaver1991}
W.~Weaver~Jr, S.~P. Timoshenko, D.~H. Young, Vibration problems in engineering,
  John Wiley \& Sons, 1991.

\bibitem{zienkiewicz2005}
O.~C. Zienkiewicz, R.~L. Taylor, The finite element method for solid and
  structural mechanics, Elsevier, 2005.

\bibitem{wang2005}
Q.~Wang, Wave propagation in carbon nanotubes via nonlocal continuum mechanics,
  Journal of Applied physics 98~(12) (2005) 124301.

\bibitem{peddieson2003}
J.~Peddieson, G.~R. Buchanan, R.~P. McNitt, Application of nonlocal continuum
  models to nanotechnology, International Journal of Engineering Science
  41~(3-5) (2003) 305--312.

\bibitem{zhang2005}
Y.~Zhang, G.~Liu, X.~Xie, Free transverse vibrations of double-walled carbon
  nanotubes using a theory of nonlocal elasticity, Physical Review B 71~(19)
  (2005) 195404.

\bibitem{navickas2006}
R.~Navickas, Technological trends of nanoelectromechanical systems, Solid State
  Phenomena 113 (2006) 7--12.

\bibitem{su2020}
Y.~Su, Z.~Zhou, Electromechanical analysis of flexoelectric nanosensors based
  on nonlocal elasticity theory, Micromachines 11~(12) (2020) 1077.

\bibitem{lim2015}
C.~Lim, G.~Zhang, J.~Reddy, A higher-order nonlocal elasticity and strain
  gradient theory and its applications in wave propagation, Journal of the
  Mechanics and Physics of Solids 78 (2015) 298--313.

\bibitem{li2016}
L.~Li, Y.~Hu, Nonlinear bending and free vibration analyses of nonlocal strain
  gradient beams made of functionally graded material, International Journal of
  Engineering Science 107 (2016) 77--97.

\bibitem{motezaker2020}
M.~Motezaker, M.~Jamali, R.~Kolahchi, Application of differential cubature
  method for nonlocal vibration, buckling and bending response of annular
  nanoplates integrated by piezoelectric layers based on surface-higher order
  nonlocal-piezoelasticity theory, Journal of Computational and Applied
  Mathematics 369 (2020) 112625.

\bibitem{duan2007}
W.~Duan, C.~M. Wang, Exact solutions for axisymmetric bending of
  micro/nanoscale circular plates based on nonlocal plate theory,
  Nanotechnology 18~(38) (2007) 385704.

\bibitem{krommer2004}
M.~Krommer, On the influence of pyroelectricity upon thermally induced
  vibrations of piezothermoelastic plates, Acta mechanica 171~(1-2) (2004)
  59--73.

\bibitem{altay2007}
G.~Altay, M.~C. D{\"o}kmeci, Variational principles for piezoelectric,
  thermopiezoelectric, and hygrothermopiezoelectric continua revisited,
  Mechanics of Advanced Materials and Structures 14~(7) (2007) 549--562.

\bibitem{yang2020}
F.~Yang, E.~Galiffi, P.~A. Huidobro, J.~Pendry, Nonlocal effects in plasmonic
  metasurfaces with almost touching surfaces, Physical Review B 101~(7) (2020)
  075434.

\bibitem{lockett1972}
F.~J. Lockett, Nonlinear viscoelastic solids, Academic Press, 1972.

\bibitem{csengul2021}
Y.~{\c{S}}eng{\"u}l, Nonlinear viscoelasticity of strain rate type: an
  overview, Proceedings of the Royal Society A 477~(2245) (2021) 20200715.

\bibitem{loret1990}
B.~Loret, J.~H. Prevost, Dynamic strain localization in elasto-(visco-) plastic
  solids, part 1. general formulation and one-dimensional examples, Computer
  Methods in Applied Mechanics and Engineering 83~(3) (1990) 247--273.

\bibitem{rivera1998}
J.~E.~M. Rivera, S.~Jiang, The thermoelastic and viscoelastic contact of two
  rods, Journal of Mathematical Analysis and Applications 217~(2) (1998)
  423--458.

\bibitem{ghajari2014}
M.~Ghajari, L.~Iannucci, P.~Curtis, A peridynamic material model for the
  analysis of dynamic crack propagation in orthotropic media, Computer Methods
  in Applied Mechanics and Engineering 276 (2014) 431--452.

\bibitem{silling2007}
S.~A. Silling, M.~Epton, O.~Weckner, J.~Xu, E.~Askari, Peridynamic states and
  constitutive modeling, Journal of Elasticity 88~(2) (2007) 151--184.

\bibitem{madenci2014}
E.~Madenci, E.~Oterkus, Peridynamic theory, in: Peridynamic theory and its
  applications, Springer, 2014, pp. 19--43.

\bibitem{oterkus2015}
S.~Oterkus, Peridynamics for the solution of multiphysics problems, Ph.D.
  thesis, The University of Arizona (2015).

\bibitem{roy2019}
P.~Roy, D.~Roy, Peridynamics model for flexoelectricity and damage, Applied
  Mathematical Modelling 68 (2019) 82--112.

\bibitem{lakes1993}
R.~Lakes, Materials with structural hierarchy, Nature 361~(6412) (1993)
  511--515.

\bibitem{kamath2006}
H.~Kamath, M.~Willatzen, R.~V. Melnik, Vibration of piezoelectric elements
  surrounded by fluid media, Ultrasonics 44~(1) (2006) 64--72.

\bibitem{wang2019}
B.~Wang, Y.~Gu, S.~Zhang, L.-Q. Chen, Flexoelectricity in solids: Progress,
  challenges, and perspectives, Progress in Materials Science 106 (2019)
  100570.

\bibitem{aifantis1992}
E.~C. Aifantis, On the role of gradients in the localization of deformation and
  fracture, International Journal of Engineering Science 30~(10) (1992)
  1279--1299.

\bibitem{krishnaswamy2020c}
J.~A. Krishnaswamy, F.~C. Buroni, R.~Melnik, L.~Rodriguez-Tembleque, A.~Saez,
  Advanced modeling of lead-free piezocomposites: The role of nonlocal and
  nonlinear effects, Composite Structures 238 (2020) 111967.

\bibitem{bouziani1997}
A.~Bouziani, Solution forte d'un probl{\`e}me mixte avec conditions non locales
  pour une classe d'{\'e}quations hyperboliques, Bulletins de l'Acad{\'e}mie
  Royale de Belgique 8~(1) (1997) 53--70.

\bibitem{merad2015}
A.~Merad, A.~Bouziani, O.~Cenap, A.~Kilicman, On solvability of the
  integrodifferential hyperbolic equation with purely nonlocal conditions, Acta
  Mathematica Scientia 35~(3) (2015) 601--609.

\bibitem{sytnyk2021}
D.~Sytnyk, R.~Melnik, Mathematical models with nonlocal initial conditions: An
  exemplification from quantum mechanics, Mathematical and Computational
  Applications 26~(4) (2021) 73.

\bibitem{ntouyas2006}
S.~Ntouyas, Nonlocal initial and boundary value problems: a survey, in:
  Handbook of differential equations: ordinary differential equations, Vol.~2,
  Elsevier, 2006, pp. 461--557.

\bibitem{raissi2019}
M.~Raissi, P.~Perdikaris, G.~E. Karniadakis, Physics-informed neural networks:
  A deep learning framework for solving forward and inverse problems involving
  nonlinear partial differential equations, Journal of Computational Physics
  378 (2019) 686--707.

\bibitem{fuks2020}
O.~Fuks, H.~A. Tchelepi, Limitations of physics informed machine learning for
  nonlinear two-phase transport in porous media, Journal of Machine Learning
  for Modeling and Computing 1~(1) (2020).

\bibitem{de2013}
S.~R. De~Groot, P.~Mazur, Non-equilibrium thermodynamics, Courier Corporation,
  2013.

\bibitem{nicolis1977}
G.~Nicolis, I.~Prigogine, Self-organization in nonequilibrium system: from
  dissipative structures to order through fluctuations NewYork London: Wiley,
  Wiley, 1977.

\bibitem{blotekjaer1970}
K.~Blotekjaer, Transport equations for electrons in two-valley semiconductors,
  IEEE Transactions on Electron Devices 17~(1) (1970) 38--47.

\bibitem{melnik2000}
R.~Melnik, H.~He, Modelling nonlocal processes in semiconductor devices with
  exponential difference schemes, Journal of Engineering Mathematics 38 (2000)
  233--263.

\bibitem{mogilner1999}
A.~Mogilner, L.~Edelstein-Keshet, A non-local model for a swarm, Journal of
  Mathematical Biology 38~(6) (1999) 534--570.

\bibitem{bastille2018}
G.~Bastille-Rousseau, D.~L. Murray, J.~A. Schaefer, M.~A. Lewis, S.~P. Mahoney,
  J.~R. Potts, Spatial scales of habitat selection decisions: implications for
  telemetry-based movement modelling, Ecography 41~(3) (2018) 437--443.

\bibitem{benhamou2014}
S.~Benhamou, Of scales and stationarity in animal movements, Ecology Letters
  17~(3) (2014) 261--272.

\bibitem{martinez2020}
R.~Martinez-Garcia, C.~H. Fleming, R.~Seppelt, W.~F. Fagan, J.~M. Calabrese,
  How range residency and long-range perception change encounter rates, Journal
  of Theoretical Biology 498 (2020) 110267.

\bibitem{potts2014}
J.~R. Potts, G.~Bastille-Rousseau, D.~L. Murray, J.~A. Schaefer, M.~A. Lewis,
  Predicting local and non-local effects of resources on animal space use using
  a mechanistic step selection model, Methods in Ecology and Evolution 5~(3)
  (2014) 253--262.

\bibitem{lee2001}
C.~Lee, M.~Hoopes, J.~Diehl, W.~Gilliland, G.~Huxel, E.~Leaver, K.~McCann,
  J.~Umbanhowar, A.~Mogilner, Non-local concepts and models in biology, Journal
  of theoretical biology 210~(2) (2001) 201--219.

\bibitem{jewell2023}
T.~J. Jewell, A.~L. Krause, P.~K. Maini, E.~A. Gaffney, Patterning of nonlocal
  transport models in biology: the impact of spatial dimension, Mathematical
  Biosciences 366 (2023) 109093.

\bibitem{giunta2021}
V.~Giunta, T.~Hillen, M.~A. Lewis, J.~R. Potts, Local and global existence for
  non-local multi-species advection-diffusion models, SIAM Journal on Applied
  Dynamical Systems 21~(3) (2022) 1686--1708.

\bibitem{carrillo2019}
J.~A. Carrillo, K.~Craig, Y.~Yao, Aggregation-diffusion equations: dynamics,
  asymptotics, and singular limits, in: Active Particles, Volume 2, Springer,
  2019, pp. 65--108.

\bibitem{topaz2006}
C.~M. Topaz, A.~L. Bertozzi, M.~A. Lewis, A nonlocal continuum model for
  biological aggregation, Bulletin of Mathematical Biology 68~(7) (2006)
  1601--1623.

\bibitem{james2015}
F.~James, N.~Vauchelet, Numerical methods for one-dimensional aggregation
  equations, SIAM Journal on Numerical Analysis 53~(2) (2015) 895--916.

\bibitem{craig2016}
K.~Craig, A.~Bertozzi, A blob method for the aggregation equation, Mathematics
  of Computation 85~(300) (2016) 1681--1717.

\bibitem{potts2019}
J.~R. Potts, M.~A. Lewis, Spatial memory and taxis-driven pattern formation in
  model ecosystems, Bulletin of Mathematical Biology 81~(7) (2019) 2725--2747.

\bibitem{watson1973}
C.~Watson, Non-local memories in physics and biology, International Journal of
  Neuroscience 6~(1) (1973) 31--33.

\bibitem{thieu2023}
T.~Thieu, R.~Melnik, Social human collective decision-making and its
  applications with brain network models, in: Crowd Dynamics, Volume 4:
  Analytics and Human Factors in Crowd Modeling, Springer, 2023, pp. 103--141.

\bibitem{contreras2025}
H.~D. Contreras, P.~Goatin, L.-M. Villada, A two-lane bidirectional nonlocal
  traffic model, Journal of Mathematical Analysis and Applications 543~(2)
  (2025) 129027.

\bibitem{srivastava2022}
A.~Srivastava, S.~Jha, et~al., Data-driven machine learning: A new approach to
  process and utilize biomedical data, in: Predictive Modeling in Biomedical
  Data Mining and Analysis, Elsevier, 2022, pp. 225--252.

\bibitem{arridge2019}
S.~Arridge, P.~Maass, O.~{\"O}ktem, C.-B. Sch{\"o}nlieb, Solving inverse
  problems using data-driven models, Acta Numerica 28 (2019) 1--174.

\bibitem{li2023}
Y.~Li, L.~Gao, S.~Hu, G.~Gui, C.-Y. Chen, Nonlocal low-rank plus deep denoising
  prior for robust image compressed sensing reconstruction, Expert Systems with
  Applications 228 (2023) 120456.

\bibitem{xie2014}
W.~Xie, B.~L. Nelson, R.~R. Barton, A bayesian framework for quantifying
  uncertainty in stochastic simulation, Operations Research 62~(6) (2014)
  1439--1452.

\bibitem{bellomo2024}
N.~Bellomo, M.~Dolfin, J.~Liao, Life and self-organization on the way to
  artificial intelligence for collective dynamics, Physics of Life Reviews 51
  (2024) 1--8.

\bibitem{boussange2023}
V.~Boussange, S.~Becker, A.~Jentzen, B.~Kuckuck, L.~Pellissier, Deep learning
  approximations for non-local nonlinear pdes with neumann boundary conditions,
  Partial Differential Equations and Applications 4~(6) (2023) 51.

\bibitem{anstine2023}
D.~M. Anstine, O.~Isayev, Machine learning interatomic potentials and
  long-range physics, The Journal of Physical Chemistry A 127~(11) (2023)
  2417--2431.

\bibitem{farajpour2024}
A.~Farajpour, W.~V. Ingman, Flexural eigenfrequency analysis of healthy and
  pathological tissues using machine learning and nonlocal viscoelasticity,
  Computers 13~(7) (2024) 179.

\bibitem{ramezanian2022}
M.~Ramezanian-Panahi, G.~Abrevaya, J.-C. Gagnon-Audet, V.~Voleti, I.~Rish,
  G.~Dumas, Generative models of brain dynamics, Frontiers in Artificial
  Intelligence 5 (2022) 807406.

\bibitem{carbone2023}
D.~Carbone, M.~Hua, S.~Coste, E.~Vanden-Eijnden, Efficient training of
  energy-based models using jarzynski equality, Advances in Neural Information
  Processing Systems 36 (2023) 52583--52614.

\bibitem{you2021}
H.~You, Y.~Yu, N.~Trask, M.~Gulian, M.~D’Elia, Data-driven learning of
  nonlocal physics from high-fidelity synthetic data, Computer Methods in
  Applied Mechanics and Engineering 374 (2021) 113553.

\bibitem{peherstorfer2016}
B.~Peherstorfer, K.~Willcox, M.~Gunzburger, Optimal model management for
  multifidelity \mbox{Monte Carlo} estimation, SIAM Journal on Scientific
  Computing 38~(5) (2016) A3163--A3194.

\bibitem{zitnik2024}
M.~Zitnik, M.~M. Li, A.~Wells, K.~Glass, D.~Morselli~Gysi, A.~Krishnan, T.~M.
  Murali, P.~Radivojac, S.~Roy, A.~Baudot, S.~Bozdag, D.~Z. Chen, L.~Cowen,
  K.~Devkota, A.~Gitter, S.~J.~C. Gosline, P.~Gu, P.~H. Guzzi, H.~Huang,
  M.~Jiang, Z.~N. Kesimoglu, M.~Koyuturk, J.~Ma, A.~R. Pico, N.~Pržulj, T.~M.
  Przytycka, B.~J. Raphael, A.~Ritz, R.~Sharan, Y.~Shen, M.~Singh, D.~K.
  Slonim, H.~Tong, X.~H. Yang, B.-J. Yoon, H.~Yu, T.~Milenković, Current and
  future directions in network biology, Bioinformatics Advances 4~(1) (2024)
  vbae099.

\bibitem{matthaus2009}
F.~Matth{\"a}us, The spread of prion diseases in the brain—models of reaction
  and transport on networks, Journal of Biological Systems 17~(04) (2009)
  623--641.

\bibitem{bressloff2014}
P.~C. Bressloff, Waves in neural media, Lecture Notes on Mathematical Modelling
  in the Life Sciences (2014) 18--19.

\bibitem{tao2018}
Y.~Tao, Q.~Sun, Q.~Du, W.~Liu, Nonlocal neural networks, nonlocal diffusion and
  nonlocal modeling, Advances in Neural Information Processing Systems 31
  (2018).

\bibitem{nalls2021}
M.~A. Nalls, C.~Blauwendraat, L.~Sargent, D.~Vitale, H.~Leonard, H.~Iwaki,
  Y.~Song, S.~Bandres-Ciga, K.~Menden, F.~Faghri, et~al., Evidence for grn
  connecting multiple neurodegenerative diseases, Brain Communications 3~(2)
  (2021) fcab095.

\bibitem{melnik2009a}
R.~V. Melnik, X.~Wei, G.~MORENO-HAGELSIEB, Nonlinear dynamics of cell cycles
  with stochastic mathematical models, Journal of Biological Systems 17~(03)
  (2009) 425--460.

\bibitem{thieu2021}
T.~K.~T. Thieu, R.~Melnik, Coupled effects of channels and synaptic dynamics in
  stochastic modelling of healthy and parkinson's-disease-affected brains, AIMS
  Bioengineering 9~(2) (2022) 213--238.

\bibitem{kim2023}
T.~Kim Thoa~Thieu, A.~Muntean, R.~Melnik, Coupled stochastic systems of
  \mbox{Skorokhod} type: Well-posedness of a mathematical model and its
  applications, Mathematical Methods in the Applied Sciences 46~(6) (2023)
  7368--7390.

\bibitem{kim2018}
J.~Kim, M.~Foo, D.~G. Bates, Computationally efficient modelling of stochastic
  spatio-temporal dynamics in biomolecular networks, Scientific Reports 8~(1)
  (2018) 3498.

\bibitem{peukert2014}
M.~Peukert, J.~Thiel, D.~Peshev, W.~Weschke, W.~Van~den Ende, H.-P. Mock,
  A.~Matros, Spatio-temporal dynamics of fructan metabolism in developing
  barley grains, The Plant Cell 26~(9) (2014) 3728--3744.

\bibitem{el2019}
M.~El-Hachem, S.~W. McCue, W.~Jin, Y.~Du, M.~J. Simpson, Revisiting the
  \mbox{Fisher--Kolmogorov--Petrovsky--Piskunov} equation to interpret the
  spreading--extinction dichotomy, Proceedings of the Royal Society A
  475~(2229) (2019) 20190378.

\bibitem{shigesada1995}
N.~Shigesada, K.~Kawasaki, Y.~Takeda, Modeling stratified diffusion in
  biological invasions, The American Naturalist 146~(2) (1995) 229--251.

\bibitem{steele1998}
J.~Steele, J.~Adams, T.~Sluckin, Modelling paleoindian dispersals, World
  Archaeology 30~(2) (1998) 286--305.

\bibitem{meszena2005}
G.~Mesz{\'e}na, M.~Gyllenberg, F.~J. Jacobs, J.~A. Metz, Link between
  population dynamics and dynamics of darwinian evolution, Physical Review
  Letters 95~(7) (2005) 078105.

\bibitem{pigolotti2007}
S.~Pigolotti, C.~L{\'o}pez, E.~Hern{\'a}ndez-Garc{\'\i}a, Species clustering in
  competitive \mbox{Lotka-Volterra} models, Physical Review Letters 98~(25)
  (2007) 258101.

\bibitem{perthame2008}
B.~Perthame, G.~Barles, Dirac concentrations in \mbox{Lotka-Volterra} parabolic
  pdes, Indiana University Mathematics Journal (2008) 3275--3301.

\bibitem{djilali2020}
S.~Djilali, Pattern formation of a diffusive predator-prey model with herd
  behavior and nonlocal prey competition, Mathematical Methods in the Applied
  Sciences 43~(5) (2020) 2233--2250.

\bibitem{calsina1994}
{\'A}.~Calsina, C.~Perell{\'o}, J.~Salda{\~n}a, Non-local reaction-diffusion
  equations modelling predator-prey coevolution, Publicacions matematiques
  (1994) 315--325.

\bibitem{saha2023}
S.~Saha, S.~Pal, R.~Melnik, The analysis of the impact of fear in the presence
  of additional food and prey refuge with nonlocal predator-prey models, arXiv
  preprint arXiv:2310.01392 (2023).

\bibitem{saha2024}
S.~Saha, S.~Pal, R.~Melnik, Nonlocal cooperative behaviour, psychological
  effects, and collective decision-making: an exemplification with
  predator-prey models, arXiv preprint arXiv:2406.10713 (2024).

\bibitem{genieys2006}
S.~G{\'e}nieys, V.~Volpert, P.~Auger, Adaptive dynamics: modelling
  \mbox{Darwin's} divergence principle, Comptes Rendus Biologies 329~(11)
  (2006) 876--879.

\bibitem{fuentes2004}
M.~Fuentes, M.~Kuperman, V.~Kenkre, Analytical considerations in the study of
  spatial patterns arising from nonlocal interaction effects, The Journal of
  Physical Chemistry B 108~(29) (2004) 10505--10508.

\bibitem{berestycki2007}
H.~Berestycki, F.~Hamel, Reaction-diffusion equations and propagation
  phenomena, Springer, 2007.

\bibitem{schenk2006}
H.~J. Schenk, Root competition: beyond resource depletion, Journal of Ecology
  94~(4) (2006) 725--739.

\bibitem{gourley2000}
S.~A. Gourley, Travelling front solutions of a nonlocal \mbox{Fisher} equation,
  Journal of Mathematical Biology 41~(3) (2000) 272--284.

\bibitem{cantrell1996}
R.~S. Cantrell, C.~Cosner, Models for predator-prey systems at multiple scales,
  SIAM Review 38~(2) (1996) 256--286.

\bibitem{billingham2003}
J.~Billingham, Dynamics of a strongly nonlocal reaction--diffusion population
  model, Nonlinearity 17~(1) (2003) 313.

\bibitem{yang2024}
F.~Yang, W.-Y. Tang, F.-Y. Yang, B.-E. Jiang, Minimal wave speed for a
  predator-prey system with nonlocal dispersal, International Journal of
  Biomathematics (2024).

\bibitem{fagioli2019}
S.~Fagioli, Y.~Jaafra, Multiple patterns formation for an aggregation/diffusion
  predator-prey system, Networks and Heterogeneous Media 16~(3) (2021)
  377--411.

\bibitem{painter2015}
K.~J. Painter, J.~Bloomfield, J.~Sherratt, A.~Gerisch, A nonlocal model for
  contact attraction and repulsion in heterogeneous cell populations, Bulletin
  of Mathematical Biology 77~(6) (2015) 1132--1165.

\bibitem{potts2016}
J.~R. Potts, M.~A. Lewis, How memory of direct animal interactions can lead to
  territorial pattern formation, Journal of the Royal Society Interface
  13~(118) (2016) 20160059.

\bibitem{bullock2017}
J.~M. Bullock, L.~Mallada~Gonz{\'a}lez, R.~Tamme, L.~G{\"o}tzenberger, S.~M.
  White, M.~P{\"a}rtel, D.~A. Hooftman, A synthesis of empirical plant
  dispersal kernels, Journal of Ecology 105~(1) (2017) 6--19.

\bibitem{allen1996}
E.~J. Allen, L.~J. Allen, X.~Gilliam, Dispersal and competition models for
  plants, Journal of Mathematical Biology 34~(4) (1996) 455--481.

\bibitem{hanski1998}
I.~Hanski, Metapopulation dynamics, Nature 396~(6706) (1998) 41--49.

\bibitem{hastings2005}
A.~Hastings, K.~Cuddington, K.~F. Davies, C.~J. Dugaw, S.~Elmendorf,
  A.~Freestone, S.~Harrison, M.~Holland, J.~Lambrinos, U.~Malvadkar, et~al.,
  The spatial spread of invasions: new developments in theory and evidence,
  Ecology Letters 8~(1) (2005) 91--101.

\bibitem{okubo2001}
A.~Okubo, S.~A. Levin, et~al., Diffusion and ecological problems: modern
  perspectives, Vol.~14, Springer, 2001.

\bibitem{yin2018}
H.~Yin, X.~Wen, Pattern formation through temporal fractional derivatives,
  Scientific Reports 8~(1) (2018) 1--9.

\bibitem{javidi2013}
M.~Javidi, N.~Nyamoradi, Dynamic analysis of a fractional order prey-predator
  interaction with harvesting, Applied mathematical modelling 37~(20-21) (2013)
  8946--8956.

\bibitem{rihan2015}
F.~Rihan, S.~Lakshmanan, A.~Hashish, R.~Rakkiyappan, E.~Ahmed, Fractional-order
  delayed predator--prey systems with \mbox{Holling type-II} functional
  response, Nonlinear Dynamics 80~(1) (2015) 777--789.

\bibitem{matouk2016}
A.~Matouk, A.~Elsadany, Dynamical analysis, stabilization and discretization of
  a chaotic fractional-order \mbox{GLV} model, Nonlinear Dynamics 85~(3) (2016)
  1597--1612.

\bibitem{sisman2021}
A.~Sisman, J.~Fransson, Fractional integral representation in statistical
  thermodynamics of confined systems, Physical Review E 104~(5) (2021) 054110.

\bibitem{bertin2003}
C.~Bertin, X.~Yang, L.~A. Weston, The role of root exudates and allelochemicals
  in the rhizosphere, Plant and Soil 256~(1) (2003) 67--83.

\bibitem{cheng2015}
F.~Cheng, Z.~Cheng, Research progress on the use of plant allelopathy in
  agriculture and the physiological and ecological mechanisms of allelopathy,
  Frontiers in Plant Science 6 (2015) 1020.

\bibitem{czaran2002}
T.~L. Cz{\'a}r{\'a}n, R.~F. Hoekstra, L.~Pagie, Chemical warfare between
  microbes promotes biodiversity, Proceedings of the National Academy of
  Sciences 99~(2) (2002) 786--790.

\bibitem{nadell2016}
C.~D. Nadell, K.~Drescher, K.~R. Foster, Spatial structure, cooperation and
  competition in biofilms, Nature Reviews Microbiology 14~(9) (2016) 589--600.

\bibitem{houston1985}
A.~I. Houston, J.~M. McNamara, A general theory of central place foraging for
  single-prey loaders, Theoretical Population Biology 28~(3) (1985) 233--262.

\bibitem{olsson2008}
O.~Olsson, J.~S. Brown, K.~L. Helf, A guide to central place effects in
  foraging, Theoretical Population Biology 74~(1) (2008) 22--33.

\bibitem{keyfitz1997}
B.~L. Keyfitz, N.~Keyfitz, The mckendrick partial differential equation and its
  uses in epidemiology and population study, Mathematical and Computer
  Modelling 26~(6) (1997) 1--9.

\bibitem{wang2015}
J.-B. Wang, W.-T. Li, F.-Y. Yang, Traveling waves in a nonlocal dispersal
  \mbox{SIR} model with nonlocal delayed transmission, Communications in
  Nonlinear Science and Numerical Simulation 27~(1-3) (2015) 136--152.

\bibitem{tian2017}
B.~Tian, R.~Yuan, Traveling waves for a diffusive \mbox{SEIR} epidemic model
  with non-local reaction and with standard incidences, Nonlinear Analysis:
  Real World Applications 37 (2017) 162--181.

\bibitem{liu2019}
H.~Liu, J.-F. Zhang, Dynamics of two time delays differential equation model to
  \mbox{HIV} latent infection, Physica A: Statistical Mechanics and its
  Applications 514 (2019) 384--395.

\bibitem{zhou2020}
J.~Zhou, Y.~Yang, C.-H. Hsu, Traveling waves for a nonlocal dispersal
  vaccination model with general incidence, Discrete \& Continuous Dynamical
  Systems-B 25~(4) (2020) 1469.

\bibitem{kang2020}
H.~Kang, S.~Ruan, X.~Yu, Age-structured population dynamics with nonlocal
  diffusion, Journal of Dynamics and Differential Equations (2020) 1--35.

\bibitem{kang2021b}
H.~Kang, S.~Ruan, Mathematical analysis on an age-structured \mbox{SIS}
  epidemic model with nonlocal diffusion, Journal of Mathematical Biology
  83~(1) (2021) 1--30.

\bibitem{xu2020}
W.-B. Xu, W.-T. Li, S.~Ruan, Spatial propagation in an epidemic model with
  nonlocal diffusion: The influences of initial data and dispersals, Science
  China Mathematics 63~(11) (2020) 2177--2206.

\bibitem{yang2019}
F.-Y. Yang, W.-T. Li, S.~Ruan, Dynamics of a nonlocal dispersal \mbox{SIS}
  epidemic model with \mbox{Neumann} boundary conditions, Journal of
  Differential Equations 267~(3) (2019) 2011--2051.

\bibitem{yang2013}
F.-Y. Yang, Y.~Li, W.-T. Li, Z.-C. Wang, Traveling waves in a nonlocal
  dispersal kermack-mckendrick epidemic model, Discrete \& Continuous Dynamical
  Systems-B 18~(7) (2013) 1969.

\bibitem{webb2022}
G.~Webb, The force of cell-cell adhesion in determining the outcome in a
  nonlocal advection diffusion model of wound healing, Mathematical Biosciences
  and Engineering: MBE 19~(9) (2022) 8689--8704.

\bibitem{bitsouni2018}
V.~Bitsouni, R.~Eftimie, Non-local parabolic and hyperbolic models for cell
  polarisation in heterogeneous cancer cell populations, Bulletin of
  Mathematical Biology 80~(10) (2018) 2600--2632.

\bibitem{swanson2008}
K.~Swanson, R.~Rostomily, E.~Alvord, A mathematical modelling tool for
  predicting survival of individual patients following resection of
  glioblastoma: a proof of principle, British Journal of Cancer 98~(1) (2008)
  113--119.

\bibitem{swanson2003}
K.~R. Swanson, C.~Bridge, J.~Murray, E.~C. Alvord~Jr, Virtual and real brain
  tumors: using mathematical modeling to quantify glioma growth and invasion,
  Journal of the Neurological Sciences 216~(1) (2003) 1--10.

\bibitem{perez2018}
J.~P{\'e}rez-Beteta, A.~Mart{\'\i}nez-Gonz{\'a}lez, V.~P{\'e}rez-Garc{\'\i}a, A
  three-dimensional computational analysis of magnetic resonance images
  characterizes the biological aggressiveness in malignant brain tumours,
  Journal of the Royal Society Interface 15~(149) (2018) 20180503.

\bibitem{vo2015}
B.~N. Vo, C.~C. Drovandi, A.~N. Pettitt, M.~J. Simpson, Quantifying uncertainty
  in parameter estimates for stochastic models of collective cell spreading
  using approximate \mbox{Bayesian} computation, Mathematical Biosciences 263
  (2015) 133--142.

\bibitem{warne2019}
D.~J. Warne, R.~E. Baker, M.~J. Simpson, Using experimental data and
  information criteria to guide model selection for reaction--diffusion
  problems in mathematical biology, Bulletin of Mathematical Biology 81~(6)
  (2019) 1760--1804.

\bibitem{tremel2009}
A.~Tremel, A.~Cai, N.~Tirtaatmadja, B.~D. Hughes, G.~W. Stevens, K.~A. Landman,
  A.~J. O’Connor, Cell migration and proliferation during monolayer formation
  and wound healing, Chemical Engineering Science 64~(2) (2009) 247--253.

\bibitem{sengers2007}
B.~G. Sengers, C.~P. Please, R.~O. Oreffo, Experimental characterization and
  computational modelling of two-dimensional cell spreading for skeletal
  regeneration, Journal of the Royal Society Interface 4~(17) (2007)
  1107--1117.

\bibitem{cai2007}
A.~Q. Cai, K.~A. Landman, B.~D. Hughes, Multi-scale modeling of a wound-healing
  cell migration assay, Journal of Theoretical Biology 245~(3) (2007) 576--594.

\bibitem{folkman1991}
J.~Folkman, D.~Hanahan, Switch to the angiogenic phenotype during
  tumorigenesis., in: Princess Takamatsu Symposia, Vol.~22, 1991, pp. 339--347.

\bibitem{d2005}
A.~d’Onofrio, A general framework for modeling tumor-immune system
  competition and immunotherapy: Mathematical analysis and biomedical
  inferences, Physica D: Nonlinear Phenomena 208~(3-4) (2005) 220--235.

\bibitem{teng2008}
M.~W. Teng, J.~B. Swann, C.~M. Koebel, R.~D. Schreiber, M.~J. Smyth,
  Immune-mediated dormancy: an equilibrium with cancer, Journal of Leukocyte
  Biology 84~(4) (2008) 988--993.

\bibitem{almog2009}
N.~Almog, L.~Ma, R.~Raychowdhury, C.~Schwager, R.~Erber, S.~Short, L.~Hlatky,
  P.~Vajkoczy, P.~E. Huber, J.~Folkman, et~al., Transcriptional switch of
  dormant tumors to fast-growing angiogenic phenotype, Cancer Research 69~(3)
  (2009) 836--844.

\bibitem{hanahan2011}
D.~Hanahan, R.~A. Weinberg, Hallmarks of cancer: the next generation, Cell
  144~(5) (2011) 646--674.

\bibitem{hedley2024}
J.~G. Hedley, K.~Coshic, A.~Aksimentiev, A.~A. Kornyshev, Electric field of dna
  in solution: who is in charge?, Physical Review X 14~(3) (2024) 031042.

\bibitem{solis2019}
J.~Sol{\'\i}s-P{\'e}rez, J.~F. G{\'o}mez-Aguilar, A.~Atangana, A fractional
  mathematical model of breast cancer competition model, Chaos, Solitons \&
  Fractals 127 (2019) 38--54.

\bibitem{yin2011}
C.~Yin, X.~Li, Anomalous diffusion of drug release from a slab matrix:
  Fractional diffusion models, International Journal of Pharmaceutics 418~(1)
  (2011) 78--87.

\bibitem{magin2010}
R.~L. Magin, Fractional calculus models of complex dynamics in biological
  tissues, Computers \& Mathematics with Applications 59~(5) (2010) 1586--1593.

\bibitem{sterner2011}
R.~W. Sterner, G.~E. Small, J.~M. Hood, The conservation of mass, Nature
  Education Knowledge 3~(10) (2011) 20.

\bibitem{pang2014}
G.~Pang, J.~Xie, Q.~Chen, Z.~Hu, Energy intake, metabolic homeostasis, and
  human health, Food Science and Human Wellness 3~(3-4) (2014) 89--103.

\bibitem{idumah2023}
G.~Idumah, E.~Somersalo, D.~Calvetti, A spatially distributed model of brain
  metabolism highlights the role of diffusion in brain energy metabolism,
  Journal of Theoretical Biology 572 (2023) 111567.

\bibitem{bohme2025}
T.~B{\"o}hme, S.~G{\"o}ttlich, A.~Neuenkirch, A nonlocal traffic flow model
  with stochastic velocity, ESAIM: Mathematical Modelling and Numerical
  Analysis 59~(1) (2025) 487--518.

\bibitem{santini2022}
L.~Santini, A.~Ben{\'\i}tez-L{\'o}pez, C.~F. Dormann, M.~A. Huijbregts,
  Population density estimates for terrestrial mammal species, Global Ecology
  and Biogeography 31~(5) (2022) 978--994.

\bibitem{edelstein1998}
L.~Edelstein-Keshet, J.~Watmough, D.~Grunbaum, Do travelling band solutions
  describe cohesive swarms? an investigation for migratory locusts, Journal of
  Mathematical Biology 36~(6) (1998) 515--549.

\bibitem{topaz2004}
C.~M. Topaz, A.~L. Bertozzi, Swarming patterns in a two-dimensional kinematic
  model for biological groups, SIAM Journal on Applied Mathematics 65~(1)
  (2004) 152--174.

\bibitem{bertozzi2009}
A.~L. Bertozzi, J.~A. Carrillo, T.~Laurent, Blow-up in multidimensional
  aggregation equations with mildly singular interaction kernels, Nonlinearity
  22~(3) (2009) 683.

\bibitem{breder1954}
C.~M. Breder, Equations descriptive of fish schools and other animal
  aggregations, Ecology 35~(3) (1954) 361--370.

\bibitem{conte2022}
M.~Conte, N.~Loy, Multi-cue kinetic model with non-local sensing for cell
  migration on a fiber network with chemotaxis, Bulletin of Mathematical
  Biology 84~(3) (2022) 1--46.

\bibitem{berg1983}
H.~C. Berg, Random walks in biology, Princeton University Press, Princeton
  (1983).

\bibitem{block1983}
S.~M. Block, J.~E. Segall, H.~C. Berg, Adaptation kinetics in bacterial
  chemotaxis, Journal of Bacteriology 154~(1) (1983) 312--323.

\bibitem{stroock1974}
D.~W. Stroock, Some stochastic processes which arise from a model of the motion
  of a bacterium, Zeitschrift f{\"u}r Wahrscheinlichkeitstheorie und verwandte
  Gebiete 28~(4) (1974) 305--315.

\bibitem{loy2020}
N.~Loy, L.~Preziosi, Kinetic models with non-local sensing determining cell
  polarization and speed according to independent cues, Journal of Mathematical
  Biology 80~(1) (2020) 373--421.

\bibitem{othmer1988}
H.~G. Othmer, S.~R. Dunbar, W.~Alt, Models of dispersal in biological systems,
  Journal of Mathematical Biology 26~(3) (1988) 263--298.

\bibitem{hillen2006}
T.~Hillen, M5 mesoscopic and macroscopic models for mesenchymal motion, Journal
  of Mathematical Biology 53~(4) (2006) 585--616.

\bibitem{Elia2024}
M.~D'Elia, M.~Gunzburger, C.~Vollmann, Nonlocal Integral Equation Continuum
  Models: Nonstandard Symmetric Interaction Neighborhoods and Finite Element
  Discretizations, SIAM, 2024.

\bibitem{gerisch2008}
A.~Gerisch, M.~A. Chaplain, Mathematical modelling of cancer cell invasion of
  tissue: local and non-local models and the effect of adhesion, Journal of
  Theoretical Biology 250~(4) (2008) 684--704.

\bibitem{painter2010}
K.~J. Painter, N.~J. Armstrong, J.~A. Sherratt, The impact of adhesion on
  cellular invasion processes in cancer and development, Journal of Theoretical
  Biology 264~(3) (2010) 1057--1067.

\bibitem{andasari2011}
V.~Andasari, A.~Gerisch, G.~Lolas, A.~P. South, M.~A. Chaplain, Mathematical
  modeling of cancer cell invasion of tissue: biological insight from
  mathematical analysis and computational simulation, Journal of Mathematical
  Biology 63~(1) (2011) 141--171.

\bibitem{chaplain2011}
M.~A. Chaplain, M.~LACHOWICZ, Z.~SZYMA{\'N}SKA, D.~Wrzosek, Mathematical
  modelling of cancer invasion: the importance of cell--cell adhesion and
  cell--matrix adhesion, Mathematical Models and Methods in Applied Sciences
  21~(04) (2011) 719--743.

\bibitem{domschke2014}
P.~Domschke, D.~Trucu, A.~Gerisch, M.~A. Chaplain, Mathematical modelling of
  cancer invasion: implications of cell adhesion variability for tumour
  infiltrative growth patterns, Journal of Theoretical Biology 361 (2014)
  41--60.

\bibitem{mogilner1995}
A.~Mogilner, Modelling spatio-angular patterns in cell biology, Ph.D. thesis,
  University of British Columbia (1995).

\bibitem{green2010}
J.~Green, S.~Waters, J.~Whiteley, L.~Edelstein-Keshet, K.~Shakesheff, H.~Byrne,
  Non-local models for the formation of hepatocyte--stellate cell aggregates,
  Journal of Theoretical Biology 267~(1) (2010) 106--120.

\bibitem{bromberek2002}
B.~Bromberek, P.~Enever, D.~Shreiber, M.~Caldwell, R.~Tranquillo, Macrophages
  influence a competition of contact guidance and chemotaxis for fibroblast
  alignment in a fibrin gel coculture assay, Experimental Cell Research 275~(2)
  (2002) 230--242.

\bibitem{provenzano2009}
P.~P. Provenzano, K.~W. Eliceiri, P.~J. Keely, Shining new light on \mbox{3D}
  cell motility and the metastatic process, Trends in Cell Biology 19~(11)
  (2009) 638--648.

\bibitem{kavallaris2007b}
N.~I. Kavallaris, T.~Suzuki, On the finite-time blow-up of a non-local
  parabolic equation describing chemotaxis, Differential Integral Equations
  20~(3) (2007) 293--308.

\bibitem{ben2000}
E.~Ben-Jacob, I.~Cohen, H.~Levine, Cooperative self-organization of
  microorganisms, Advances in Physics 49~(4) (2000) 395--554.

\bibitem{camazine2020}
S.~Camazine, J.-L. Deneubourg, N.~R. Franks, J.~Sneyd, G.~Theraula,
  E.~Bonabeau, Self-organization in biological systems, Princeton University
  Press, 2020.

\bibitem{parrish1997}
J.~K. Parrish, W.~M. Hamner, Animal groups in three dimensions: how species
  aggregate, Cambridge University Press, 1997.

\bibitem{parrish1999}
J.~K. Parrish, L.~Edelstein-Keshet, Complexity, pattern, and evolutionary
  trade-offs in animal aggregation, Science 284~(5411) (1999) 99--101.

\bibitem{wang2017}
Z.~L. Wang, On maxwell's displacement current for energy and sensors: the
  origin of nanogenerators, Materials Today 20~(2) (2017) 74--82.

\bibitem{buroni2020}
J.~L. Buroni, F.~C. Buroni, A.~P. Cisilino, R.~Melnik,
  L.~Rodr{\'\i}guez-Tembleque, A.~S{\'a}ez, Analytical expressions to estimate
  the effective piezoelectric tensor of a textured polycrystal for any crystal
  symmetry, Mechanics of Materials 151 (2020) 103604.

\bibitem{krishnaswamy2019}
J.~A. Krishnaswamy, F.~C. Buroni, E.~Garc{\'\i}a-Mac{\'\i}as, R.~Melnik,
  L.~Rodriguez-Tembleque, A.~Saez, Design of lead-free \mbox{PVDF/CNT/BaTiO3}
  piezocomposites for sensing and energy harvesting: the role of
  polycrystallinity, nanoadditives, and anisotropy, Smart Materials and
  Structures 29~(1) (2019) 015021.

\bibitem{krishnaswamy2019a}
J.~A. Krishnaswamy, F.~C. Buroni, F.~Garcia-Sanchez, R.~Melnik,
  L.~Rodriguez-Tembleque, A.~Saez, Lead-free piezocomposites with
  \mbox{CNT-modified} matrices: Accounting for agglomerations and molecular
  defects, Composite Structures 224 (2019) 111033.

\bibitem{krishnaswamy2019b}
J.~A. Krishnaswamy, F.~C. Buroni, F.~Garcia-Sanchez, R.~Melnik,
  L.~Rodriguez-Tembleque, A.~Saez, Improving the performance of lead-free
  piezoelectric composites by using polycrystalline inclusions and tuning the
  dielectric matrix environment, Smart Materials and Structures 28~(7) (2019)
  075032.

\bibitem{krishnaswamy2020}
J.~A. Krishnaswamy, L.~Rodriguez-Tembleque, R.~Melnik, F.~C. Buroni, A.~Saez,
  Size dependent electro-elastic enhancement in geometrically anisotropic
  lead-free piezocomposites, International Journal of Mechanical Sciences 182
  (2020) 105745.

\bibitem{wang2018}
D.~Wang, R.~Melnik, L.~Wang, Material influence in newly proposed ferroelectric
  energy harvesters, Journal of Intelligent Material Systems and Structures
  29~(16) (2018) 3305--3316.

\bibitem{krishnaswamy2020a}
J.~A. Krishnaswamy, F.~C. Buroni, R.~Melnik, L.~Rodriguez-Tembleque, A.~Saez,
  Design of polymeric auxetic matrices for improved mechanical coupling in
  lead-free piezocomposites, Smart Materials and Structures 29~(5) (2020)
  054002.

\bibitem{krishnaswamy2020b}
J.~A. Krishnaswamy, F.~C. Buroni, E.~Garc{\'\i}a-Mac{\'\i}as, R.~Melnik,
  L.~Rodriguez-Tembleque, A.~Saez, Design of nano-modified pvdf matrices for
  lead-free piezocomposites: Graphene vs carbon nanotube nano-additions,
  Mechanics of Materials 142 (2020) 103275.

\bibitem{krishnaswamy2021}
J.~A. Krishnaswamy, F.~C. Buroni, R.~Melnik, L.~Rodriguez-Tembleque, A.~Saez,
  Multiscale design of nanoengineered matrices for lead-free piezocomposites:
  Improved performance via controlling auxeticity and anisotropy, Composite
  Structures 255 (2021) 112909.

\bibitem{krishnaswamy2023}
J.~A. Krishnaswamy, F.~C. Buroni, R.~Melnik, L.~Rodriguez-Tembleque, A.~Saez,
  Flexoelectric enhancement in lead-free piezocomposites with graded inclusion
  concentrations and porous matrices, Computers \& Structures 289 (2023)
  107176.

\bibitem{buroni2024}
J.~L. Buroni, R.~Melnik, L.~Rodr{\'\i}guez-Tembleque, A.~S{\'a}ez, F.~C.
  Buroni, Closed-form expressions for computing flexoelectric coefficients in
  textured polycrystalline dielectrics, Applied Mathematical Modelling 125
  (2024) 375--389.

\bibitem{zou2021}
Y.~Zou, L.~Bo, Z.~Li, Recent progress in human body energy harvesting for smart
  bioelectronic system, Fundamental Research 1~(3) (2021) 364--382.

\bibitem{yang2013a}
Y.~Yang, H.~Zhang, Z.-H. Lin, Y.~S. Zhou, Q.~Jing, Y.~Su, J.~Yang, J.~Chen,
  C.~Hu, Z.~L. Wang, Human skin based triboelectric nanogenerators for
  harvesting biomechanical energy and as self-powered active tactile sensor
  system, ACS Nano 7~(10) (2013) 9213--9222.

\bibitem{balitsky1996}
I.~Balitsky, Operator expansion for high-energy scattering, Nuclear Physics B
  463~(1) (1996) 99--157.

\bibitem{fukushima2006}
K.~Fukushima, Deriving the
  \mbox{Jalilian-Marian--Iancu--McLerran--Weigert--Leonidov--Kovner} equation
  with classical and quantum source terms, Nuclear Physics A 775~(1-2) (2006)
  69--88.

\bibitem{ikeda1965}
N.~Ikeda, M.~Nagasawa, S.~Watanabe, On branching \mbox{Markov} processes,
  Proceedings of the Japan Academy 41~(9) (1965) 816--821.

\bibitem{ikeda1968}
N.~Ikeda, M.~Nagasawa, S.~Watanabe, Branching \mbox{Markov} processes \mbox{I},
  Journal of Mathematics of Kyoto University 8~(2) (1968) 233--278.

\bibitem{mckean1975}
H.~P. McKean, Application of \mbox{Brownian} motion to the equation of
  \mbox{Kolmogorov-Petrovskii-Piskunov}, Communications on Pure and Applied
  Mathematics 28~(3) (1975) 323--331.

\bibitem{binney1987}
J.~Binney, S.~Tremaine, Galactic Dynamics, , Princeton University Press,
  Princeton, New Jersey, 1987.

\bibitem{spergel2000}
D.~N. Spergel, P.~J. Steinhardt, Observational evidence for self-interacting
  cold dark matter, Physical Review Letters 84~(17) (2000) 3760.

\bibitem{bernoff2016}
A.~J. Bernoff, C.~M. Topaz, Biological aggregation driven by social and
  environmental factors: A nonlocal model and its degenerate
  \mbox{Cahn--Hilliard} approximation, SIAM Journal on Applied Dynamical
  Systems 15~(3) (2016) 1528--1562.

\bibitem{bertozzi2007}
A.~L. Bertozzi, T.~Laurent, Finite-time blow-up of solutions of an aggregation
  equation in \mbox{$\mathbb{R}^n$}, Communications in Mathematical Physics
  274~(3) (2007) 717--735.

\bibitem{bertozzi2012}
A.~L. Bertozzi, J.~B. Garnett, T.~Laurent, Characterization of radially
  symmetric finite time blowup in multidimensional aggregation equations, SIAM
  Journal on Mathematical Analysis 44~(2) (2012) 651--681.

\bibitem{bernoff2011}
A.~J. Bernoff, C.~M. Topaz, A primer of swarm equilibria, SIAM Journal on
  Applied Dynamical Systems 10~(1) (2011) 212--250.

\bibitem{carrillo2014}
J.~A. Carrillo, M.~Chipot, Y.~Huang, On global minimizers of
  repulsive--attractive power-law interaction energies, Philosophical
  Transactions of the Royal Society A: Mathematical, Physical and Engineering
  Sciences 372~(2028) (2014) 20130399.

\bibitem{simione2015}
R.~Simione, D.~Slep{\v{c}}ev, I.~Topaloglu, Existence of ground states of
  nonlocal-interaction energies, Journal of Statistical Physics 159~(4) (2015)
  972--986.

\bibitem{bedrossian2011}
J.~Bedrossian, Global minimizers for free energies of subcritical aggregation
  equations with degenerate diffusion, Applied Mathematics Letters 24~(11)
  (2011) 1927--1932.

\bibitem{burger2014}
M.~Burger, R.~Fetecau, Y.~Huang, Stationary states and asymptotic behavior of
  aggregation models with nonlinear local repulsion, SIAM Journal on Applied
  Dynamical Systems 13~(1) (2014) 397--424.

\bibitem{gilboa2009}
G.~Gilboa, S.~Osher, Nonlocal operators with applications to image processing,
  Multiscale Modeling \& Simulation 7~(3) (2009) 1005--1028.

\bibitem{kavallaris2018}
N.~I. Kavallaris, T.~Suzuki, Non-local partial differential equations for
  engineering and biology, Mathematical Modeling and Analysis (Mathematics for
  Industry 31 (2018).

\bibitem{kavallaris2002}
N.~I. Kavallaris, D.~Tzanetis, Blow-up and stability of a nonlocal
  diffusion-convection problem arising in \mbox{Ohmic} heating of foods,
  Differential and Integral Equations 15~(3) (2002) 271--288.

\bibitem{kavallaris2004}
N.~I. Kavallaris, D.~E. TZANETIS, Behaviour of a non-local reactive-convective
  problem with variable velocity in ohmic heating of food, Nonlocal Elliptic
  and Parabolic Problems (2004) 189--198.

\bibitem{kavallaris2004a}
N.~I. Kavallaris, Asymptotic behaviour and blow-up for a nonlinear diffusion
  problem with a non-local source term, Proceedings of the Edinburgh
  Mathematical Society 47~(2) (2004) 375--395.

\bibitem{kavallaris2007}
N.~I. Kavallaris, T.~Nadzieja, On the blow-up of the non-local thermistor
  problem, Proceedings of the Edinburgh Mathematical Society 50~(2) (2007)
  389--409.

\bibitem{romano2018}
G.~Romano, R.~Luciano, R.~Barretta, M.~Diaco, Nonlocal integral elasticity in
  nanostructures, mixtures, boundary effects and limit behaviours, Continuum
  Mechanics and Thermodynamics 30 (2018) 641--655.

\bibitem{barretta2023}
R.~Barretta, F.~Marotti~de Sciarra, M.~S. Vaccaro, Nonlocal elasticity for
  nanostructures: a review of recent achievements, Encyclopedia 3~(1) (2023)
  279--310.

\bibitem{varghese2015}
J.~Z. Varghese, R.~G. Boone, et~al., Overview of autonomous vehicle sensors and
  systems, in: International Conference on Operations Excellence and Service
  Engineering, sn, 2015, pp. 178--191.

\bibitem{wu2021}
Y.~Wu, Y.~Ma, H.~Zheng, S.~Ramakrishna, Piezoelectric materials for flexible
  and wearable electronics: A review, Materials \& Design 211 (2021) 110164.

\bibitem{thaheld2003}
F.~Thaheld, Biological nonlocality and the mind--brain interaction problem:
  comments on a new empirical approach, BioSystems 70~(1) (2003) 35--41.

\bibitem{de2009}
F.~M. de~Sciarra, A nonlocal model with strain-based damage, International
  Journal of Solids and Structures 46~(22-23) (2009) 4107--4122.

\bibitem{kavallaris2015}
N.~I. Kavallaris, Y.~Yan, A time discretization scheme for a nonlocal
  degenerate problem modelling resistance spot welding, Mathematical Modelling
  of Natural Phenomena 10~(6) (2015) 90--112.

\bibitem{melaibari2022}
A.~Melaibari, A.~A. Abdelrahman, M.~A. Hamed, A.~W. Abdalla, M.~A. Eltaher,
  Dynamic analysis of a piezoelectrically layered perforated nonlocal strain
  gradient nanobeam with flexoelectricity, Mathematics 10~(15) (2022) 2614.

\bibitem{zhang2005a}
X.~Zhang, P.~Sharma, Size dependency of strain in arbitrary shaped anisotropic
  embedded quantum dots due to nonlocal dispersive effects, Physical Review B
  72~(19) (2005) 195345.

\bibitem{yang2006}
J.~Yang, A review of a few topics in piezoelectricity, Appl. Mech. Rev. 59
  (2006) 335--345.

\bibitem{chen2019}
Z.~Chen, S.~Niazi, F.~Bobaru, A peridynamic model for brittle damage and
  fracture in porous materials, International Journal of Rock Mechanics and
  Mining Sciences 122 (2019) 104059.

\bibitem{gillies2011}
T.~E. Gillies, C.~Cabernard, Cell division orientation in animals, Current
  Biology 21~(15) (2011) R599--R609.

\bibitem{sellitto2012}
A.~Sellitto, F.~Alvarez, D.~Jou, Geometrical dependence of thermal conductivity
  in elliptical and rectangular nanowires, International Journal of Heat and
  Mass Transfer 55~(11-12) (2012) 3114--3120.

\bibitem{sobolev1994}
S.~Sobolev, Equations of transfer in non-local media, International Journal of
  Heat and Mass Transfer 37~(14) (1994) 2175--2182.

\bibitem{sobolev2012}
S.~Sobolev, Rapid colloidal solidifications under local nonequilibrium
  diffusion conditions, Physics Letters A 376~(47-48) (2012) 3563--3566.

\bibitem{sobolev1998}
S.~Sobolev, Y.~M. Mikhailov, Diffusion of low-molecular-mass substances in
  glassy polymers described on the basis of extended nonequilibrium
  thermodynamics, Polymer Science. Series B 40~(3-4) (1998) 108--111.

\bibitem{kavallaris2007a}
N.~I. Kavallaris, A.~Lacey, C.~Nikolopoulos, C.~Voong, Behaviour of a non-local
  equation modelling linear friction welding, IMA journal of applied
  mathematics 72~(5) (2007) 597--616.

\bibitem{mcmahon2009}
J.~M. McMahon, S.~K. Gray, G.~C. Schatz, Nonlocal optical response of metal
  nanostructures with arbitrary shape, Physical review letters 103~(9) (2009)
  097403.

\bibitem{kavallaris2015a}
N.~I. Kavallaris, Explosive solutions of a stochastic non-local
  reaction--diffusion equation arising in shear band formation, Mathematical
  Methods in the Applied Sciences 38~(16) (2015) 3564--3574.

\bibitem{abouelregal2022}
A.~E. Abouelregal, M.~Alesemi, Fractional moore-gibson-thompson heat transfer
  model with nonlocal and nonsingular kernels of a rotating viscoelastic
  annular cylinder with changeable thermal properties, Plos one 17~(6) (2022)
  e0269862.

\bibitem{abouelregal2025}
A.~E. Abouelregal, M.~Marin, A.~{\"O}chsner, A modified spatiotemporal nonlocal
  thermoelasticity theory with higher-order phase delays for a viscoelastic
  micropolar medium exposed to short-pulse laser excitation, Continuum
  Mechanics and Thermodynamics 37~(1) (2025) 15.

\bibitem{sobolev1993}
S.~Sobolev, Two-temperature discrete model for nonlocal heat conduction,
  Journal de Physique III 3~(12) (1993) 2261--2269.

\bibitem{sobolev1997}
S.~L. Sobolev, Local non-equilibrium transport models, Physics-Uspekhi 40~(10)
  (1997) 1043.

\bibitem{sobolev2014}
S.~Sobolev, Nonlocal diffusion models: Application to rapid solidification of
  binary mixtures, International Journal of Heat and Mass Transfer 71 (2014)
  295--302.

\bibitem{jou1996}
D.~Jou, J.~Casas-V{\'a}zquez, G.~Lebon, D.~Jou, J.~Casas-V{\'a}zquez, G.~Lebon,
  Extended irreversible thermodynamics, Springer, 1996.

\bibitem{joseph1989}
D.~D. Joseph, L.~Preziosi, Heat waves, Reviews of modern physics 61~(1) (1989)
  41.

\bibitem{millennium2005}
M.~Millennium~ecosystem assessment, Ecosystems and human well-being, Vol.~5,
  Island press Washington, DC, 2005.

\bibitem{marchello2024}
R.~Marchello, A.~Colombi, L.~Preziosi, C.~Giverso, A non local model for cell
  migration in response to mechanical stimuli, Mathematical Biosciences 368
  (2024) 109124.

\bibitem{colombi2021}
A.~Colombi, S.~Falletta, M.~Scianna, L.~Scuderi, An integro-differential
  non-local model for cell migration and its efficient numerical solution,
  Mathematics and Computers in Simulation 180 (2021) 179--204.

\bibitem{dankulov2019}
M.~M. Dankulov, B.~Tadi{\'c}, R.~Melnik, Spectral properties of hyperbolic
  nanonetworks with tunable aggregation of simplexes, Physical Review E 100~(1)
  (2019) 012309.

\bibitem{paliy2009}
M.~Paliy, R.~Melnik, B.~A. Shapiro, Molecular dynamics study of the \mbox{RNA}
  ring nanostructure: a phenomenon of self-stabilization, Physical Biology
  6~(4) (2009) 046003.

\bibitem{paliy2010}
M.~Paliy, R.~Melnik, B.~A. Shapiro, Coarse-graining \mbox{RNA} nanostructures
  for molecular dynamics simulations, Physical Biology 7~(3) (2010) 036001.

\bibitem{badu2014}
S.~Badu, R.~Melnik, M.~Paliy, S.~Prabhakar, A.~Sebetci, B.~Shapiro, Modeling of
  \mbox{RNA} nanotubes using molecular dynamics simulation, European Biophysics
  Journal 43 (2014) 555--564.

\bibitem{badu2020}
S.~Badu, R.~Melnik, S.~Singh, Mathematical and computational models of rna
  nanoclusters and their applications in data-driven environments, Molecular
  Simulation 46~(14) (2020) 1094--1115.

\bibitem{badu2020a}
S.~Badu, S.~Prabhakar, R.~Melnik, S.~Singh, Atomistic to continuum model for
  studying mechanical properties of \mbox{RNA} nanotubes, Computer Methods in
  Biomechanics and Biomedical Engineering 23~(8) (2020) 396--407.

\bibitem{badu2020b}
S.~Badu, S.~Prabhakar, R.~Melnik, Coarse-grained models of \mbox{RNA} nanotubes
  for large time scale studies in biomedical applications, Biomedicines 8~(7)
  (2020) 195.

\bibitem{yang2007c}
J.~X.-D. Yang, R.~V. Melnik, A new constitutive model for the analysis of
  semi-flexible polymers with internal viscosity, in: Computational
  Science--ICCS 2007: 7th International Conference, Beijing, China, May 27-30,
  2007, Proceedings, Part I 7, Springer, 2007, pp. 834--841.

\bibitem{yang2008}
X.-D. Yang, R.~Melnik, Dynamic analysis of polymeric fluid in shear flow for
  dumbbell model with internal viscosity, Journal of Central South University
  of Technology 15~(Suppl 1) (2008) 17--20.

\bibitem{melnik2002}
R.~Melnik, Mathematical and computer modelling of coupled reaction kinetics and
  heat transfer in processing polymeric materials, Modelling \& Simulation in
  Materials Science and Engineering 10 (2002) 341--358.

\bibitem{melnik2003d}
R.~V. Melnik, Computationally efficient algorithms for modelling thermal
  degradation and spiking phenomena in polymeric materials, Computers \&
  Chemical Engineering 27~(10) (2003) 1473--1484.

\bibitem{melnik2003e}
R.~Melnik, A.~Uhlherr, J.~Hodgkin, F.~De~Hoog, Distance geometry algorithms in
  molecular modelling of polymer and composite systems, Computers \&
  Mathematics with Applications 45~(1-3) (2003) 515--534.

\bibitem{nicolis1971}
G.~Nicolis, I.~Prigogine, Fluctuations in nonequilibrium systems, Proceedings
  of the National Academy of Sciences 68~(9) (1971) 2102--2107.

\bibitem{zhabotinsky2007}
A.~M. Zhabotinsky, \mbox{Belousov-Zhabotinsky} reaction, Scholarpedia 2~(9)
  (2007) 1435.

\bibitem{hasty2000}
J.~Hasty, J.~Pradines, M.~Dolnik, J.~J. Collins, Noise-based switches and
  amplifiers for gene expression, Proceedings of the National Academy of
  Sciences 97~(5) (2000) 2075--2080.

\bibitem{mee2012}
M.~T. Mee, H.~H. Wang, Engineering ecosystems and synthetic ecologies,
  Molecular BioSystems 8~(10) (2012) 2470--2483.

\bibitem{schenk2022}
C.~Schenk, V.~H. Schulz, Existence, uniqueness, and numerical modeling of wine
  fermentation based on integro-differential equations, SIAM Journal on Applied
  Mathematics 82~(4) (2022) 1220--1245.

\bibitem{oguz2013}
C.~Oguz, T.~Laomettachit, K.~C. Chen, L.~T. Watson, W.~T. Baumann, J.~J. Tyson,
  Optimization and model reduction in the high dimensional parameter space of a
  budding yeast cell cycle model, BMC Systems Biology 7 (2013) 1--17.

\bibitem{zhou2022}
L.~Zhou, S.~Zhu, Z.~Zhu, S.~Yu, X.~Xie, Improved peridynamic model and its
  application to crack propagation in rocks, Royal Society Open Science 9~(10)
  (2022) 221013.

\bibitem{chu2018}
D.~Chu, M.~Prokopenko, J.~C.~J. Ray, Computation by natural systems (2018).

\bibitem{thieu2022}
T.~K.~T. Thieu, R.~Melnik, Effects of random inputs and short-term synaptic
  plasticity in a lif conductance model for working memory applications, in:
  International Work-Conference on Bioinformatics and Biomedical Engineering,
  Springer, 2022, pp. 59--72.

\bibitem{thieu2022a}
T.~K.~T. Thieu, R.~Melnik, Effects of noise on leaky integrate-and-fire neuron
  models for neuromorphic computing applications, in: International Conference
  on Computational Science and Its Applications, Springer, 2022, pp. 3--18.

\bibitem{ying2022}
Z.~Ying, H.~Zhang, L.~Wang, R.~Melnik, Propulsion optimization of a
  jellyfish-inspired robot based on a nonintrusive reduced-order model with
  proper orthogonal decomposition, Bioinspiration \& Biomimetics 17~(4) (2022)
  046005.

\bibitem{ying2022a}
Z.~Ying, L.~Wang, R.~Melnik, Parameter optimization of the bio-inspired robot
  propulsion through the deep learning based reduced order fluid-structure
  interaction model, Ocean Engineering 255 (2022) 111436.

\bibitem{ying2023}
Z.~Ying, H.~Zhang, L.~Wang, R.~Melnik, A two-dimensional hydrodynamics
  prediction framework for mantle-undulated propulsion robot using multiple
  proper orthogonal decomposition and long short term memory neural network,
  Bioinspiration \& Biomimetics 19~(1) (2023) 016005.

\bibitem{mahapatra2006}
D.~R. Mahapatra, R.~Melnik, Modelling and analysis of collagen piezoelectricity
  in human cornea, Dynamics of Continuous, Discrete and Impulsive Systems
  Series A: Mathematical Analysis 13 (2006) 377--384.

\bibitem{singh2022}
S.~Singh, R.~Melnik, Auxeticity in biosystems: an exemplification of its
  effects on the mechanobiology of heterogeneous living cells, Computer Methods
  in Biomechanics and Biomedical Engineering 25~(5) (2022) 521--535.

\bibitem{melnik1998}
R.~Melnik, K.~Melnik, A note on the class of weakly coupled problems of
  non-stationary piezoelectricity, Communications in numerical methods in
  engineering 14~(9) (1998) 839--847.

\bibitem{melnik2000a}
R.~Melnik, K.~Melnik, Modelling dynamics of piezoelectric solids in the
  two-dimensional case, Applied Mathematical Modelling 24~(3) (2000) 147--163.

\bibitem{melnik2000b}
R.~V. Melnik, Generalised solutions, discrete models and energy estimates for a
  \mbox{2D} problem of coupled field theory, Applied Mathematics and
  Computation 107~(1) (2000) 27--55.

\bibitem{melnik2001}
R.~V. Melnik, Computational analysis of coupled physical fields in
  piezothermoelastic media, Computer Physics Communications 142~(1-3) (2001)
  231--237.

\bibitem{melnik2003}
R.~V. Melnik, Numerical analysis of dynamic characteristics of coupled
  piezoelectric systems in acoustic media, Mathematics and Computers in
  Simulation 61~(3-6) (2003) 497--507.

\bibitem{melnik2003a}
R.~Melnik, Modelling coupled dynamics: Piezoelectric elements under changing
  temperature conditions, International Communications in Heat and Mass
  Transfer 30~(1) (2003) 83--92.

\bibitem{melnik2000f}
R.~Melnik, A.~Roberts, K.~Thomas, Computing dynamics of copper-based \mbox{SMA}
  via centre manifold reduction of \mbox{3D} models, Computational Materials
  Science 18~(3-4) (2000) 255--268.

\bibitem{melnik2004}
R.~V. Melnik, A.~H. Roberts, Computational models for multi-scale coupled
  dynamic problems, Future Generation Computer Systems 20~(3) (2004) 453--464.

\bibitem{matus2004}
P.~Matus, R.~V. Melnik, L.~Wang, I.~Rybak, Applications of fully conservative
  schemes in nonlinear thermoelasticity: modelling shape memory materials,
  Mathematics and Computers in Simulation 65~(4-5) (2004) 489--509.

\bibitem{wang2004}
L.~Wang, R.~V. Melnik, Dynamics of shape memory alloys patches, Materials
  Science and Engineering: A 378~(1-2) (2004) 470--474.

\bibitem{melnik2004a}
R.~Melnik, K.~Zotsenko, Mixed electroelastic waves and \mbox{CFL} stability
  conditions in computational piezoelectricity, Applied Numerical Mathematics
  48~(1) (2004) 41--62.

\bibitem{wang2006c}
L.~Wang, R.~V. Melnik, Mechanically induced phase combination in shape memory
  alloys by chebyshev collocation methods, Materials Science and Engineering: A
  438 (2006) 427--430.

\bibitem{wang2006d}
L.~Wang, R.~Melnik, Dynamics of \mbox{SMA} patches with mechanically induced
  transformations, Discrete and Continuous Dynamical Systems 15~(4) (2006)
  1237--1252.

\bibitem{wang2007}
L.~X. Wang, R.~V. Melnik, Numerical model for vibration damping resulting from
  the first-order phase transformations, Applied Mathematical Modelling 31~(9)
  (2007) 2008--2018.

\bibitem{wang2007a}
L.~X. Wang, R.~V. Melnik, Model reduction applied to square to rectangular
  martensitic transformations using proper orthogonal decomposition, Applied
  Numerical Mathematics 57~(5-7) (2007) 510--520.

\bibitem{mahapatra2007a}
D.~R. Mahapatra, R.~V. Melnik, Finite element approach to modelling evolution
  of \mbox{3D} shape memory materials, Mathematics and Computers in Simulation
  76~(1-3) (2007) 141--148.

\bibitem{wang2007c}
L.~X. Wang, R.~V. Melnik, Finite volume analysis of nonlinear thermo-mechanical
  dynamics of shape memory alloys, Heat and Mass Transfer 43 (2007) 535--546.

\bibitem{xiao2008}
Y.~Xiao, K.~Bhattacharya, A continuum theory of deformable, semiconducting
  ferroelectrics, Archive for Rational Mechanics and Analysis 189 (2008)
  59--95.

\bibitem{wang2009}
L.~Wang, R.~V. Melnik, Control of coupled hysteretic dynamics of ferroelectric
  materials with a \mbox{Landau-type} differential model and feedback
  linearization, Smart Materials and Structures 18~(7) (2009) 074011.

\bibitem{wang2010}
L.~X. Wang, R.~V. Melnik, Low dimensional approximations to ferroelastic
  dynamics and hysteretic behavior due to phase transformations, Journal of
  Applied Mechanics 77~(3) (2010) 031015.

\bibitem{wang2011}
L.~Wang, R.~Melnik, F.~Lv, Stress induced polarization switching and coupled
  hysteretic dynamics in ferroelectric materials, Frontiers of Mechanical
  Engineering 6 (2011) 287--291.

\bibitem{he2018}
X.~He, D.~Wang, L.~Wang, R.~Melnik, Modelling of creep hysteresis in
  ferroelectrics, Philosophical Magazine 98~(14) (2018) 1256--1271.

\bibitem{wang2016}
D.~Wang, L.~Wang, R.~Melnik, A differential algebraic approach for the modeling
  of polycrystalline ferromagnetic hysteresis with minor loops and frequency
  dependence, Journal of Magnetism and Magnetic Materials 410 (2016) 144--149.

\bibitem{he2020}
X.~He, H.~Du, D.~Wang, L.~Wang, R.~Melnik, Modelling ageing phenomenon in
  ferroelectrics via a landau-type phenomenological model, Smart Materials and
  Structures 30~(1) (2020) 015017.

\bibitem{du2020}
Q.~Du, B.~Engquist, X.~Tian, Multiscale modeling, homogenization and nonlocal
  effects: Mathematical and computational issues, Contemporary Mathematics 754
  (2020) 115--140.

\bibitem{du2020b}
H.~Du, X.~He, L.~Wang, R.~Melnik, Analysis of shape memory alloy vibrator using
  harmonic balance method, Applied Physics A 126 (2020) 568.

\bibitem{he2020a}
X.~He, H.~Du, Z.~Tong, D.~Wang, L.~Wang, R.~Melnik, A dynamic hysteresis model
  based on landau phenomenological theory of fatigue phenomenon in
  ferroelectrics, Materials Today Communications 25 (2020) 101479.

\bibitem{wang2017b}
D.~Wang, L.~Wang, R.~Melnik, A hysteresis model for ferroelectric ceramics with
  mechanism for minor loops, Physics Letters A 381~(4) (2017) 344--350.

\bibitem{melnik2001a}
R.~Melnik, A.~Roberts, K.~Thomas, Coupled thermomechanical dynamics of phase
  transitions in shape memory alloys and related hysteresis phenomena,
  Mechanics Research Communications 28~(6) (2001) 637--651.

\bibitem{melnik2003b}
R.~V. Melnik, A.~Roberts, Modelling nonlinear dynamics of shape-memory-alloys
  with approximate models of coupled thermoelasticity, ZAMM-Journal of Applied
  Mathematics and Mechanics/Zeitschrift f{\"u}r Angewandte Mathematik und
  Mechanik: Applied Mathematics and Mechanics 83~(2) (2003) 93--104.

\bibitem{mahapatra2006a}
D.~R. Mahapatra, R.~Melnik, Finite element analysis of phase transformation
  dynamics in shape memory alloys with a consistent landau-ginzburg free energy
  model, Mechanics of Advanced Materials and Structures 13~(6) (2006) 443--455.

\bibitem{wang2006}
L.-x. Wang, R.~V. Melnik, Differential-algebraic approach to coupled problems
  of dynamic thermoelasticity, Applied Mathematics and Mechanics 27 (2006)
  1185--1196.

\bibitem{wang2007b}
L.~X. Wang, R.~V. Melnik, Thermo-mechanical wave propagations in shape memory
  alloy rod with phase transformations, Mechanics of Advanced Materials and
  Structures 14~(8) (2007) 665--676.

\bibitem{mahapatra2007}
D.~R. Mahapatra, R.~Melnik, Finite element modelling and simulation of phase
  transformations in shape memory alloy thin films, International Journal for
  Multiscale Computational Engineering 5~(1) (2007).

\bibitem{wang2008}
L.~Wang, R.~Melnik, Modifying macroscale variant combinations in a
  two-dimensional structure using mechanical loadings during thermally induced
  transformation, Materials Science and Engineering: A 481 (2008) 190--193.

\bibitem{wang2008a}
L.~Wang, R.~V. Melnik, Simulation of phase combinations in shape memory alloys
  patches by hybrid optimization methods, Applied Numerical Mathematics 58~(4)
  (2008) 511--524.

\bibitem{dhote2012}
R.~Dhote, R.~Melnik, J.~Zu, Dynamic thermo-mechanical coupling and size effects
  in finite shape memory alloy nanostructures, Computational Materials Science
  63 (2012) 105--117.

\bibitem{wang2012}
L.~Wang, R.~V. Melnik, Nonlinear dynamics of shape memory alloy oscillators in
  tuning structural vibration frequencies, Mechatronics 22~(8) (2012)
  1085--1096.

\bibitem{dhote2013}
R.~P. Dhote, M.~Fabrizio, R.~Melnik, J.~Zu, Hysteresis phenomena in shape
  memory alloys by non-isothermal \mbox{Ginzburg--Landau} models,
  Communications in Nonlinear Science and Numerical Simulation 18~(9) (2013)
  2549--2561.

\bibitem{dhote2015}
R.~Dhote, H.~Gomez, R.~Melnik, J.~Zu, 3d coupled thermo-mechanical phase-field
  modeling of shape memory alloy dynamics via isogeometric analysis, Computers
  \& Structures 154 (2015) 48--58.

\bibitem{dhote2015a}
R.~P. Dhote, H.~G{\'o}mez, R.~N. Melnik, J.~Zu, Shape memory alloy
  nanostructures with coupled dynamic thermo-mechanical effects, Computer
  Physics Communications 192 (2015) 48--53.

\bibitem{dhote2016}
R.~Dhote, H.~Gomez, R.~Melnik, J.~Zu, et~al., Effect of aspect ratio and
  boundary conditions in modeling shape memory alloy nanostructures with 3d
  coupled dynamic phase-field theories, Mathematical Problems in Engineering
  2016 (2016).

\bibitem{du2021}
H.~Du, Y.~Han, L.~Wang, R.~Melnik, A differential model for the hysteresis in
  magnetic shape memory alloys and its application of feedback linearization,
  Applied Physics A 127~(6) (2021) 432.

\bibitem{han2022}
Y.~Han, H.~Du, L.~Wang, R.~Melnik, A phenomenological model for
  thermally-induced hysteresis in polycrystalline shape memory alloys with
  internal loops, Journal of Intelligent Material Systems and Structures 33~(9)
  (2022) 1170--1181.

\bibitem{han2022a}
Y.~Han, L.~Wang, R.~Melnik, Phenomenological modeling for magneto-mechanical
  couplings of martensitic variant reorientation in ferromagnetic shape memory
  alloys, Applied Physics A 128~(12) (2022) 1066.

\bibitem{fu2019}
J.~Fu, J.~Liu, H.~Tian, Y.~Li, Y.~Bao, Z.~Fang, H.~Lu, Dual attention network
  for scene segmentation, in: Proceedings of the IEEE/CVF conference on
  computer vision and pattern recognition, 2019, pp. 3146--3154.

\bibitem{conte2023}
M.~Conte, N.~Loy, A non-local kinetic model for cell migration: a study of the
  interplay between contact guidance and steric hindrance, SIAM Journal on
  Applied Mathematics (2023) S429--S451.

\bibitem{metzler1994}
R.~Metzler, W.~G. Gl{\"o}ckle, T.~F. Nonnenmacher, Fractional model equation
  for anomalous diffusion, Physica A: Statistical Mechanics and its
  Applications 211~(1) (1994) 13--24.

\bibitem{fallahpour2022}
R.~Fallahpour, R.~Melnik, A molecular dynamics study of nanowire resonator
  bio-object detection, Journal of Mechanics in Medicine and Biology 22~(01)
  (2022) 2250003.

\bibitem{mahapatra2012}
D.~R. Mahapatra, M.~Willatzen, R.~Melnik, B.~Lassen, Modeling heterostructures
  with schr{\"o}dinger--poisson--navier iterative schemes, effect of carrier
  charge, and influence of electromechanical coupling, Nano 7~(04) (2012)
  1250031.

\bibitem{bahrami2010}
M.~Bahrami-Samani, S.~R. Patil, R.~Melnik, Higher-order nonlinear
  electromechanical effects in wurtzite \mbox{GaN/AlN} quantum dots, Journal of
  Physics: Condensed Matter 22~(49) (2010) 495301.

\bibitem{prabhakar2010}
S.~Prabhakar, R.~Melnik, Influence of electromechanical effects and wetting
  layers on band structures of \mbox{AlN/GaN} quantum dots and spin control,
  Journal of Applied Physics 108~(6) (2010).

\bibitem{patil2009}
S.~R. Patil, R.~V. Melnik, Thermoelectromechanical effects in quantum dots,
  Nanotechnology 20~(12) (2009) 125402.

\bibitem{melnik2007}
R.~Melnik, R.~Mahapatra, Coupled effects in quantum dot nanostructures with
  nonlinear strain and bridging modelling scales, Computers \& Structures
  85~(11-14) (2007) 698--711.

\bibitem{badu2020c}
S.~Badu, R.~Melnik, S.~Singh, Analysis of photosynthetic systems and their
  applications with mathematical and computational models, Applied Sciences
  10~(19) (2020) 6821.

\bibitem{badu2020d}
S.~Badu, S.~Prabhakar, R.~Melnik, Component spectroscopic properties of
  light-harvesting complexes with \mbox{DFT} calculations, Biocell 44~(3)
  (2020) 279.

\bibitem{badu2015}
S.~Badu, R.~Melnik, S.~Prabhakar, \mbox{RNA} nanostructures in physiological
  solutions: Multiscale modeling and applications, Physics of Liquid Matter:
  Modern Problems: Proceedings, Kyiv, Ukraine, 23-27 May 2014 (2015) 337--355.

\bibitem{badu2015a}
S.~R. Badu, R.~Melnik, S.~Prabhakar, Transport properties of \mbox{RNA}
  nanotubes using molecular dynamics simulation, in: Bioinformatics and
  Biomedical Engineering: Third International Conference, IWBBIO 2015, Granada,
  Spain, April 15-17, 2015. Proceedings, Part II 3, Springer, 2015, pp.
  578--583.

\bibitem{badu2015b}
S.~Badu, R.~Melnik, S.~Prabhakar, Studying properties of \mbox{RNA} nanotubes
  via molecular dynamics, in: Nanosensors, Biosensors, and Info-Tech Sensors
  and Systems 2015, Vol. 9434, SPIE, 2015, pp. 66--71.

\bibitem{melnik2002b}
R.~Melnik, D.~R. Jenkins, On computational control of flow in airblast
  atomisers for pulmonary drug delivery, International Journal of Pharmaceutics
  239~(1-2) (2002) 23--35.

\bibitem{yang2007b}
X.-D. Yang, R.~V. Melnik, Effect of internal viscosity on brownian dynamics of
  \mbox{DNA} molecules in shear flow, Computational Biology and Chemistry
  31~(2) (2007) 110--114.

\bibitem{sinha2007}
N.~Sinha, D.~R. Mahapatra, J.~Yeow, R.~Melnik, D.~Jaffray, Carbon nanotube thin
  film field emitting diode: Understanding the system response based on
  multiphysics modeling, Journal of Computational and Theoretical Nanoscience
  4~(3) (2007) 535--549.

\bibitem{sinha2007a}
N.~Sinha, D.~R. Mahapatra, Y.~Sun, J.~Yeow, R.~Melnik, D.~Jaffray,
  Electromechanical interactions in a carbon nanotube based thin film field
  emitting diode, Nanotechnology 19~(2) (2007) 025701.

\bibitem{mahapatra2008}
D.~R. Mahapatra, N.~Sinha, J.~Yeow, R.~Melnik, Field emission from strained
  carbon nanotubes on cathode substrate, Applied Surface Science 255~(5) (2008)
  1959--1966.

\bibitem{roy2009}
D.~Roy~Mahapatra, S.~Anand, N.~Sinha, R.~Melnik, Enhancing field emission from
  a carbon nanotube array by lateral control of electrodynamic force field,
  Molecular Simulation 35~(6) (2009) 512--519.

\bibitem{mahapatra2011}
D.~R. Mahapatra, N.~Sinha, R.~Melnik, Degradation and failure of field emitting
  carbon nanotube arrays, Journal of Nanoscience and Nanotechnology 11~(5)
  (2011) 3911--3915.

\bibitem{yavarian2023}
M.~Yavarian, R.~Melnik, Z.~L. Mi{\v{s}}kovi{\'c}, Modeling of charging dynamics
  in electrochemical systems with a graphene electrode, Journal of
  Electroanalytical Chemistry 946 (2023) 117711.

\bibitem{singh2022a}
S.~Singh, R.~Melnik, Coupled multiphysics modelling of sensors for chemical,
  biomedical, and environmental applications with focus on smart materials and
  low-dimensional nanostructures, Chemosensors 10~(5) (2022) 157.

\bibitem{fallahpour2021}
R.~Fallahpour, R.~Melnik, Nonlinear vibration analysis of nanowire resonators
  for ultra-high resolution mass sensing, Measurement 175 (2021) 109136.

\bibitem{lew2004}
L.~Lew Yan~Voon, B.~Lassen, R.~Melnik, M.~Willatzen, Prediction of barrier
  localization in modulated nanowires, Journal of Applied Physics 96~(8) (2004)
  4660--4662.

\bibitem{lassen2004}
B.~Lassen, L.~L.~Y. Voon, M.~Willatzen, R.~Melnik, Exact envelope-function
  theory versus symmetrized \mbox{Hamiltonian} for quantum wires: a comparison,
  Solid State Communications 132~(3-4) (2004) 141--149.

\bibitem{willatzen2004}
M.~Willatzen, R.~V. Melnik, C.~Galeriu, L.~L.~Y. Voon, Quantum confinement
  phenomena in nanowire superlattice structures, Mathematics and Computers in
  Simulation 65~(4-5) (2004) 385--397.

\bibitem{galeriu2004}
C.~Galeriu, L.~L.~Y. Voon, R.~Melnik, M.~Willatzen, Modeling a nanowire
  superlattice using the finite difference method in cylindrical polar
  coordinates, Computer Physics Communications 157~(2) (2004) 147--159.

\bibitem{radulovic2006}
N.~Radulovic, M.~Willatzen, R.~V. Melnik, L.~C. Lew Yan~Voon, Influence of the
  metal contact size on the electron dynamics and transport inside the
  semiconductor heterostructure nanowire, Journal of Computational and
  Theoretical Nanoscience 3~(4) (2006) 551--559.

\bibitem{lassen2006a}
B.~Lassen, M.~Willatzen, R.~Melnik, L.~L.~Y. Voon, Electronic structure of
  free-standing \mbox{InP and InAs} nanowires, Journal of Materials Research
  21~(11) (2006) 2927--2935.

\bibitem{patil2009a}
S.~R. Patil, R.~Melnik, Thermopiezoelectric effects on optoelectronic
  properties of \mbox{CdTe/ZnTe} quantum wires, Physica Status Solidi (a)
  206~(5) (2009) 960--964.

\bibitem{patil2009b}
S.~R. Patil, R.~V. Melnik, Coupled electromechanical effects in \mbox{II--VI}
  group finite length semiconductor nanowires, Journal of Physics D: Applied
  Physics 42~(14) (2009) 145113.

\bibitem{alvaro2013}
M.~Alvaro, L.~Bonilla, M.~Carretero, R.~Melnik, S.~Prabhakar, Transport in
  semiconductor nanowire superlattices described by coupled quantum mechanical
  and kinetic models, Journal of Physics: Condensed Matter 25~(33) (2013)
  335301.

\bibitem{prabhakar2013}
S.~Prabhakar, R.~Melnik, L.~L. Bonilla, Coupled multiphysics, barrier
  localization, and critical radius effects in embedded nanowire superlattices,
  Journal of Applied Physics 113~(24) (2013).

\bibitem{melnik2003f}
R.~Melnik, M.~Willatzen, Bandstructures of conical quantum dots with wetting
  layers, Nanotechnology 15~(1) (2003) 1.

\bibitem{melnik2005}
R.~Melnik, D.~Strunin, A.~Roberts, Nonlinear analysis of rubber-based polymeric
  materials with thermal relaxation models, Numerical Heat Transfer, Part A:
  Applications 47~(6) (2005) 549--569.

\bibitem{lew2005}
L.~Lew Yan~Voon, C.~Galeriu, B.~Lassen, M.~Willatzen, R.~Melnik, Electronic
  structure of wurtzite quantum dots with cylindrical symmetry, Applied Physics
  Letters 87~(4) (2005).

\bibitem{prabhakar2012}
S.~Prabhakar, E.~Takhtamirov, R.~Melnik, Coupled multiphysics models for the
  analysis of the conduction and valence band eigenenergies in cylindrical
  quantum dots, Acta Physica Polonica A 121~(1) (2012) 85--88.

\bibitem{prabhakar2012a}
S.~Prabhakar, R.~V. Melnik, P.~Neittaanm{\"a}ki, T.~Tiihonen, Coupled
  electromechanical effects in wurtzite quantum dots with wetting layers in
  gate controlled electric fields: The multiband case, Physica E:
  Low-dimensional Systems and Nanostructures 46 (2012) 97--104.

\bibitem{prabhakar2013a}
S.~Prabhakar, R.~V. Melnik, P.~Neittaanm{\"a}ki, T.~Tiihonen, Coupled
  magneto-thermo-electromechanical effects and electronic properties of quantum
  dots, Journal of Computational and Theoretical Nanoscience 10~(3) (2013)
  534--547.

\bibitem{lassen2006}
B.~Lassen, M.~Willatzen, R.~Melnik, Inclusion of nonlinear strain effects in
  the hamiltonian for nanoscale semiconductor structures, Journal of
  Computational and Theoretical Nanoscience 3~(4) (2006) 588--597.

\bibitem{willatzen2006}
M.~Willatzen, B.~Lassen, L.~Lew Yan~Voon, R.~V. Melnik, Dynamic coupling of
  piezoelectric effects, spontaneous polarization, and strain in
  lattice-mismatched semiconductor quantum-well heterostructures, Journal of
  Applied Physics 100~(2) (2006).

\bibitem{lassen2005}
B.~Lassen, M.~Willatzen, R.~Melnik, L.~Yan~Voon, A general treatment of
  deformation effects in \mbox{Hamiltonians} for inhomogeneous crystalline
  materials, Journal of Mathematical Physics 46~(11) (2005).

\bibitem{lassen2005a}
B.~Lassen, R.~Melnik, M.~Willatzen, L.~L.~Y. Voon, Non-linear strain theory for
  low-dimensional semiconductor structures, Nonlinear Analysis: Theory, Methods
  \& Applications 63~(5-7) (2005) e1607--e1617.

\bibitem{prabhakar2015}
S.~Prabhakar, R.~Melnik, Relaxation of electron--hole spins in strained
  graphene nanoribbons, Journal of Physics: Condensed Matter 27~(43) (2015)
  435801.

\bibitem{prabhakar2016}
S.~Prabhakar, R.~Melnik, L.~Bonilla, Pseudospin lifetime in relaxed-shape
  armchair graphene nanoribbons due to in-plane phonon modes, Physical Review B
  93~(11) (2016) 115417.

\bibitem{prabhakar2017}
S.~Prabhakar, R.~Melnik, L.~Bonilla, Strain engineering of graphene
  nanoribbons: pseudomagnetic versus external magnetic fields, The European
  Physical Journal B 90 (2017) 1--8.

\bibitem{prabhakar2019}
S.~Prabhakar, R.~Melnik, Ab-initio calculations of strain induced relaxed shape
  armchair graphene nanoribbon, Physica E: Low-dimensional Systems and
  Nanostructures 114 (2019) 113648.

\bibitem{leon2022}
C.~Le{\'o}n, R.~Melnik, Machine learning for shape memory graphene nanoribbons
  and applications in biomedical engineering, Bioengineering 9~(3) (2022) 90.

\bibitem{prabhakar2014}
S.~Prabhakar, R.~Melnik, L.~L. Bonilla, S.~Badu, Thermoelectromechanical
  effects in relaxed-shape graphene and band structures of graphene quantum
  dots, Physical Review B 90~(20) (2014) 205418.

\bibitem{prabhakar2015a}
S.~Prabhakar, R.~Melnik, L.~L. Bonilla, S.~Badu, Relaxed shape of graphene
  sheet due to ripples, Quantum Matter 4~(4) (2015) 305--307.

\bibitem{prabhakar2013b}
S.~Prabhakar, R.~Melnik, L.~L. Bonilla, J.~E. Raynolds, Spin echo dynamics
  under an applied drift field in graphene nanoribbon superlattices, Applied
  Physics Letters 103~(23) (2013).

\bibitem{prabhakar2022}
S.~Prabhakar, R.~Melnik, First-principles calculations of electrical
  conductivities of edge-modified graphene nanoribbons: Strain effect, Physica
  E: Low-dimensional Systems and Nanostructures 142 (2022) 115267.

\bibitem{lovinger2008}
D.~M. Lovinger, Communication networks in the brain: neurons, receptors,
  neurotransmitters, and alcohol, Alcohol Research \& Health 31~(3) (2008) 196.

\bibitem{seguin2023}
C.~Seguin, O.~Sporns, A.~Zalesky, Brain network communication: concepts, models
  and applications, Nature Reviews Neuroscience 24~(9) (2023) 557--574.

\bibitem{papo2024}
D.~Papo, J.~Buld{\'u}, Does the brain behave like a (complex) network? i.
  dynamics, Physics of Life Reviews (2024).

\bibitem{heffernan2017}
R.~Heffernan, Y.~Yang, K.~Paliwal, Y.~Zhou, Capturing non-local interactions by
  long short-term memory bidirectional recurrent neural networks for improving
  prediction of protein secondary structure, backbone angles, contact numbers
  and solvent accessibility, Bioinformatics 33~(18) (2017) 2842--2849.

\bibitem{owolabi2023}
K.~M. Owolabi, S.~Jain, Spatial patterns through diffusion-driven instability
  in modified predator--prey models with chaotic behaviors, Chaos, Solitons \&
  Fractals 174 (2023) 113839.

\bibitem{obembe2020}
A.~D. Obembe, A fractional diffusion model for single-well simulation in
  geological media, Journal of Petroleum Science and Engineering 191 (2020)
  107162.

\bibitem{lang2013}
J.~M. Lang, M.~E. Benbow, Species interactions and competition, Nature
  Education Knowledge 4~(4) (2013) 8.

\bibitem{berrio2021}
L.~Berrio-Giraldo, C.~Villegas-Palacio, S.~Arango-Aramburo, Understating
  complex interactions in socio-ecological systems using system dynamics: a
  case in the tropical andes, Journal of Environmental Management 291 (2021)
  112675.

\bibitem{jungel2022}
A.~J{\"u}ngel, S.~Portisch, A.~Zurek, Nonlocal cross-diffusion systems for
  multi-species populations and networks, Nonlinear Analysis 219 (2022) 112800.

\bibitem{doumic2024}
M.~Doumic, S.~Hecht, B.~Perthame, D.~Peurichard, Multispecies cross-diffusions:
  from a nonlocal mean-field to a porous medium system without self-diffusion,
  Journal of Differential Equations 389 (2024) 228--256.

\bibitem{colombo2024}
E.~H. Colombo, R.~Martinez-Garcia, J.~M. Calabrese, C.~L{\'o}pez,
  E.~Hern{\'a}ndez-Garc{\'\i}a, Pulsed interactions unify reaction--diffusion
  and spatial nonlocal models for biological pattern formation, Journal of
  Statistical Mechanics: Theory and Experiment 2024~(3) (2024) 034001.

\bibitem{venturino2024}
E.~Venturino, S.~Ani{\c{t}}a, D.~Mezzanotte, D.~Occorsio, A high order
  numerical scheme for a nonlinear nonlocal reaction--diffusion model arising
  in population theory, Journal of Computational and Applied Mathematics (2024)
  116082.

\bibitem{berestycki2024}
H.~Berestycki, J.-M. Roquejoffre, L.~Rossi, Biological invasions and epidemics
  with nonlocal diffusion along a line, Mathematical Medicine and Biology: A
  Journal of the IMA (2024) dqae014.

\bibitem{zhang2008}
S.~Zhang, W.~Dai, H.~Wang, R.~V. Melnik, A finite difference method for
  studying thermal deformation in a 3d thin film exposed to ultrashort pulsed
  lasers, International journal of heat and mass transfer 51~(7-8) (2008)
  1979--1995.

\bibitem{wang2006a}
H.~Wang, W.~Dai, R.~Melnik, A finite difference method for studying thermal
  deformation in a double-layered thin film exposed to ultrashort pulsed
  lasers, International Journal of Thermal Sciences 45~(12) (2006) 1179--1196.

\bibitem{wang2006b}
H.~Wang, W.~Dai, R.~Nassar, R.~Melnik, A finite difference method for studying
  thermal deformation in a thin film exposed to ultrashort-pulsed lasers,
  International Journal of Heat and Mass Transfer 49~(15-16) (2006) 2712--2723.

\bibitem{singh2021}
S.~Singh, R.~Melnik, Fluid--structure interaction and \mbox{non-Fourier}
  effects in coupled electro-thermo-mechanical models for cardiac ablation,
  Fluids 6~(8) (2021) 294.

\bibitem{singh2022b}
S.~Singh, P.~Saccomandi, R.~Melnik, Three-phase-lag bio-heat transfer model of
  cardiac ablation, Fluids 7~(5) (2022) 180.

\bibitem{singh2020d}
S.~Singh, R.~Melnik, Computational modeling of cardiac ablation incorporating
  electrothermomechanical interactions, Journal of Engineering and Science in
  Medical Diagnostics and Therapy 3~(4) (2020) 041004.

\bibitem{singh2020e}
S.~Singh, R.~Melnik, Computational model of radiofrequency ablation of cardiac
  tissues incorporating thermo-electro-mechanical interactions, in: ASME
  International Mechanical Engineering Congress and Exposition, Vol. 84522,
  American Society of Mechanical Engineers, 2020, p. V005T05A029.

\bibitem{singh2019}
S.~Singh, R.~Melnik, Effects of heterogeneous surroundings on the efficacy of
  continuous radiofrequency for pain relief, in: Proceedings of the 3rd
  International Conference on Vision, Image and Signal Processing, 2019, pp.
  1--5.

\bibitem{singh2019a}
S.~Singh, R.~Melnik, Radiofrequency ablation for treating chronic pain of
  bones: Effects of nerve locations, in: Bioinformatics and Biomedical
  Engineering: 7th International Work-Conference, IWBBIO 2019, Granada, Spain,
  May 8-10, 2019, Proceedings, Part II 7, Springer, 2019, pp. 418--429.

\bibitem{singh2019b}
S.~Singh, R.~Melnik, Computational analysis of pulsed radiofrequency ablation
  in treating chronic pain, in: Computational Science--ICCS 2019: 19th
  International Conference, Faro, Portugal, June 12--14, 2019, Proceedings,
  Part IV 19, Springer, 2019, pp. 436--450.

\bibitem{singh2020f}
S.~Singh, R.~Melnik, Domain heterogeneity in radiofrequency therapies for pain
  relief: a computational study with coupled models, Bioengineering 7~(2)
  (2020) 35.

\bibitem{dayan2005}
P.~Dayan, L.~F. Abbott, Theoretical neuroscience: computational and
  mathematical modeling of neural systems, MIT press, 2005.

\bibitem{shaheen2022}
H.~Shaheen, S.~Pal, R.~Melnik, Multiscale co-simulation of deep brain
  stimulation with brain networks in neurodegenerative disorders, Brain
  Multiphysics 3 (2022) 100058.

\bibitem{shaheen2023}
H.~Shaheen, S.~Pal, R.~Melnik, Astrocytic clearance and fragmentation of toxic
  proteins in alzheimer’s disease on large-scale brain networks, Physica D:
  Nonlinear Phenomena 454 (2023) 133839.

\bibitem{sevimoglu2014}
T.~Sevimoglu, K.~Y. Arga, The role of protein interaction networks in systems
  biomedicine, Computational and structural biotechnology journal 11~(18)
  (2014) 22--27.

\bibitem{lee2022}
M.~D. Lee, C.~Buckley, X.~Zhang, L.~Louhivuori, P.~Uhl{\'e}n, C.~Wilson, J.~G.
  McCarron, Small-world connectivity dictates collective endothelial cell
  signaling, Proceedings of the National Academy of Sciences 119~(18) (2022)
  e2118927119.

\bibitem{falco2021}
J.~Falco, A.~Agosti, I.~G. Vetrano, A.~Bizzi, F.~Restelli, M.~Broggi,
  M.~Schiariti, F.~DiMeco, P.~Ferroli, P.~Ciarletta, et~al., In silico
  mathematical modelling for glioblastoma: a critical review and a
  patient-specific case, Journal of Clinical Medicine 10~(10) (2021) 2169.

\bibitem{melnik2008}
R.~V. Melnik, Markov chain network training and conservation law
  approximations: Linking microscopic and macroscopic models for evolution,
  Applied Mathematics and Computation 199~(1) (2008) 315--333.

\bibitem{newman2018}
M.~Newman, Networks, Oxford university press, 2018.

\bibitem{dankulov2015}
M.~M. Dankulov, R.~Melnik, B.~Tadi{\'c}, The dynamics of meaningful social
  interactions and the emergence of collective knowledge, Scientific Reports
  5~(1) (2015) 12197.

\bibitem{andjelkovic2016}
M.~Andjelkovi{\'c}, B.~Tadi{\'c}, M.~Mitrovi{\'c}~Dankulov, M.~Rajkovi{\'c},
  R.~Melnik, Topology of innovation spaces in the knowledge networks emerging
  through questions-and-answers, PloS One 11~(5) (2016) e0154655.

\bibitem{christakis2007}
N.~A. Christakis, J.~H. Fowler, The spread of obesity in a large social network
  over 32 years, New England Journal of Medicine 357~(4) (2007) 370--379.

\bibitem{tadic2021}
B.~Tadi{\'c}, R.~Melnik, Microscopic dynamics modeling unravels the role of
  asymptomatic virus carriers in \mbox{SARS-CoV-2} epidemics at the interplay
  between biological and social factors, Computers in Biology and Medicine 133
  (2021) 104422.

\bibitem{tadic2020}
B.~Tadi{\'c}, R.~Melnik, Modeling latent infection transmissions through
  biosocial stochastic dynamics, PloS one 15~(10) (2020) e0241163.

\bibitem{mitrovic2022}
M.~Mitrovi{\'c}~Dankulov, B.~Tadi{\'c}, R.~Melnik, Analysis of worldwide
  time-series data reveals some universal patterns of evolution of the
  \mbox{SARS-CoV-2} pandemic, Frontiers in Physics 10 (2022) 936618.

\bibitem{tadic2019}
B.~Tadi{\'c}, M.~Andjelkovi{\'c}, R.~Melnik, Functional geometry of human
  connectomes, Scientific Reports 9~(1) (2019) 12060.

\bibitem{andjelkovic2020}
M.~Andjelkovi{\'c}, B.~Tadi{\'c}, R.~Melnik, The topology of higher-order
  complexes associated with brain hubs in human connectomes, Scientific Reports
  10~(1) (2020) 17320.

\bibitem{kramer2014}
A.~D. Kramer, J.~E. Guillory, J.~T. Hancock, Experimental evidence of
  massive-scale emotional contagion through social networks, Proceedings of the
  National academy of Sciences of the United States of America 111~(24) (2014)
  8788.

\bibitem{sharma2018}
M.~Sharma, S.~P. Khurana, Biomedical engineering: The recent trends, in: Omics
  Technologies and Bio-Engineering, Elsevier, 2018, pp. 323--336.

\bibitem{el2022}
R.~A. El-Nabulsi, W.~Anukool, Nonlocal thermal effects on biological tissues
  and tumors, Thermal Science and Engineering Progress 34 (2022) 101424.

\bibitem{zhang2023}
Y.~Zhang, A.~Cloninger, B.~Li, X.~Tian, A neural network kernel decomposition
  for learning multiple steady states in parameterized dynamical systems, arXiv
  preprint arXiv:2312.10315 (2023).

\bibitem{alber2019}
M.~Alber, A.~Buganza~Tepole, W.~R. Cannon, S.~De, S.~Dura-Bernal,
  K.~Garikipati, G.~Karniadakis, W.~W. Lytton, P.~Perdikaris, L.~Petzold,
  et~al., Integrating machine learning and multiscale modeling—perspectives,
  challenges, and opportunities in the biological, biomedical, and behavioral
  sciences, NPJ Digital Medicine 2~(1) (2019) 115.

\bibitem{ghosh2020}
A.~Ghosh, W.~Xu, N.~Gupta, D.~H. Gracias, Active matter therapeutics, Nano
  Today 31 (2020) 100836.

\bibitem{gompper2020}
G.~Gompper, R.~G. Winkler, T.~Speck, A.~Solon, C.~Nardini, F.~Peruani,
  H.~L{\"o}wen, R.~Golestanian, U.~B. Kaupp, L.~Alvarez, et~al., The 2020
  motile active matter roadmap, Journal of Physics: Condensed Matter 32~(19)
  (2020) 193001.

\bibitem{vicsek1995}
T.~Vicsek, A.~Czir{\'o}k, E.~Ben-Jacob, I.~Cohen, O.~Shochet, Novel type of
  phase transition in a system of self-driven particles, Physical Review
  Letters 75~(6) (1995) 1226.

\bibitem{paperin2011}
G.~Paperin, D.~G. Green, S.~Sadedin, Dual-phase evolution in complex adaptive
  systems, Journal of the Royal Society Interface 8~(58) (2011) 609--629.

\bibitem{hubsch2023}
T.~H{\"u}bsch, D.~Minic, K.~Nikolic, S.~Pajevic, On the emergent “quantum”
  theory in complex adaptive systems, Annals of Physics 464 (2024) 169641.

\bibitem{gear2003}
C.~W. Gear, J.~M. Hyman, P.~G. Kevrekidid, I.~G. Kevrekidis, O.~Runborg,
  C.~Theodoropoulos, Equation-free, coarse-grained multiscale computation:
  Enabling mocroscopic simulators to perform system-level analysis,
  Communications in Mathematical Sciences 1~(4) (2003) 715 -- 762.

\bibitem{strunin2000}
D.~Strunin, R.~Melnik, A.~Roberts, Numerical modelling of thermoelastic
  processes using nonlinear theories with thermal relaxation time, ANZIAM
  Journal 42 (2000) C1356--C1378.

\bibitem{strunin2001}
D.~Strunin, R.~Melnik, A.~Roberts, Coupled thermomechanical waves in hyperbolic
  thermoelasticity, Journal of Thermal Stresses 24~(2) (2001) 121--140.

\bibitem{hillen2002}
T.~Hillen, Hyperbolic models for chemosensitive movement, Mathematical Models
  and Methods in Applied Sciences 12~(07) (2002) 1007--1034.

\bibitem{melnik2002a}
R.~Melnik, A.~Roberts, K.~Thomas, Phase transitions in shape memory alloys with
  hyperbolic heat conduction and differential-algebraic models, Computational
  Mechanics 29 (2002) 16--26.

\bibitem{singh2019d}
S.~Singh, R.~Melnik, Coupled thermo-electro-mechanical models for thermal
  ablation of biological tissues and heat relaxation time effects, Physics in
  Medicine \& Biology 64~(24) (2019) 245008.

\bibitem{liang2019}
F.~Liang, X.~Yang, W.~Zhang, Y.~Qian, R.~Melnik, et~al., Parametric vibration
  analysis of pipes conveying fluid by nonlinear normal modes and a numerical
  iterative approach, Advances in Applied Mathematics and Mechanics 11~(1)
  (2019) 38--52.

\bibitem{nilsson1998}
C.~Nilsson, On nonlocal rate-independent plasticity, International Journal of
  Plasticity 14~(6) (1998) 551--575.

\bibitem{shizawa1999}
K.~Shizawa, H.~Zbib, A thermodynamical theory of gradient elastoplasticity with
  dislocation density tensor. i: Fundamentals, International Journal of
  Plasticity 15~(9) (1999) 899--938.

\bibitem{lion2000}
A.~Lion, Constitutive modelling in finite thermoviscoplasticity: a physical
  approach based on nonlinear rheological models, International Journal of
  Plasticity 16~(5) (2000) 469--494.

\bibitem{provatas2011}
N.~Provatas, K.~Elder, Phase-field methods in materials science and
  engineering, John Wiley \& Sons, 2011.

\bibitem{dhote2014}
R.~Dhote, M.~Fabrizio, R.~Melnik, J.~Zu, A three-dimensional non-isothermal
  \mbox{Ginzburg--Landau} phase-field model for shape memory alloys, Modelling
  and Simulation in Materials Science and Engineering 22~(8) (2014) 085011.

\bibitem{dhote2014a}
R.~Dhote, H.~Gomez, R.~Melnik, J.~Zu, Isogeometric analysis of a dynamic
  thermo-mechanical phase-field model applied to shape memory alloys,
  Computational Mechanics 53 (2014) 1235--1250.

\bibitem{dhote2014b}
R.~Dhote, R.~Melnik, J.~Zu, Dynamic multi-axial behavior of shape memory alloy
  nanowires with coupled thermo-mechanical phase-field models, Meccanica 49
  (2014) 1561--1575.

\bibitem{wang2017a}
D.~Wang, L.~Wang, R.~Melnik, Vibration energy harvesting based on
  stress-induced polarization switching: a phase field approach, Smart
  Materials and Structures 26~(6) (2017) 065022.

\bibitem{wang2018a}
D.~Wang, H.~Du, L.~Wang, R.~Melnik, A phase field approach for the fully
  coupled thermo-electro-mechanical dynamics of nanoscale ferroelectric
  actuators, Smart Materials and Structures 27~(5) (2018) 055012.

\bibitem{scholl2001}
E.~Sch{\"o}ll, Nonlinear spatio-temporal dynamics and chaos in semiconductors,
  10, Cambridge University Press, 2001.

\bibitem{konieczny2023}
L.~Konieczny, I.~Roterman-Konieczna, P.~Sp{\'o}lnik, Regulation in biological
  systems, in: Systems Biology: Functional Strategies of Living Organisms,
  Springer, 2023, pp. 159--203.

\bibitem{sun2017}
H.~Sun, Z.~Li, Y.~Zhang, W.~Chen, Fractional and fractal derivative models for
  transient anomalous diffusion: Model comparison, Chaos, Solitons \& Fractals
  102 (2017) 346--353.

\bibitem{scheidweiler2020}
D.~Scheidweiler, F.~Miele, H.~Peter, T.~J. Battin, P.~de~Anna, Trait-specific
  dispersal of bacteria in heterogeneous porous environments: from pore to
  porous medium scale, Journal of The Royal Society Interface 17~(164) (2020)
  20200046.

\bibitem{buttenschon2020}
A.~Buttensch{\"o}n, L.~Edelstein-Keshet, Bridging from single to collective
  cell migration: A review of models and links to experiments, PLoS
  Computational Biology 16~(12) (2020) e1008411.

\bibitem{li2021}
J.~Li, T.~Xiang, L.~He, Modeling epidemic spread in transportation networks: A
  review, Journal of Traffic and Transportation Engineering (English Edition)
  8~(2) (2021) 139--152.

\bibitem{senk2022}
J.~Senk, B.~Kriener, M.~Djurfeldt, N.~Voges, H.-J. Jiang, L.~Sch{\"u}ttler,
  G.~Gramelsberger, M.~Diesmann, H.~E. Plesser, S.~J. van Albada, Connectivity
  concepts in neuronal network modeling, PLoS Computational Biology 18~(9)
  (2022) e1010086.

\bibitem{drapaca2016}
C.~Drapaca, Fractional calculus in neuronal electromechanics, Journal of
  Mechanics of Materials and Structures 12~(1) (2016) 35--55.

\bibitem{vascon2020}
F.~Vascon, M.~Gasparotto, M.~Giacomello, L.~Cendron, E.~Bergantino,
  F.~Filippini, I.~Righetto, Protein electrostatics: From computational and
  structural analysis to discovery of functional fingerprints and
  biotechnological design, Computational and Structural Biotechnology Journal
  18 (2020) 1774--1789.

\bibitem{li2004}
J.~Li, J.~Zhang, W.~Ge, X.~Liu, Multi-scale methodology for complex systems,
  Chemical Engineering Science 59~(8-9) (2004) 1687--1700.

\bibitem{ramstead2019}
M.~J. Ramstead, A.~Constant, P.~B. Badcock, K.~J. Friston, Variational ecology
  and the physics of sentient systems, Physics of Life Reviews 31 (2019)
  188--205.

\bibitem{giardina2008}
I.~Giardina, Collective behavior in animal groups: theoretical models and
  empirical studies, HFSP Journal 2~(4) (2008) 205--219.

\bibitem{du2019}
Q.~Du, Nonlocal Modeling, Analysis, and Computation: Nonlocal Modeling,
  Analysis, and Computation, SIAM, 2019.

\bibitem{you2022}
H.~You, Y.~Yu, Nonlocal models for complex physical responses: Analysis and
  applications, Ph.D. thesis, Lehigh University (2022).

\end{thebibliography}
% \printbibliography

\end{document}